\documentclass[prl,aps, superscriptaddress,twocolumn]{revtex4-2}
\usepackage{amsmath,amssymb}
\usepackage{graphicx}
\usepackage{multirow}
\usepackage{wasysym}
\usepackage{amsfonts}
\usepackage{amsthm}
\usepackage{bm}
\usepackage{enumerate}
\usepackage{enumitem}
\usepackage{color}
\usepackage[resetlabels]{multibib}
\usepackage{epstopdf}
\usepackage{latexsym}
\usepackage[breaklinks,colorlinks = true,linkcolor = red,urlcolor=cyan,citecolor=red]{hyperref}
\usepackage[caption=false,singlelinecheck=false]{subfig}
\usepackage[normalem]{ulem}
\usepackage{booktabs}
\usepackage{mathtools}

\usepackage{tikz}
\usetikzlibrary{fit,matrix}
\usepackage{multirow}

\usepackage{times}
\newcommand{\bea}{\begin{eqnarray}}
\newcommand{\eea}{\end{eqnarray}}
\newcommand{\be}{\begin{eqnarray}}
\newcommand{\ee}{\end{eqnarray}}
\newcommand{\bw}{\begin{widetext}}
\newcommand{\ew}{\end{widetext}}

\newcommand{\rcv}{
\begin{tikzpicture}[baseline=-0.5*\rd]
\draw[line width=0.5mm, red] (0,0) circle (1*\rd);
\end{tikzpicture}
}

\newcommand{\rcvv}[1]{
\begin{tikzpicture}[baseline=-0.5*\rd]
\draw[line width=0.5mm, red] (0,0) circle (1*\rd);\draw (0,0) node {\footnotesize $#1$};
\end{tikzpicture}
}

\newcommand{\bsv}{
\begin{tikzpicture}[baseline=-0.5*\rd]
\draw[line width=0.5mm, blue] (-1*\rd,-1*\rd) rectangle ++(2*\rd,2*\rd);\draw[line width=0.5mm, blue] (-1*\rd,-1*\rd) rectangle ++(2*\rd,2*\rd);\draw (0,0) node {\footnotesize $0$};
\end{tikzpicture}
}

\newcommand{\bsvnz}{
\begin{tikzpicture}[baseline=-0.5*\rd]
\draw[line width=0.5mm, blue] (-1*\rd,-1*\rd) rectangle ++(2*\rd,2*\rd);\draw[line width=0.5mm, blue] (-1*\rd,-1*\rd) rectangle ++(2*\rd,2*\rd);
\end{tikzpicture}
}

\def\rd{0.425em}

\newsavebox{\measurebox}

\DeclareMathOperator{\len}{len}

\begin{document}

\title{Graph theoretical proof of nonintegrability in quantum many-body systems :
 Application to the PXP model}
\author{HaRu K. Park}
\email{haru.k.park@kaist.ac.kr}
\affiliation{Department of Physics, Korea Advanced Institute of Science and Technology, Daejeon, 34141, Korea}
\author{SungBin Lee}
\email{sungbin@kaist.ac.kr}
\affiliation{Department of Physics, Korea Advanced Institute of Science and Technology, Daejeon, 34141, Korea}
\date{\today}

\begin{abstract}
A rigorous proof of integrability or non-integrability in quantum many-body systems is among the most challenging tasks, as it involves demonstrating the presence or absence of local conserved quantities and deciphering the complex dynamics of the system. In this paper, we establish a graph-theoretical analysis as a comprehensive framework for proving the non-integrability of quantum systems. Exemplifying the PXP model, which is widely believed to be non-integrable,  this work rigorously proves the absence of local conserved quantities, thereby confirming its non-integrability. 
This proof for the PXP model gives several important messages not only that the system is non-integrable, but also the quantum many body scaring observed in the model is not associate with the existence of local conserved quantities. 
From a graph-theoretical perspective, we also highlight its advantage, even in integrable systems, as the classification of local conserved quantities can be achieved by simply counting the number of isolated loops in the graphs.
 Our new approach is broadly applicable for establishing proofs of (non-)integrability in other quantum many-body systems, significantly simplifying the process of proving nonintegrability and giving numerous potential applications.
\end{abstract}

\vspace{\baselineskip}

\maketitle

\textit{Introduction} -- Integrability is characterized by the presence of infinitely many local conserved quantities that fully determine a system's dynamics, thereby preventing quantum thermalization\cite{PhysRevLett.98.050405,pozsgay2013generalized,vidmar2016generalized,PhysRevLett.45.379,zamolodchikov1978relativistic,PhysRevD.11.2088,faddeev1982integrable,zhang2005yang}.  Various methods\cite{bethe1931theorie, yang1967some,baxter1978solvable,baxter2016exactly, takhtadzhan1979quantum} have been used to rigorously demonstrate the integrability of several quantum systems\cite{baxter1972one,takhtadzhan1979quantum,nozawa2020explicit, PhysRevLett.56.2453,PhysRevLett.131.256704, PhysRevLett.98.160409}. 
However, there are limited studies and developed methods focused on non-integrability, specifically for showing the absence of local conserved quantities. 
The most general approaches for claiming the absence of local conserved quantities involve investigating level statistics \cite{PhysRevLett.54.1350} or demonstrating the absence of three-site support conserved quantities \cite{M_P_Grabowski_1995}. However, since these methods are based on conjectures, their conclusions are sometimes subject to debate.
 Only recently have rigorous proofs of non-integrability in spin-$1/2$ models begun to be explored\cite{shiraishi2019proof,PhysRevB.109.035123}. 

Despite the lack of rigorous proof for non-integrability, the demand for such proofs is growing, particularly for systems exhibiting exotic dynamical features. A prominent example is the Quantum Many-Body Scar (QMBS) system, which typically thermalizes for most initial states but exhibits unusual short-time revivals of fidelity when the initial state is a special product state.\cite{alhambra2020revivals}. One of the pioneering models exhibiting QMBS is the PXP model\cite{turner2018weak}. Numerous theoretical studies have investigated modified versions of the PXP model\cite{PhysRevB.101.165139,bluvstein2021controlling,PhysRevResearch.5.023010,desaules2024ergodicity} and other distinct models\cite{PhysRevLett.123.147201,PhysRevB.101.195131,PhysRevB.101.024306} as potential QMBS systems. Thus, it is crucial to explore the nonintegrability of such systems, to determine if local conserved quantities are responsible for QMBS.

The contributions of this letter are twofold. First, we rigorously prove that the exotic dynamics of the PXP model
are not associated with local conserved quantities by showing their absence. To the best of our knowledge, this is the first rigorous proof of the non-integrability of a QMBS model. 
Second, we introduce a novel graph theoretical approach that not only simplifies the proof of nonintegrability but is also applicable for identifying local conserved quantities in integrable models.

\medskip
\textit{Model and Notations.} --- The Hamiltonian of 1-dimensional PXP model is following.
\begin{equation}\label{eq:pxp}
H=\sum_j P_j X_{j+1} P_{j+2}.
\end{equation}
Here $P\! \coloneq\!(I-Z)/2$ is a projection operator onto the ground state, and $X,Z$ are standard Pauli matrices. We assume periodic boundary condition, i.e., $A_{L+j}=A_j$ for any operator $A$. This model is associated with the Rydberg atom chain model, where an atom excited to a Rydberg state prevents its neighboring atoms from being excited due to a strong dipole interaction, a phenomenon known as the Rydberg blockade. One can verify that the subspace consisting of states with no two consecutive sites in the excited state is invariant under the action of $H$. Since we are interested in the dynamics within this constrained subspace, we refer to a quantity $C$ as \textit{trivial} when it vanishes within this subspace.

To discuss local conserved quantities, we first need to define what it means for an operator to be local. An operator $C$ is said to have \textit{length $l$}, denoted $\len (C)=l$, if it can be expressed as a sum of terms, $C=\sum_\alpha c_\alpha$, where each term $c_\alpha$ acts on at most $l$ consecutive sites. For example, the Hamiltonian $H$ has $\len (H)=3$. An operator $C$ is considered \textit{local} if $\len (C)<L/2$, where $L$ is the system size. This definition is reasonable because when $\len (C)\geq L/2$, the operator acts on a substantial portion of the system, losing its local nature.\cite{shiraishi2019proof,PhysRevB.109.035123}

\medskip
\textit{Main Result.} --- We prove that \textit{if $C$ is a local quantity of length $l\geq 4$ satisfying $[H,C]=0$, then $C$ is trivial.} Conserved quantities of length less than $4$ are excluded, as these correspond to insignificant cases, such as the Hamiltonian itself. This rigorous proof of the nonintegrability of the PXP model demonstrates that no local conserved quantity governs its dynamics and offers key insights that the QMBS phenomenon in this model is not associated with local conserved quantities.

\medskip
\textit{Strategy.} --- We begin by focusing on local quantities that exhibit translational invariance\footnote{In \cite{sm}, we show that our arguments extend to operators that are not translationally invariant.}. A general translationally invariant local operator of length $k$ can be expressed as,
\begin{equation}\label{eq:general_form}
C=\sum_{l=1}^k \sum_{\bm{A^l}} \sum_{j=1}^L q(\bm{A}^l)\left\{\bm{A}^l\right\}_j,
\end{equation}
where $q(\bm{A}^l)$ are coefficients corresponding to each Pauli string $\bm{A}^l$ of length $l$. Here, $\bm{A}^l$, referred to as a \textit{Pauli string of length $l$}, denotes a sequence of $l$ operators consisting of Pauli matrices and identity matrices, where the sequence neither starts nor ends with an identity matrix. The notation $\left\{ \bm{A}^l \right\}_j$ indicates that the first operator in the Pauli string $\bm{A}^l$ acts on the $j$-th site of the system, the second operator acts on the $j+1$-th site, and so on. The site index is omitted when it is either clear from context or not significant.

It is useful to represent $H$ in terms of Pauli strings: $H=\frac{1}{4}\sum_j \{ZXZ\}_j - \{XZ\}_j -\{ZX\}_j + \{X\}_j$. Next, consider the commutator between  $H$ and $C$ with $\len(C)=k$. Since both $C$ and $H$ are translationally invariant, the commutator $[H,C]$ also inherits this property and can be expressed as: 
\begin{equation}\label{eq:commutator}
[H,C]=\sum_{l=1}^{k+2} \sum_{\bm{B^l}} \sum_{j=1}^L p(\bm{B}^l)\left\{\bm{B}^l\right\}_j,
\end{equation}
where each $B^l$ represents a Pauli string of length at most $k+2$ and $p(\bm{B}^l)$ represents its coefficient. This is because the commutator of a Pauli string of length $k$ from $C$ and a Pauli string of length $3$ from $H$, will result in a Pauli string of length at most $k+2$.

Our goal is to show there is no local quantity $C$ that satisfies $[H,C]=0$, i.e., no  local conserved quantity. This is equivalent to find $q(\bm{A}^l) =0$ for all $\bm{A}^l$ under the condition $p(\bm{B}^l)=0$ for any $\bm{B}^l$. By substituting Eq.\eqref{eq:general_form} into Eq.\eqref{eq:commutator} and setting $p(\bm{B}^l)=0$, we obtain a system of linear equations involving the parameters $q(\bm{A}^l)$. Then the solution determines the values of $q(\bm{A}^l)$.

The main drawback of this brute-force approach is the overwhelming time complexity. Consider Pauli strings of length $k$ in $C$, denoted as $\bm{A}^k$. Then, one should consider $\mathcal{O}(4^k)$ independent coefficients $q_{\bm{A}^k}$.
 Solving a system of linear equations with $n$ parameters and $n$ equations using Gaussian elimination requires $\mathcal{O}(n^3)$ time complexity\cite{chapra1987numerical}, which means that the total time complexity of our problem is at least $\mathcal{O}(64^k)$. As $k$ increases, this computational burden grows exponentially.
A key insight of our proof is that this complexity can be dramatically reduced to a constant $\mathcal{O}(1)$, independent to $k$, by introducing a graph theoretical approach.

\medskip
\textit{Graph Theoretical Approach.} --- In this section we define how to transform the map $C\mapsto [H,C]$ into a graph structure, which we refer to as a \textit{commutator graph} of $H$. We also introduce three special subgraph structures, namely the ''promising path", ''loop" and ''quasi-promising path".  It is important to note that these concepts are applicable to general quantum many-body systems.

The \textit{commutator graph} of $H$ for quantities of length $k$ is defined as follows. The Pauli strings $\bm{A}^l$ in Eq.\eqref{eq:general_form} are represented by red circled vertices \rcv, while $\bm{B}^m$ from $[H,C]$  in Eq.\eqref{eq:commutator} are depicted as blue squared vertices $\bsvnz$. An edge connects $\rcv$ to $\bsvnz$ for a vertex $\bm{A}^l$ and a  vertex $\bm{B}^m$, if the Pauli string $\bm{B}^m$ appears from the commutator $[H,\bm{A}^l]$ with a nonzero coefficient. The edge is labeled with a weight corresponding to this coefficient.

Once the graph is constructed, the linear equations relating the coefficients of each Pauli string, represented by the vertices, can be derived as follows. First, $p(\bm{B}^l)=0$ for all $\bm{B}^l$ due to $[H,C]=0$, is indicated as $0$ inside each blue vertex $\bsvnz$, i.e. $\bsv$. Additionally, for every $\bsv$, the sum of the coefficients of its neighboring $\rcv$ vertices, multiplied by the weight of their respective edges, must also be zero. Consequently, the commutator graph provides sufficient information to identify the local conserved quantity of length $k$.

\begin{figure}[h!]
\captionsetup{justification=centering}
\centering
\includegraphics[width=1\linewidth]{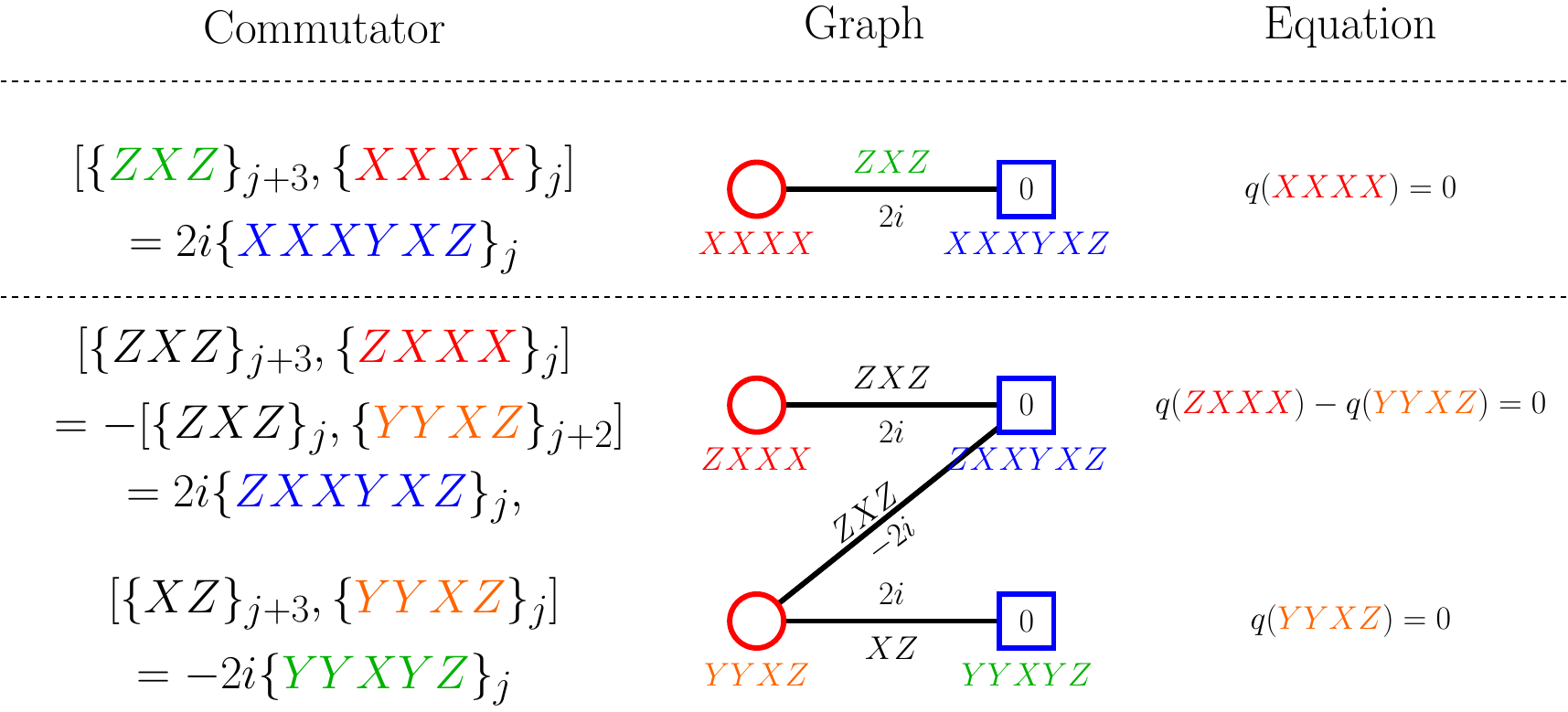}
\caption{Illustration of commutators, their corresponding subgraphs, and linear equations of coefficients of Pauli strings. 
(Left) Examples of commutators between Pauli strings from $H$ and Pauli strings of $\bm{A}^l$. (Middle) Subgraphs of each commutator where $\bm{A}^l$ is represented as $\protect\rcv$  and $\bm{B}^m$ as $\protect\bsv$, with solid lines corresponding the Pauli strings from $H$. $\pm 2i$ are the coefficients from each commutator. (Right) Linear equations of the corresponding coefficients. }
  \label{fig:graph_examples}
\end{figure}

Fig.\ref{fig:graph_examples} illustrates examples of 
(left) commutators between an operator string from $H$ and a local operator string $\bm{A}^l$ of length $4$, 
(middle) their corresponding subgraphs, where $\bm{A}^l$ is represented as $\rcv$ and $\bm{B}^m$ as $\bsv$, and
(right) the relationship between $q(\bm{A}^l)$ assuming $q( \bm{B}^m)=0$. 
For example, two Pauli strings, $ZXXX$ and $YYXZ$, are related to the string $ZXXYXZ$, through their commutator with $H$. The two $\rcv$'s, representing each Pauli string, are connected to $\bsv$, resulting in $q(ZXXX) - q(YYXZ) = 0$. Solving all these equations results in the coefficients of every $\rcv$ vertex to be zero, $q(XXXX)=q(ZXXX) =  q(YYXZ) =0$. Consequently, every $\rcv$ vertex in the subgraphs of Fig. \ref{fig:graph_examples} has a zero coefficient.
The key characteristic of these subgraphs, shown in the middle of Fig.\ref{fig:graph_examples}, is that among $n$ number of $\bsv$ vertices present in the subgraph, all $n-1$ number of $\bsv$ vertices has two neighboring $\rcv$, and only one of them has a single neighbor $\rcv$. Motivated by this observation, if a subgraph satisfies this property, then we refer to it as a \textit{promising path.}\footnote{Note that, independent to our work, the primitive concept of our promising path approach has been appeared in Ref.\cite{fukai2024study,fukai2024proof}, where the concept is systemized and applicable to general Hamiltonians in our work.} Later, we will show that every $\rcv$ vertex in a promising path has zero coefficients.

\begin{figure}[h!]
\captionsetup{justification=centering}
\centering
\includegraphics[width=1\linewidth]{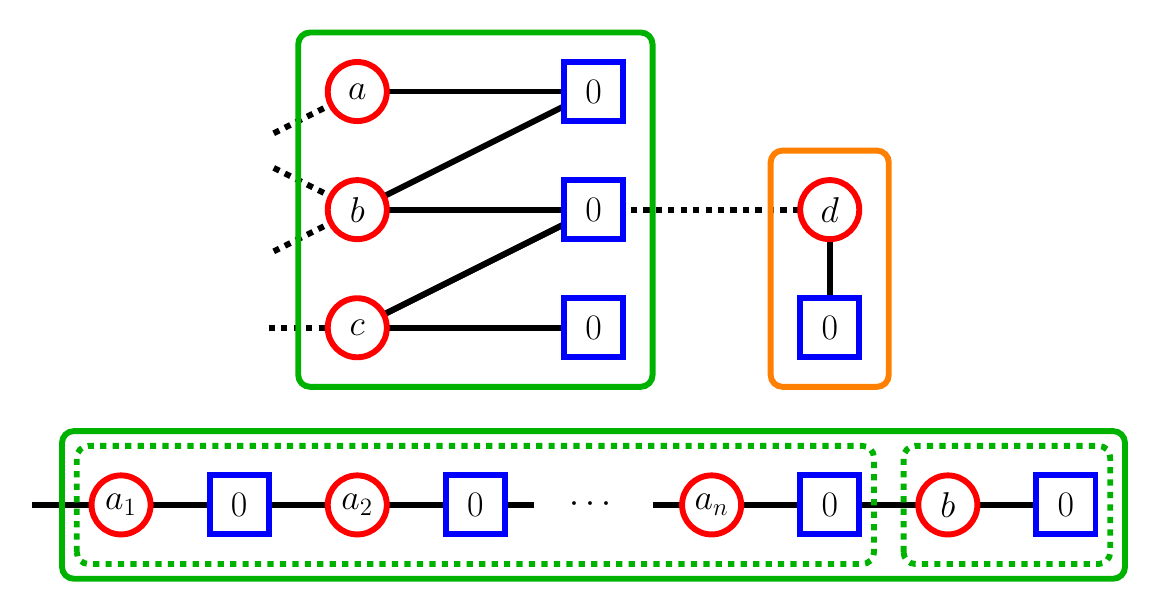}
\caption{Promising path subgraphs. The presence of additional neighboring red circled vertex does not change the fact that the subgraph outlined in green is a promising path. In the top-left green box, the subgraph is a promising path because the coefficient of the red circled vertex labeled $d$ is zero, allowing us to disregard the dashed line in the subgraph structure. The lower part of the figure demonstsrates that, for a promising path, all coefficients must be zero.}
\label{fig:ppath}
\end{figure}

Fig.\ref{fig:ppath} illustrates examples of promising paths. 
The subgraph inside the orange box is clearly a promising path, resulting in the coefficient of $\rcv$ indicated as $d$ to be 0. Therefore $\rcvv{d}$ with $d=0$ is not considered as a neighbor of $\bsv$, and is thus represented as a dashed line. 
Consequently, for the $\bsv$ vertex in the middle of the left green box, only the two neighbors labeled $\rcvv{b}$ and $\rcvv{c}$ need to be considered ignoring $\rcvv{d}$, and it makes the subgraph in the left green box a promising path as well.

The following important lemma holds: \textit{All coefficients of $\rcv$ vertices in a promising path must be zero.} We prove this by induction on the number $n$ of $\rcv$ vertices in the promising path. The base case, $n=1$, is straightforward, as shown in the first example in Fig.\ref{fig:graph_examples}. Consider the  subgraph of promising path in the lower part of Fig.\ref{fig:ppath}, which contains $n+1$ number of $\rcv$ vertices, labeled as $a_1,a_2, ... , a_n$ and $b$. The subgraph in the right dashed box gives $b=0$. Consequently, the subgraph in the left dashed box forms a promising path with $n$ number of $\rcv$ vertices. By the induction hypothesis, all coefficients $a_j$ must be zero.

This lemma is simple yet powerful, since if one simply identifies a promising path, all $\rcv$ in the path becomes zero. Furthermore, even if a path includes $\bsv$ vertices that are connected by three or more neighboring $\rcv$ vertices, (one example is shown in the upper part of Fig.\ref{fig:ppath}, where one $\bsv$ is connected by $\rcvv{b}, \rcvv{c}$, and $\rcvv{d}$), the path remains a promising path as long as all the branches also form other promising paths.

\begin{figure}[h!]
\captionsetup{justification=centering}
\centering
\includegraphics[width=0.8\linewidth]{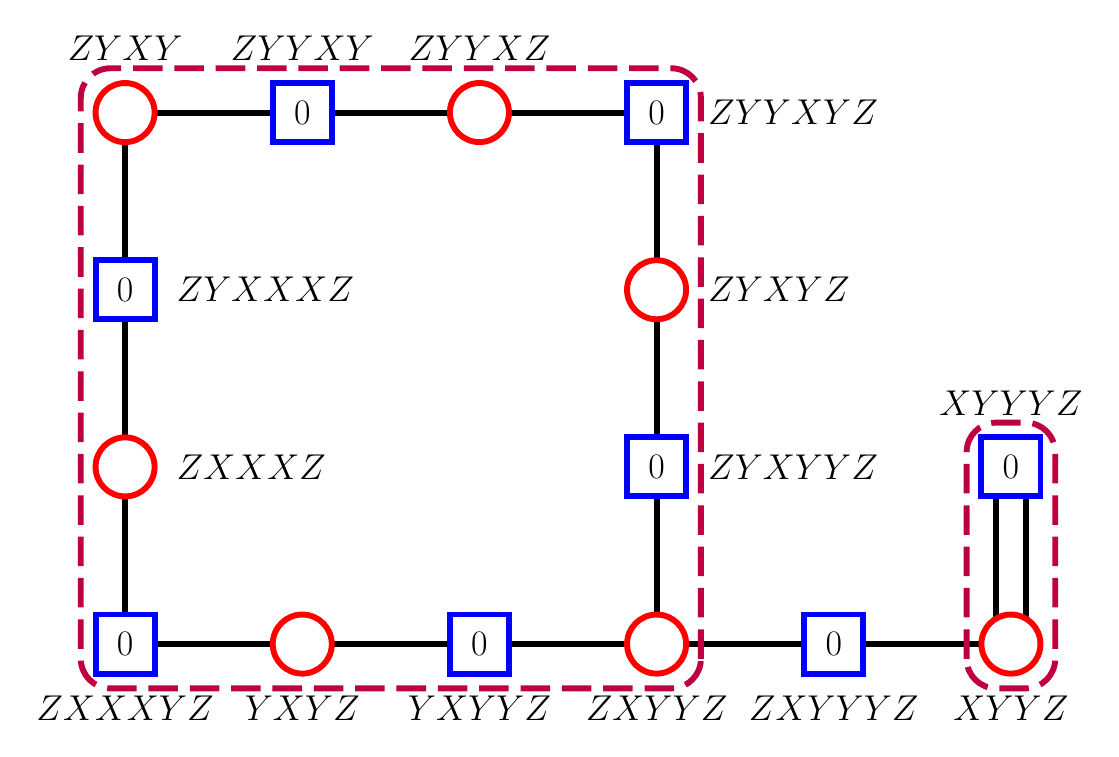}
\caption{Examples of loop structures in the commutator graph of the PXP Hamiltonian for quantities of length $k=5$. For each $\protect\bsv$ vertex, all its neighboring $\protect\rcv$ vertices are shown. The weights of the edges are omitted for simplicity.}
\label{fig:exception_2_exmp}
\end{figure}

The main challenge arises when a loop structure appears in subgraphs. A \textit{loop} refers to a path in the commutator graph that starts and ends at the same vertex, including closed walks\footnote{A \textit{closed walk} in a graph is an alternative sequence of vertices $v_1,v_2,\cdots,v_n$ and edges $(v_1\rightarrow v_2),(v_2\rightarrow v_3),\cdots, (v_{n-1}\rightarrow v_n)$, where the first vertex is the same as the last, i.e., $v_1=v_n$.},
where every $\bsv$ vertex has exactly two neighboring $\rcv$. Fig.\ref{fig:exception_2_exmp} shows one example of loop structures that is present for the Pauli strings with length $5$.

When constructing the commutator graph of an integrable system, the conserved quantities often correspond to loop structures\cite{sm}. However, the loop structure appears in general quantum many-body systems, and thus one needs to show all coefficient of $\rcv$ vertices inside the loop also vanish, to prove the nonintegrability of the systems.

We can infer that the coefficient of each $\rcv$ vertex within the loop is unique, except for an overall scaling factor. Therefore, fixing the coefficient of any single $\rcv$ vertex in a loop determines the coefficients of all other $\rcv$ vertices in the loop. This constraint reduces the degrees of freedom to $1$, and one can always simplify a loop into a single $\rcv$ vertex.

\begin{figure}[h!]
\captionsetup{justification=centering}
\centering
\includegraphics[width=1\linewidth]{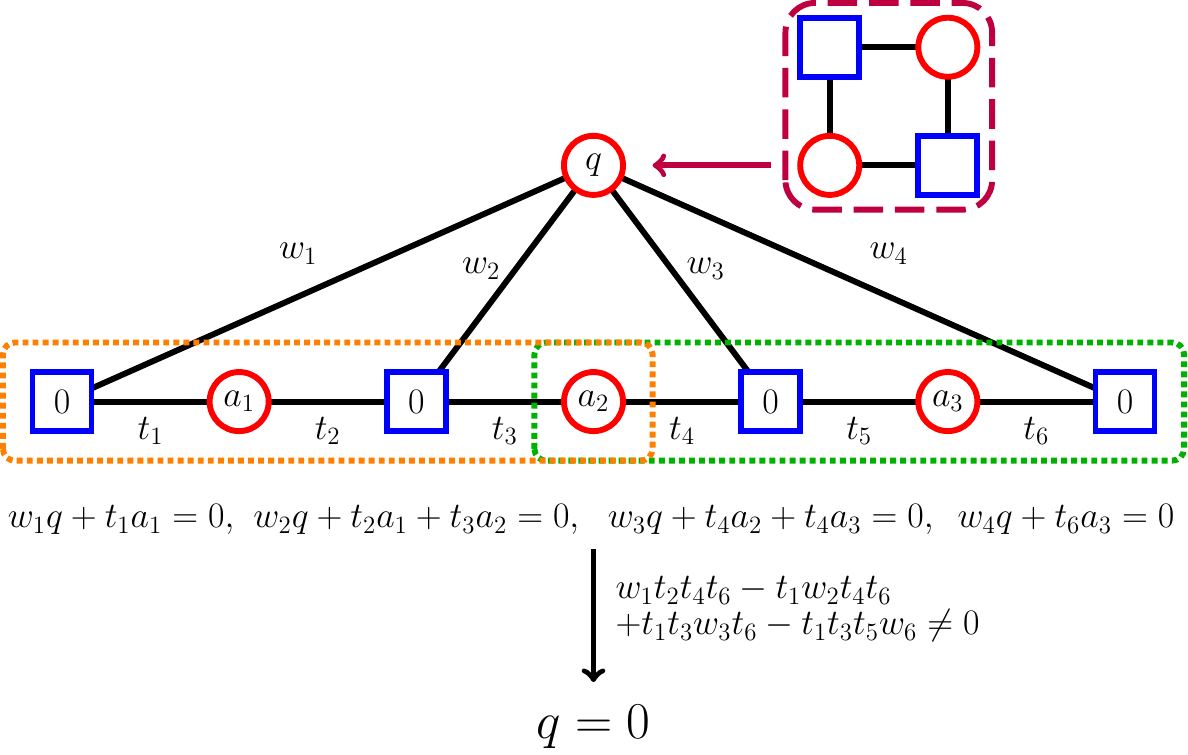}
\caption{Quasi-promising paths and the derivation of zero coefficients for all Pauli strings in a loop structure. The subgraphs within the orange and green dashed boxes form promising paths when $\protect\rcvv{q}$ is removed, thus referred to as quasi-promising paths. Here this $\protect\rcvv{q}$ stands for a specific loop structure. These quasi-promising paths yield the equations below, causing the coefficients of the Pauli strings within the loop structure, labeled as $q$, to be zero, except for very fine tuned parameter space.}
\label{fig:qpp}
\end{figure}

By simplifying the loop into a single $\rcv$ vertex, the most straightforward way to show that its coefficient must be zero is to identify a specific subgraph structure, called a \textit{quasi-promising path}. A  \textit{quasi-promising path} is a subgraph that becomes a promising path when a specific vertex is removed. Fig.\ref{fig:qpp} shows two different quasi-promising paths, highlighted as the orange dashed box and the green dashed box. Each becomes a promising path when $\rcvv{q}$ is ignored. We refer to such  $\rcvv{q}$ as a \textit{disturbing vertex}. (Note that a single (quasi-)promising path should contain only one $\bsv$ that is connected by one neighboring $\rcv$, as illustrated in Fig.\ref{fig:ppath}.
Since $\bsv$ with one neighboring $\rcv$ appears at both ends, one should define two (quasi-)promising paths.) 

In Fig.\ref{fig:qpp}, the equations obtained by the subgraph indicates $q=0$, unless a very fine-tuned condition, given next to the arrow, is satisfied. In general, if a vertex $\rcvv{a}$ is included in two different quasi-promising paths, which become promising paths by removing $\rcvv{q}$, then one can conclude all coefficients of the vertices in the loop must be zero.

By adopting the graph theory concepts as described above, the time complexity in determining the absence of local conserved quantities is being reduced using (quasi-) promising paths, loop structures and etc. This approach thus facilitates the proof of nonintegrablility of various quantum many-body systems. Below, we provide the proof of nonintegrability for the PXP model using graph theoretical approach. The application of this graph analysis to other spin-$1/2$ models are discussed in \cite{sm}.

\medskip
\textit{Proof of nonintegrability for the PXP model: Outline.} --- Adopting a graph-theoretical approach, we introduce three theorems that directly support the proof of our main result. Based on these theorems, the outline of the proofs is provided here, with details available in \cite{sm}.

\textit{Theorem 1.} In the commutator graph of the PXP Hamiltonian for quantities of length $k$, every length-$k$ Pauli string has zero coefficient, except for the following categories: (1) Pauli strings that start with $ZZ\cdots$ or $ZI\cdots$ and end with $\cdots ZZ$ or $\cdots IZ$, which correspond to trivial operators, i.e., operators that vanish within the subspace of states where no two consecutive sites are in the excited states; (2) Pauli string that start and end with $Z$, with every operator in between being either $X$ or $Y$. The Pauli strings in this case are classified into two disconnected loops, $L_o$ and $L_e$, where $L_o$(resp. $L_e$) consists of Pauli strings with an odd(resp. even) number of $X$ operators.

\textit{Proof.} For every length-$k$ Pauli string, except those in the categories (1) and (2), a promising path containing it can be found similarly to those shown in Fig.\ref{fig:graph_examples}, thus its coefficient is zero.

For Pauli strings in category (1), consider the length $k$ local quantity $C$ in this category. A simple calculation shows that the length $k$ Pauli strings, $ZZ\cdots$ and $ZI\cdots$, which differ only by the second operator in their sequences, must have the same coefficient; denote this coefficient as $a$. Now, take the operator $Q=(Z+I)/2$, which projects onto the excited state. Then the operator string containing $QQ=\frac{1}{4}(ZZ+ZI+IZ+II)$ is trivial. Therefore, defining the modified operator $C'\coloneqq C-2a\cdot QQ\cdots$, $C'$ is a nontrivial conserved quantity of $H$ if and only if $C$ is also a nontrivial conserved quantity, hence replacing $C$ with $C'$ does not affect the proof. However, $C'$ does not contain any $ZZ\cdots$ or $ZI\cdots$ Pauli strings by definition, and thus it is sufficient to search for local conserved quantities which Pauli strings that do not fall into category (1). 

For Pauli strings in category (2), we consider three properties. (A) The parity of the number of $X$ operators is the same for all $\rcv$ and $\bsv$ vertices within a connected graph. (B) For any Pauli strings in category (2) with $n\geq 2$ number of $X$ operators, there is always a loop that includes this string and a Pauli string with $n-2$ number of $X$ operators. (C) All Pauli strings in category (2) with a single $X$ operator between $Z$ operators form a loop. Fig.\ref{fig:exception_2_exmp} satisfies all these properties, indicating their general validity. In Fig.\ref{fig:exception_2_exmp}, recall that category (2) only restricts the form of length $k$ Pauli strings, and thus some Pauli strings of length $k-1=4$ are in a loop, although they do not starts or ends with $Z$. However, every length $k=5$ Pauli strings satisfies all the properties (A), (B), and (C).

Now consider two Pauli strings $\bm{A}^k_1$ and $\bm{A}^k_2$ in category (2). If both have an even number of $X$ operators, then by applying the property (B) repetitively, $\bm{A}^k_1$ is in the same loop as $ZYYY\cdots YYZ$, and so is $\bm{A}^k_2$ and they belong to the same loop, $L_e$.
Similarly, if $\bm{A}^k_1$ and $\bm{A}^k_2$ both have an odd number of $X$ operators, then by applying the property (B), $\bm{A}^k_1$ is in the same loop as one of the Pauli strings in category (2) with a single $X$ between $Z$ operators, i.e. one of $ZXYY\cdots YYZ, ZYXY\cdots YYZ, \cdots, ZYYY\cdots YXZ$, and so is $\bm{A}^k_2$. By the property (C), all the Pauli strings in category (2) with a single $X$ between $Z$ operators are in the same loop, and hence $\bm{A}^k_1$ and $\bm{A}^k_2$ belong to the same loop, $L_o$. Finally, if the parity of the number of $X$ operators in these Pauli strings is different, then by the property (A) there is no path between them, and hence $L_o$ and $L_e$ are disconnected. $\square$

\textit{Theorem 2.} For every Pauli string with an even number of $X$ operators, classified in category (2) of Theorem 1.,  its coefficient must be zero. 

\textit{Proof.} It is enough to show that one of such Pauli strings in the loop is  zero, since the loop degrees of freedom is $1$ as discussed above. Denote $(A)^{(t)}$ as $t$ times repetition of $A$ operator. Since $[\{Z(Y)^{k-2}Z\}_j,\{XZ\}_{j+k-1}] = [\{ZX\}_j, \{Z(Y)^{k-2}Z\}_{j+1}]=\{Z(Y)^{k-1}Z\}_{j}$, we conclude $q(Z(Y)^{k-2}Z)+q(Z(Y)^{k-2}Z)=2q(Z(Y)^{k-2}Z)=0$. $\square$

\medskip
\textit{Theorem 3.} In $C$ with $\len(C)=k$, there is a $\rcv$ vertex corresponding to a Pauli string of length $k-2$, that is part of two quasi-promising paths sharing the same disturbing vertex.

\textit{Proof.} 
Consider $k\geq 7$; for $k=4,5,6$ one can easily check this statement case-by-case. Now, by ignoring the disturbing vertex, if we find a $\rcv$ vertex that is part of two different promising paths, the proof is complete by the definition of a quasi-promising path.  Denote $\bm{B}^{k-1}_{a}\coloneqq X(Y)^{(a)}Z(Y)^{(k-a-4)}Z$ and $\bm{A}^{k-2}_{a}\coloneqq X(Y)^{(a-1)}Z(Y)^{(k-a-4)}Z$. Every $\bm{B}_a^{k-1}$ Pauli string has neighboring $\rcv$ vertices in the loop $L_o$, since $[\{X(Y)^{(k-3)}Z\}_j, \{X\}_{j+a+1}] =- [\{X(Y)^{(a)}XYX(Y)^{(k-a-5)}Z\}_j,\{ZXZ\}_{j+a}] = 2i\{\bm{B}^{k-1}_a\}_{j}$, where one can show that a length $k-1$ Pauli strings obtained by removing a $Z$ operator from edge of a length $k$ Pauli string in $L_o$ should also be included in $L_o$. Furthermore, the following two commutators $[\{XZ\}_j,\{\bm{A}^{k-2}_{a}\}_{j+1}]=2i\{\bm{B}_{a}^{k-1}\}_j$ and $[\{XZ\}_{j+k-2},\{\bm{A}^{k-2}_{a+1}\}_j]=-2i\{\bm{B}_{a}^{k-1}\}_j$ show that there are two $\rcv$ neighbors of $\bsv$ vertex $\bm{B}_a^{k-1}$, which are $\bm{A}_{a}^{k-2}$ and $\bm{A}_{a+1}^{k-2}$, and one $\rcv$ neighbor $\bm{A}^{k-2}_1$ for $\bm{B}_{0}$. 
Also, one can show that (with some cancelation) the $\bsv$ vertex $\bm{B}^{k}_F \coloneqq  ZX(Y)^{(k-5)}ZYZ$ has only one neighboring $\rcv$ vertex, $\bm{A}^{k-2}_{k-6}$. Except the operators $\bm{A}^{k-2}_a$ and the operators in the loop $L_o$, there are no other neighbors of Pauli strings $\bm{B}^{k-1}_a$ and $\bm{B}^{k}_{F}$. Therefore, for $\rcvv{\delta}$ which represents  the loop structure $L_o$, with vertices $\rcv$ representing $\bm{A}^{k-2}_a$, and vertices $\bsv$ representing $\bm{B}^{k-1}_a$ and $\bm{B}^{k}_{F}$, they form a subgraph which is very similar to Fig.\ref{fig:qpp}, as illustrated below.
\begin{equation*}
\begin{tikzpicture}

\draw[line width=0.5mm] (0*\rd,1) -- (-20*\rd,0*\rd);
\draw[line width=0.5mm] (0*\rd,1) -- (-10*\rd,0*\rd);
\draw[line width=0.5mm] (0*\rd,1) -- (10*\rd,0*\rd);
\draw[line width=0.5mm] (0*\rd,1) -- (20*\rd,0*\rd);
\draw[line width=0.5mm] (-20*\rd,0*\rd) -- (-2.5*\rd,0*\rd);
\draw (0,0) node {\footnotesize $\cdots$};
\draw[line width=0.5mm] (20*\rd,0*\rd) -- (2.5*\rd,0*\rd);

\draw (-20*\rd,-0.5) node {\footnotesize $\bm{B}^k_F$};
\draw (-15*\rd,-0.5) node {\footnotesize $\bm{A}^{k-2}_{k-5}$};
\draw (-10*\rd,-0.5) node {\footnotesize $\bm{B}^{k-1}_{k-6}$};
\draw (-5*\rd,-0.5) node {\footnotesize $\bm{A}^{k-2}_{k-6}$};

\draw (20*\rd,-0.5) node {\footnotesize $\bm{B}^{k-1}_0$};
\draw (15*\rd,-0.5) node {\footnotesize $\bm{A}^{k-2}_{1}$};
\draw (10*\rd,-0.5) node {\footnotesize $\bm{B}^{k-1}_{1}$};
\draw (5*\rd,-0.5) node {\footnotesize $\bm{A}^{k-2}_{2}$};

\draw[line width=0.5mm, blue, fill=white] (-21*\rd,-1*\rd) rectangle ++(2*\rd,2*\rd);
\draw[line width=0.5mm, blue, fill=white] (-21*\rd,-1*\rd) rectangle ++(2*\rd,2*\rd);
\draw (-20*\rd,0) node {\footnotesize $0$};

\draw[line width=0.5mm, red, fill=white] (-15*\rd,0) circle (1*\rd);

\draw[line width=0.5mm, blue, fill=white] (-11*\rd,-1*\rd) rectangle ++(2*\rd,2*\rd);
\draw[line width=0.5mm, blue, fill=white] (-11*\rd,-1*\rd) rectangle ++(2*\rd,2*\rd);
\draw (-10*\rd,0) node {\footnotesize $0$};

\draw[line width=0.5mm, red, fill=white] (-5*\rd,0) circle (1*\rd);

\draw[line width=0.5mm, blue, fill=white] (19*\rd,-1*\rd) rectangle ++(2*\rd,2*\rd);
\draw[line width=0.5mm, blue, fill=white] (19*\rd,-1*\rd) rectangle ++(2*\rd,2*\rd);
\draw (20*\rd,0) node {\footnotesize $0$};

\draw[line width=0.5mm, red, fill=white] (15*\rd,0) circle (1*\rd);

\draw[line width=0.5mm, blue, fill=white] (9*\rd,-1*\rd) rectangle ++(2*\rd,2*\rd);
\draw[line width=0.5mm, blue, fill=white] (9*\rd,-1*\rd) rectangle ++(2*\rd,2*\rd);
\draw (10*\rd,0) node {\footnotesize $0$};

\draw[line width=0.5mm, red, fill=white] (5*\rd,0) circle (1*\rd);

\draw[line width=0.5mm, red, fill=white] (0*\rd,1) circle (1*\rd);
\draw (0*\rd,0.65) node {\footnotesize $L_o$};
\draw (0*\rd,1) node {\footnotesize $\delta$};

\end{tikzpicture}
\end{equation*}

Choosing $\rcvv{\delta}$ as a disturbing vertex and ignoring it makes the vertex $\bm{A}^{k-2}_{k-6}$ becomes part of two promising paths. Hence the statement is proven. $\square$

\medskip
\textit{Proof Summary.} --- Based on Theorems 1, 2, and 3, we have shown that if $[H,C]=0$ and $C$ is a nontrivial local conserved quantity of length $k\geq 4$, then every length $k$ Pauli string in $C$ must have a zero coefficient. This contradicts the assumption that $C$ is of length $k$. Therefore, no such $C$ exists, completing our proof.

\medskip
\textit{Discussion.} ---
The PXP model has long been considered non-integrable based on numerical energy level statistics, though conserved quantities were not definitively excluded. Our work provides a rigorous proof confirming its non-integrability and the absence of local conserved quantities, also demonstrating that quantum many-body scarring in this model is not associated with such quantities. To prove it, we develop a novel graph-theoretical approach applicable to general quantum many-body systems. This method offers several advantages, including reducing time complexity for nonintegrability proof and converting the identification of conserved quantities into a problem of analyzing graph structures. This framework also utilizes cohomology theory to analyze Pauli strings within loops.\footnote{Private communication with H. Katsura}.

The generality of the graph-theoretical method makes it applicable to higher spin models with additional spin-exchange interactions, such as the AKLT model\footnote{HaRu K. Park and SungBin Lee, in preparation}. Although the absence of conserved quantities in the PXP model has been demonstrated, this does not imply that all QMBS models lack conserved quantities. In this context, it would be also interesting to explore the spin-$1$ $XY$ model, a QMBS system exhibiting perfect revivals\cite{PhysRevLett.123.147201}, which we leave for future work.


\medskip
We thank Naoto Shiraishi and Hosho Katsura for valuable discussions. We also thank Jaeho Han for helpful comments on the paper. This research was supported by National Research Foundation Grant (2021R1A2C109306013).

\bibliography{biblio.bib}

\end{document}


\title{Proof of the nonintegrability of the PXP model and general spin-$1/2$ systems:\\ Supplement Material}
\author{HaRu K. Park, SungBin Lee}
\maketitle

\tableofcontents

\section{Classifying the Pauli Strings: Finding Promising Paths}
In the main text we introduced the concepts of the commutator graph and the promising path. In this section, we provide a detailed explanation of how to find a promising path that includes a given vertex. This method is quite general and can be applied to various spin-$1/2$ models.

\textbf{Algorithm for Finding the Promising Path.} --- For a given $\rcv$ vertex $A$, a promising path that contains it can be found using a series of simple inductive steps.

\textbf{Step 1:} Begin by scanning all neighboring $\bsv$ vertices, and select those with either one or two neighboring vertices.

\textbf{Step 2-1:} If a neighboring $\bsv$ vertex with only one connection is found, a promising path for vertex $A$ is identified, and the algorithm concludes.
 
\textbf{Step 2-2:} If only $\bsv$ vertices with two neighboring vertices are present, select their adjacent $\rcv$ vertices, excluding any previously chosen vertices (in this case, $A$). Label these vertices as $B_1,\cdots,B_n$. 

Next, apply the process inductively: for each of the vertices $B_1,\cdots,B_n$, repeat Step 1 by scanning their neighboring $\bsv$ vertices, selecting those with one or two neighbors. If a $\bsv$ vertex with only one neighbor is found (Step 2-1), the algorithm terminates. If only $\bsv$ vertices with two neighbors are found (Step 2-2), identify their neighboring $\rcv$ vertices, excluding the previously chosen vertices ($A, B_1,\cdots,B_n$), and label them $C_1, C_2,\cdots, C_m$. Repeat the process with these newly identified $\rcv$ vertices $(C_1,\cdots,C_m)$ by returning to Step 1. Continue this iterative process until Step 2-1 applies, concluding the algorithm.

\begin{figure}[h!] 
\captionsetup{justification=centering}
  \begin{minipage}[b]{0.5\linewidth}
    \centering
    \subfloat[Subgraph of the commutator graph for quantities of length $5$.]{\includegraphics[width=0.75\linewidth]{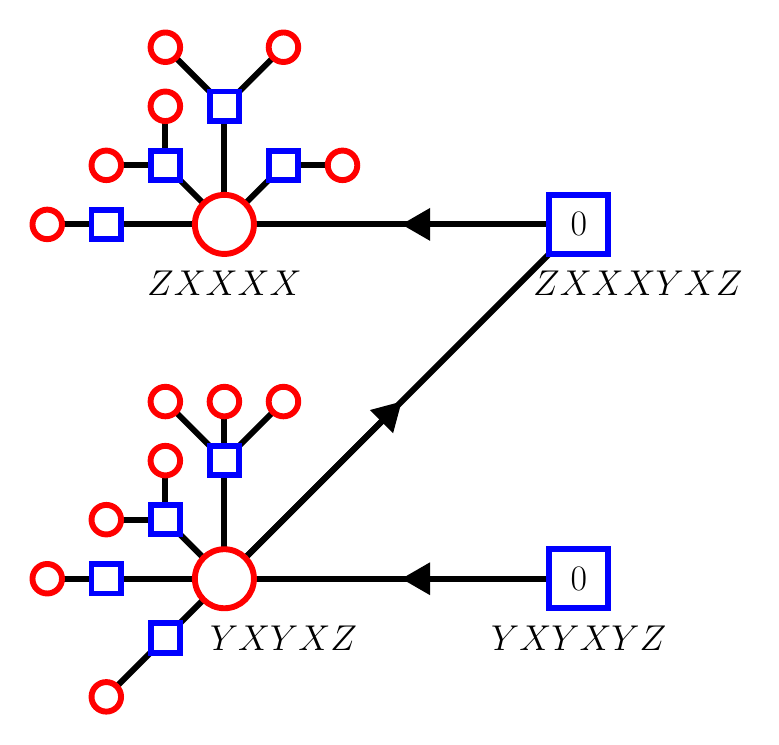}\label{fig:ppath}}

    \vspace{4ex}
  \end{minipage}
  \begin{minipage}[b]{0.5\linewidth}
    \centering
    \subfloat[Step 1.]{\includegraphics[width=0.75\linewidth]{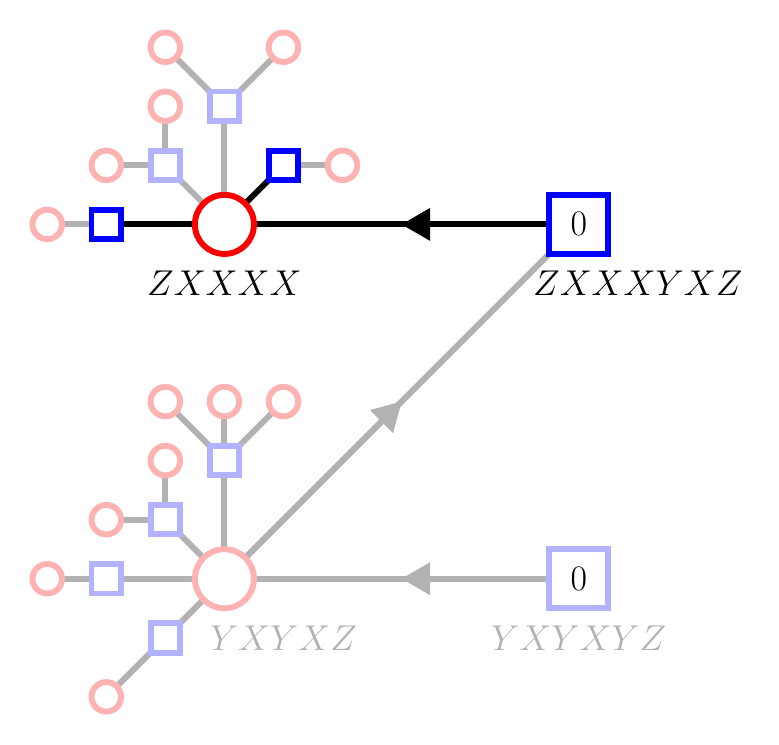}
    \label{fig:ppath_1} }
    \vspace{4ex}
  \end{minipage} 
  \begin{minipage}[b]{0.5\linewidth}
    \centering
    \subfloat[Step 2-2.]{\includegraphics[width=0.75\linewidth]{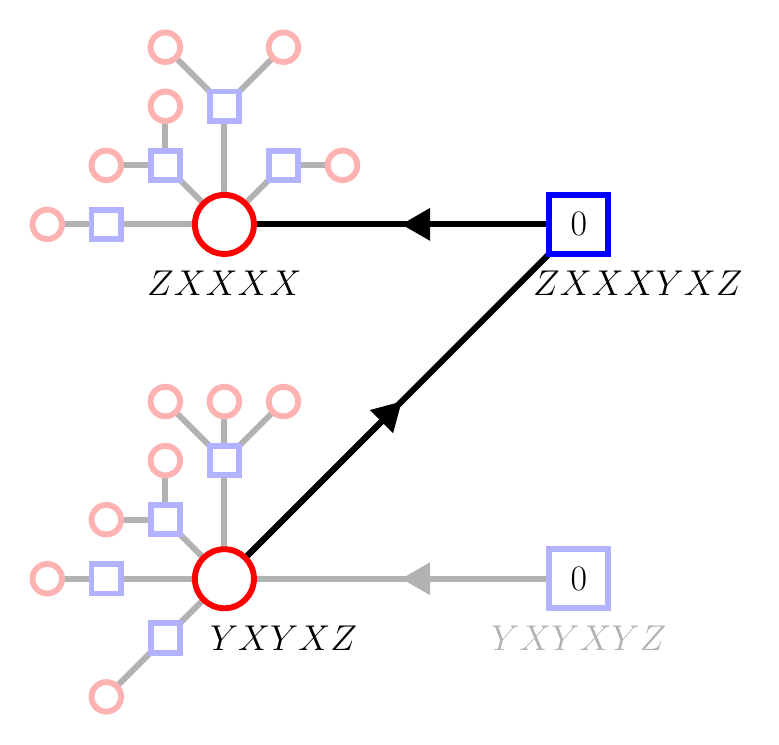}
    \label{fig:ppath_2} }
  \end{minipage}
  \begin{minipage}[b]{0.5\linewidth}
    \centering
    \subfloat[Step 1 and Step 2-1.]{\includegraphics[width=0.75\linewidth]{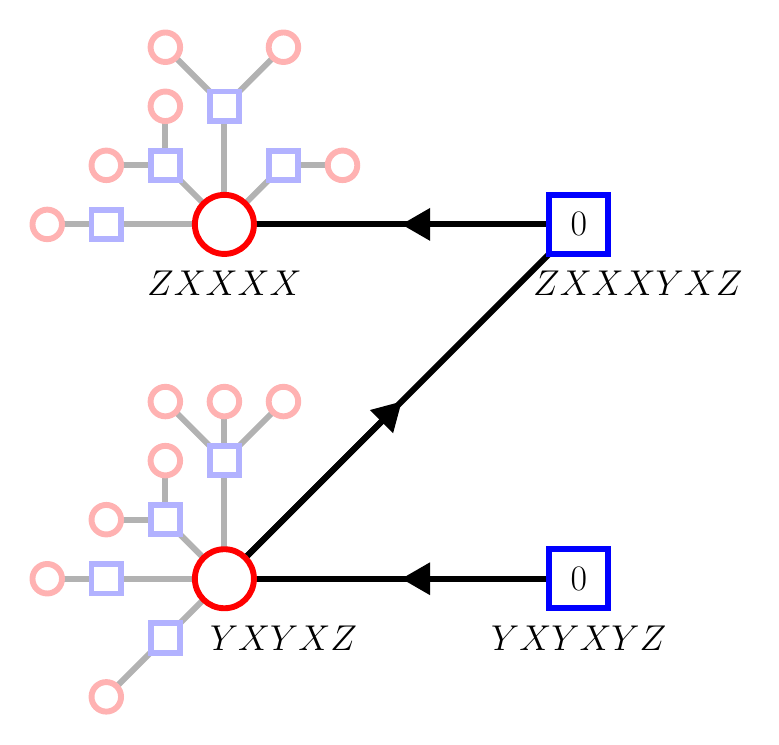}
    \label{fig:ppath_3} }
  \end{minipage} 
  \caption{\ref{sub@fig:ppath} Subgraph of the commutator graph for quantities of length $5$. We are going to find the promising path starting from the $\protect\rcv$ vertex labeled $ZXXXX$. \ref{sub@fig:ppath_1} Step 1 of the algorithm: All $\protect\bsv$ vertices with one or two neighbors are scanned. \ref{sub@fig:ppath_2} Step 2-2 of the algorithm: Since no $\protect\bsv$ vertex with only one neighbor is found, we scan the neighbors of the selected $\protect\bsv$ vertices. For simplicity, only the $\protect\rcv$ vertex labeled $YXYXZ$, which leads to a promising path, is highlighted. \ref{sub@fig:ppath_3} Step 1 and Step 2-1 of the algorithm. When scanning the $\protect\bsv$ neighbors of the $\protect\rcv$ vertex labeled $YXYXZ$, we find its $\protect\bsv$ neighbor labeled $YXYXYZ$, which has only one neighbor. This completes the algorithm.}
  \label{fig:ppath_exmps} 
\end{figure}

Fig.\ref{fig:ppath_exmps} demonstrates how to find a promising path that includes the $\rcv$ vertex labeled $ZXXXX$ within the subgraph of the commutator graph of PXP model for quantities of length $5$. In the PXP model, the weights of edges are restricted to $\pm 2i$, which are represented by the direction of the arrows: arrows pointing towards $\bsv$ vertices correspond to weight $+2i$, and arrows pointing towards $\rcv$ vertices correspond to weight $-2i$. Due to the properties of the promising path, this implies that $q(ZXXXX)=0$ for any length-$5$ quantity $C$ that satisfies $[C,H]=0$.

\begin{figure}[t!] 
\captionsetup{justification=centering}
  \begin{minipage}[b]{0.5\linewidth}
    \centering
    \subfloat[Exception 1]{\includegraphics[width=1\linewidth]{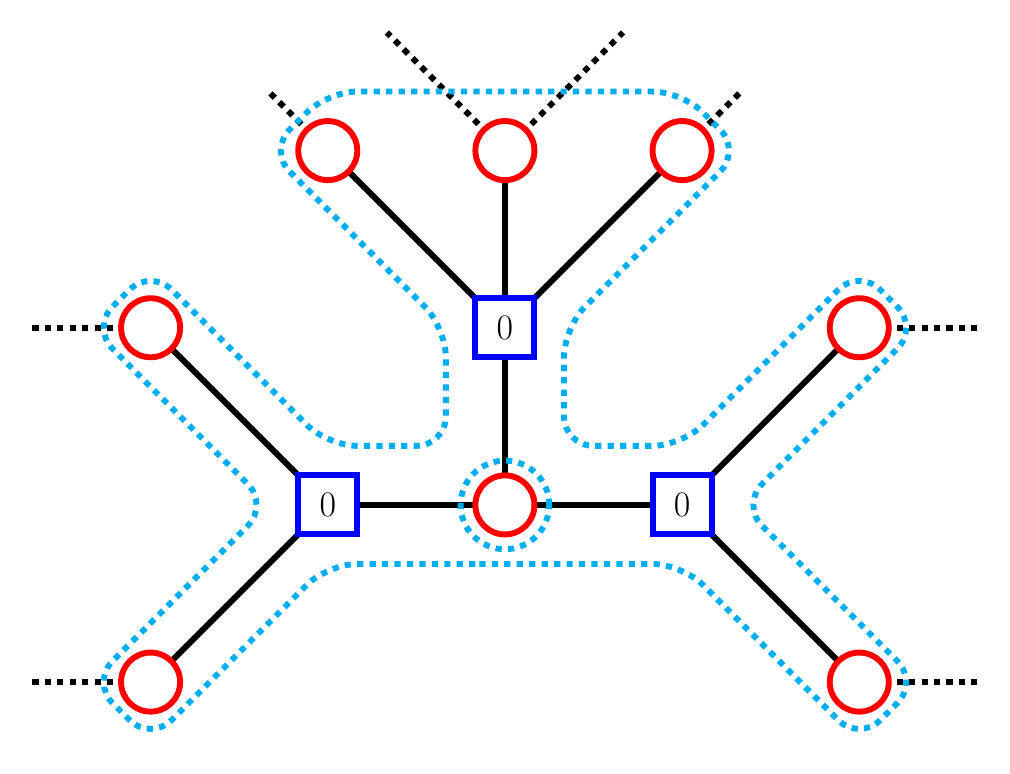}
    \label{fig:exception_1} }
    \vspace{4ex}
  \end{minipage}
  \begin{minipage}[b]{0.5\linewidth}
    \centering
    \subfloat[Exception 2]{\includegraphics[width=1\linewidth]{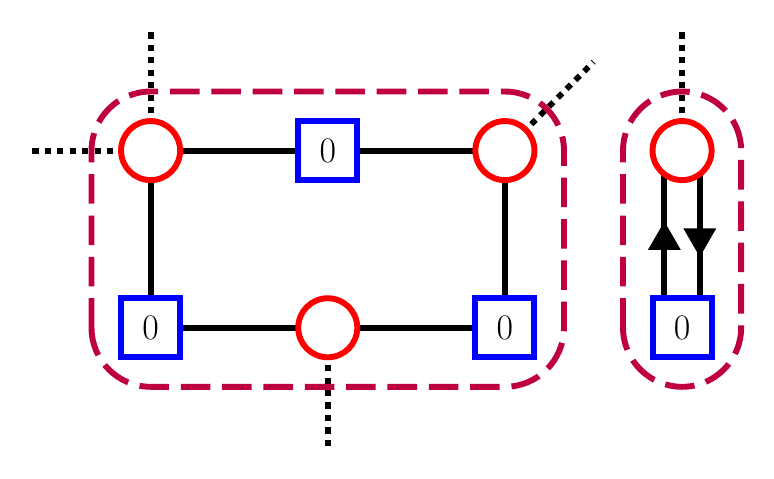}
    \label{fig:exception_2} }
    \vspace{4ex}
  \end{minipage} 
  \begin{minipage}[b]{0.5\linewidth}
    \centering
    \subfloat[Exception 1: example]{\includegraphics[width=1\linewidth]{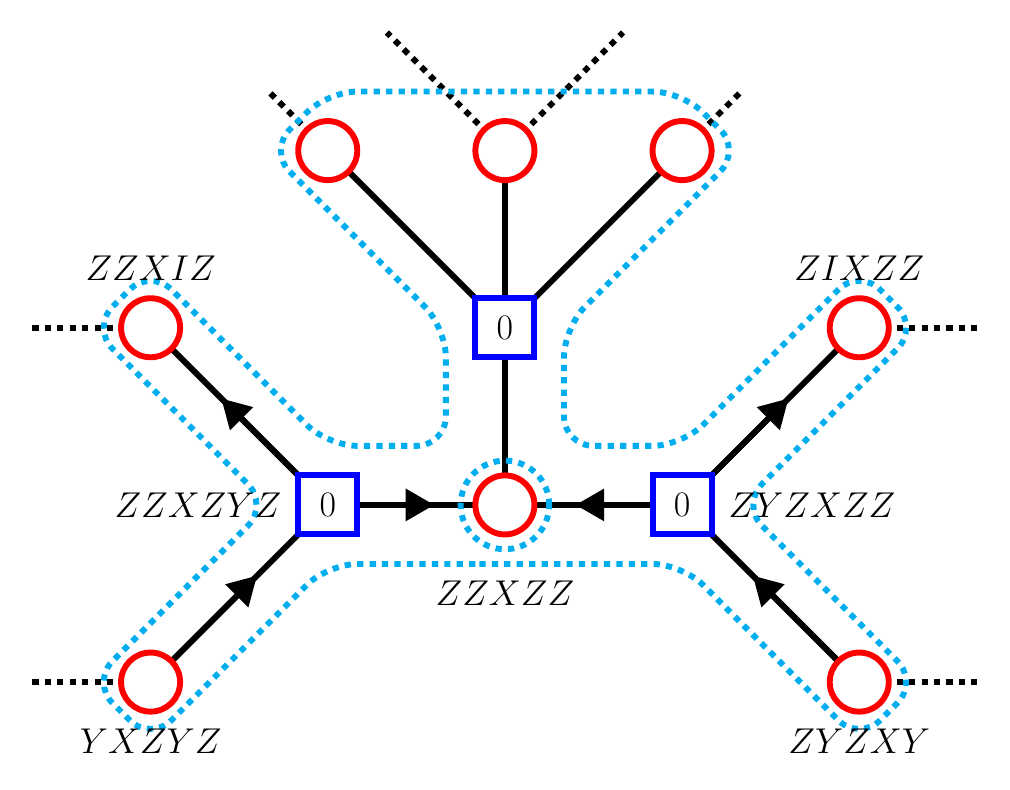}
    \label{fig:exception_1_exmp} }
  \end{minipage}
  \begin{minipage}[b]{0.5\linewidth}
    \centering
    \subfloat[Exception 2: example]{\includegraphics[width=1\linewidth]{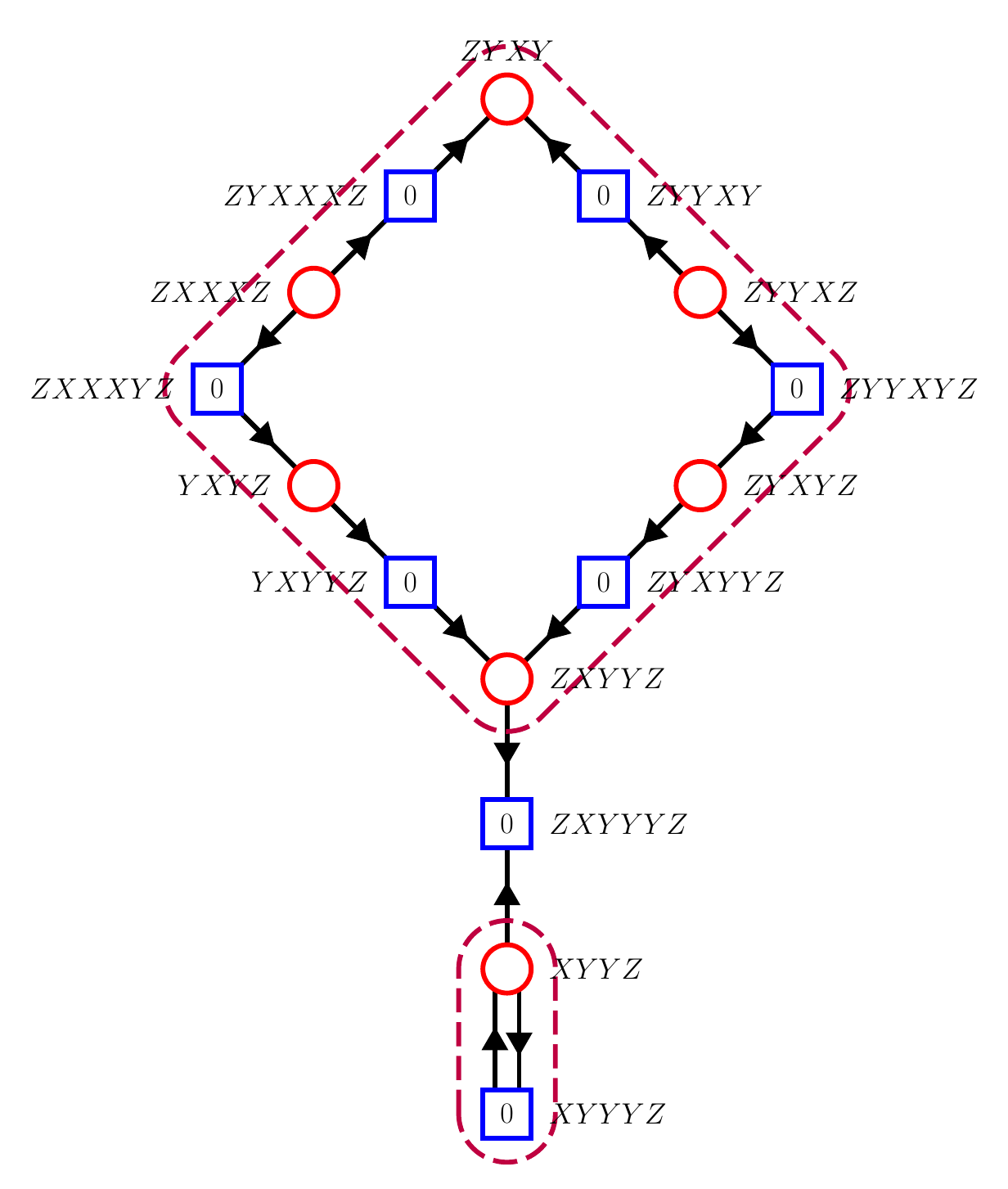}
    \label{fig:exception_2_exmp} }
  \end{minipage} 
  \caption{\ref{sub@fig:exception_1} First exception where a promising path cannot be found. For the central $\protect\rcv$ vertex, all neighboring $\protect\bsv$ vertices have more than two adjacent $\protect\rcv$ vertices, preventing the execution of Step 1. \ref{sub@fig:exception_2} Second exception where a promising path cannot be found. Although all $\protect\bsv$ vertices have one or two neighboring $\protect\rcv$ vertices, none of them have \textit{exactly} one neighboring $\protect\rcv$ vertex, preventing the execution Step 2-1. Additionally, after sufficient iterations, there are no unchosen vertices remaining, violating Step 2-2. \ref{sub@fig:exception_1_exmp} Example of Exception 1 and \ref{sub@fig:exception_2_exmp} example of Exception 2 in the commutator graph for quantities of length $k=5$.}
  \label{fig:exception_exmps} 
\end{figure}

\textbf{When the Algorithm Fails: Exceptions.} --- The commutator graph for quantities of fixed length $k$ contains finitely many vertices. Since no vertex is selected more than once during the process, the algorithm always terminates. While this method successfully identifies promising paths for many $\rcv$ vertices, it is not universally successful. There are exceptional cases where a promising path cannot be found, and these exceptions can be classified into two categories, which we will now explain. Fig.\ref{fig:exception_exmps} p[rovides a visual outline of these exceptional cases.

\textbf{Exception 1.} --- Fig. \ref{fig:exception_1} demonstrates a situation where no neighboring $\bsv$ vertex with one or two neighbors can be found. For the central $\rcv$ vertex in the figure, all its neighboring $\bsv$ vertices have more than two neighbors, violating Step 1 and making it impossible to find a promising path without additional information.

Exception 1 occurs frequently when dealing with $\bsv$ Pauli strings of length $\leq k$. In this case, when attempting to compute the commutator between $\rcv$ Pauli strings and the Hamiltonian strings(Pauli strings from the Hamiltonian) that generates the $\bsv$ Pauli string, there are no constraints on the position of the Pauli strings in the Hamiltonian. This contrasts with the situation for $\bsv$ vertices labeled by Pauli strings of length $>k$, where the position of the Pauli strings in the Hamiltonian is restricted to the edges of the $\rcv$ Pauli string. Therefore, to avoid Exception 1, focusing on $\bsv$ Pauli strings of length $>k$ can be an effective strategy. This approach works well for some Hamiltonians, such as the $XYZ+h$ model\cite{shiraishi2019proof}, where concentrating on $\bsv$ Pauli strings of length $>k$ completely eliminates the Exception 1 scenario. However, in the PXP model, this strategy does not apply. Fig. \ref{fig:exception_1_exmp} provides an example where, even when focusing on $\bsv$ Pauli strings of length $k+1=6$, it is still impossible to find a promising path that includes the $\rcv$ vertex labeled $ZZXZZ$. 

We will soon show that each Exception 1 $\rcv$ type vertex either does indeed have a promising path or falls under the Exception 2 case, which we will explain next.

\textbf{Exception 2.} --- In Fig \ref{fig:exception_2}, alternating $\rcv$ and $\bsv$ vertices form a loop, as defined in the main text. This occurs because after sufficient iterations, in Step 2-2, when we attempt to choose a $\rcv$ vertex which has not been previously selected, it becomes impossible since both neighbors of the $\bsv$ vertex have already been chosen. This prevents the identification of a promising path.

Fig. \ref{fig:exception_2_exmp}, representing the same subgraph discussed in the main text, shows an example where the $\rcv$ vertex labeled $ZXYYZ$ falls under Exception 2, becoming part of a loop in the commutator graph for quantities of length $k=5$. As discussed in the main text, to determine the coefficients of $\rcv$ vertices in the loop structure, we introduce the concept of a quasi-promising path.

After identifying the exceptional cases in our algorithm, a natural question arises: how can we determine whether a given Pauli string has a promising path or belongs to Exception 1 or 2? In the following section, we classify the Pauli strings for PXP Hamiltonian. This classification is both simple and systematic, and it has the potential to be generalized to other Hamiltonians as well.

Before moving forward, we emphasize that this categorization applies to every spin-$1/2$ system. For example, in \cite{shiraishi2019proof}, it is shown that every Pauli string that is not a ``doubling-product operator" has zero coefficients. The Pauli strings that are not ``doubling-product operators" correspond to Pauli strings with a promising path, while the ``doubling-product operators" correspond to Exception 2 cases.

\section{Theorem 1 in the Main Text}
\begin{table}[t!]
\centering
\begin{tabular}{clcrl}
\toprule
Pauli String Type& \multicolumn{3}{c}{\begin{tabular}[c]{@{}c@{}}Initial and Final Operators\\ (Length $k$)\end{tabular}} & \multicolumn{1}{c}{\begin{tabular}[c]{@{}c@{}}Resolved by\\ \textbf{Lemma}...\end{tabular}}  \\ \midrule
\multirow{3}{*}{\begin{tabular}[c]{@{}c@{}}\\ \\ Simple cases\end{tabular}} & $\begin{bmatrix}X\\Y\end{bmatrix}$&$\cdots$&$ \begin{bmatrix}X\\Y\end{bmatrix}$ & \textbf{1} \\ \cmidrule{2-5} &                                                                                                                                  $Z\begin{bmatrix}Y\\Z\\I\end{bmatrix}$&$\cdots$&$ \begin{bmatrix}X\\Y\end{bmatrix}$ & \textbf{2} \\ \cmidrule{2-5} & $ZX$&$\cdots$&$\begin{bmatrix}X\\Y\end{bmatrix}$ & \textbf{3} \\ \midrule
\begin{tabular}[c]{@{}c@{}}\textbf{Exception 1}:\\ \textbf{Category 1}\end{tabular} & $Z\begin{bmatrix}Z\\I\end{bmatrix}$&$\cdots$&$ \begin{bmatrix}Z\\I\end{bmatrix}Z$ & \textbf{8}                                                                 \\ \midrule
\begin{tabular}[c]{@{}c@{}}\textbf{Exception 1}:\\ \textbf{Category 2}\end{tabular} & $ZY\begin{bmatrix}Z\\I\end{bmatrix}$&$\cdots$&$ \begin{bmatrix}X\\Y\end{bmatrix}Z$ & \textbf{5}                                                                 \\ \midrule
\begin{tabular}[c]{@{}c@{}}\textbf{Exception 1},\textbf{2}:\\ \textbf{Category 3}\end{tabular} & $ZX\begin{bmatrix}X\\Y\end{bmatrix}$&$\cdots$&$ \begin{bmatrix}X\\Y\end{bmatrix}Z$ & \textbf{4}, \textbf{5}, \textbf{6} and \textbf{7} \\ \midrule
\textbf{Exception 2} & $ZY\begin{bmatrix}X\\Y\end{bmatrix}$&$\cdots$&$ \begin{bmatrix}X\\Y\end{bmatrix}Z$ & \textbf{5}, \textbf{6} and \textbf{7} \\ \midrule
\multirow{2}{*}{\begin{tabular}[c]{@{}c@{}}\\Simple cases\end{tabular}} & $ZX\begin{bmatrix}Z\\I\end{bmatrix}$&$\cdots$&$ \begin{bmatrix}X\\Y\end{bmatrix}Z$& \multirow{2}{*}{\begin{tabular}[c]{@{}c@{}}\\ \textbf{5}\end{tabular}} \\ \cmidrule{2-4}
&$Z\begin{bmatrix}Z\\I\end{bmatrix}$&$\cdots$&$\begin{bmatrix}X\\Y\end{bmatrix}Z$&\\ \bottomrule
\end{tabular}
\caption{Pauli string types, their initial and final Pauli operator sequences, and how they are resolved(i.e. shown to have a zero coefficient or gives trivial conserved quantity). The middle column of the table represents the initial and final operators of length $k$ Pauli string or its reflected one, where the Pauli operators written vertically represents the possible choices in the initial or final Pauli operator sequence. For example, the third row includes the operator strings $ZX\cdots X$, $ZX\cdots Y$, $X\cdots XZ$, and $Y\cdots XZ$. }
\label{table:proof_table}
\end{table}

\textbf{Outline.} --- In this section, we categorize the length $k$ Pauli strings as we have done in Theorem 1 of the main text, showing that in the commutator graph of the PXP model for quantities of length $k$, each $\rcv$ vertex either has promising path of falls under Exception 1 or Exception 2. For each case, we demonstrate that the coefficient of each vertex must be zero, using the methods briefly discussed in the main text. We present lemmas to classify the length-$k$ Pauli strings, with a summary provided in Table \ref{table:proof_table}.

Pauli strings that do not start or end with the operator $Z$ can be shown to have a promising path, as proven in \textbf{Lemmas 1}, \textbf{2}, and \textbf{3}. Therefore, we can narrow our focus to Pauli strings that begin and end with $Z$. There are also some simple cases within this group that can easily be proven to have a promising path, and for simplicity, we will address these Pauli strings alongside others that start and end with $Z$ operators.

For Pauli strings that start and end with $Z$ operators, it is possible to classify them into the ones that fall under Exception 1 or Exception 2. Generally, the reason for Exception 1 can be categorized into three casees, each requiring a different approach. Pauli strings that start with $ZZ\cdots$ or $ZI\cdots$ and end with $\cdots ZZ$ or $\cdots IZ$ are special in that they always relate to the trivial quantities, so they require separate treatment, which we address in \textbf{Lemma 8}. All other Pauli strings can be categorized using a single strategy, supported by \textbf{Lemmas 4} and \textbf{5}.

Finally, we need to resolve Exception 2 vertices, which are always part of a loop, using the quasi-promising path. \textbf{Lemma 6} and \textbf{7} outline how to handle these Pauli strings.

\textbf{Vertical notation.} --- Before presenting the lemmas and theorems, we introduce the concept of vertical notation. The standard description of the commutator relation between two Pauli strings, such as $[\{XXXX\}_j,\{ZX\}_{j+3}]=-2i\{XXXYX\}_{j}$, is not very  intuitive since it does not clearly show the positions where each individual Pauli operator in the strings acts. Instead, we use vertical notation, which visually aligns the operators as follows:
\begin{equation*}
\begin{array}{rrrrrrr}
 &X&X&X&X& \\
 & & & &Z&X\\\hline
 &X&X&X&Y&X
\end{array}
\end{equation*}
This format is more intuitive, as it allows us to easily see the relative positions where each Pauli operator acts. In this notation, we omit the specific position index $j$ and the commutator coefficient commutator $-2i$, focusing only on the operators themselves.

\textbf{Simple cases.} --- We begin with the simple case of Pauli strings that do not start and end with the operator $Z$.

\begin{theorem}\label{thm:1}
In the commutator graph for quantities of length $k$, every Pauli string of length $k$ that does not start and end with $Z$ is part of a promising path.
\end{theorem}

\begin{proof} Let $A_1A_2\cdots A_k$ be a Pauli string where $A_1 \neq Z$ and $A_k\neq Z$. Consider the following commutator relation:
\begin{equation}\label{eq:thm1proof1}
\begin{array}{rrrrrrr}
 &A_1&A_2&\cdots&A_k& & \\
 & & & &Z&X&Z\\\hline
 &A_1&A_2&\cdots&\overline{A_k}&X&Z
\end{array}
\end{equation}
Here, $\overline{X}=Y$ and $\overline{Y}=X$. This shows that the $\rcv$ vertex $A_1A_2\cdots A_k$ is connected to the $\bsv$ vertex $A_1A_2\cdots \overline{A_k} XZ$.

Next, we check for other neighboring $\rcv$ vertices of  $A_1A_2\cdots \overline{A_k} XZ$. Since we are considering quantities of length $k$, the only possible commutator giving $A_1A_2\cdots \overline{A_k}XZ$ takes the following form:
\begin{equation}\label{eq:thm1proof2}
\begin{array}{rrrrrrrr}
 & & &?&\cdots&A_k&X&Z\\
 &Z&X&Z& & & & \\\hline
 &A_1&A_2&A_3&\cdots&\overline{A_k}&X&Z
\end{array}
\end{equation}
This shows that $A_1=Z$, which is a contradiction because we assumed $A_1\neq Z$. Therefore, the only neighbor of the $\bsv$ vertex $A_1A_2\cdots \overline{A_k} XZ$ is the $\rcv$ vertex $A_1A_2\cdots A_k$, which satisfies the promising path condition.
\end{proof}

\begin{theorem}\label{thm:2}
In the commutator graph for quantities of length $k$, every Pauli string of length $k$ that starts with $ZY, ZZ,$ or $ZI$ and does not end with $Z$ (or their reflected forms) is part of a promising path.
\end{theorem}

\begin{proof}
Without loss of generality, let $Z A_2\cdots A_k$ be a Pauli string where $A_2\neq X$ and $A_k\neq Z$. Since the PXP Hamiltonian is mirror-symmetric, the same theorem holds for the reflected Pauli strings. Consider the following commutator relation:
\begin{equation}\label{eq:thm2proof1}
\begin{array}{rrrrrrr}
 &Z&A_2&\cdots&A_k& & \\
 & & & &Z&X&Z\\\hline
 &Z&A_2&\cdots&\overline{A_k}&X&Z
\end{array}
\end{equation}
Here, $\overline{X}=Y$ and $\overline{Y}=X$. This shows that the $\rcv$ Pauli string $ZA_2\cdots A_k$ is connected to the $\bsv$ Pauli string $ZA_2\cdots \overline{A_k} XZ$.

If there were another neighboring $\rcv$ Pauli string connected to the $\bsv$ Pauli string, the commutator relation would have the following form:
\begin{equation}\label{eq:thm2proof2}
\begin{array}{rrrrrrrr}
 & &B_2&?&\cdots&\overline{A_k}&X&Z\\
 &Z&X&Z& & & & \\\hline
  &Z&A_2&A_3&\cdots&A_k&X&Z
\end{array}
\end{equation}
However, since $A_2\neq X$, it follows that $B_2\neq I$. As a result, the commutator in Equation \ref{eq:thm2proof2} would generate a Pauli string of length $k+1$ $\rcv$ Pauli string, which is outside the scope of our graph. Therefore, there is no other neighboring $\rcv$ Pauli string connected to the $\bsv$ Pauli string, and thus the Pauli string $ZA_2\cdots A_k$ is part of a promising path.
\end{proof}

\begin{theorem}\label{thm:3}
In the commutator graph for a conserved quantity of length $k$, every Pauli string of length $k$ that starts with $ZX$ and does not end with $Z$, or its reflected form, has a promising path.
\end{theorem}

\begin{proof}
Consider a Pauli string $ZXA_3\cdots A_k$ with $A_k\neq Z$. We begin with the following commutator relation:
\begin{equation}\label{eq:thm3proof1}
\begin{array}{rrrrrrrr}
 & &Z&X&A_3&\cdots&A_{k}\\
 &Z&X& &   &      & & \\\hline
 &Z&Y&X&A_3&\cdots&A_{k}
\end{array}
\end{equation}
Thus, the $\rcv$ Pauli string $ZXA_3\cdots A_{k}$ and the $\bsv$ Pauli string $ZYXA_3\cdots A_{k}$ are connected. Since the $\bsv$ Pauli string has length $k+1$, if there were another $\rcv$ Pauli string connected to it, the only possible form would be\footnote{Since $A_k\neq Z$, the Hamiltonian string $XZ$ cannot be used.}:
\begin{equation}\label{eq:thm3proof2}
\begin{array}{rrrrrrrrr}
 &Z&Y&X&A_3&\cdots&A_{k-2}&B_{k}&\\
 & & & &   &      & &    Z&X\\\hline
  &Z&Y&X&A_3&\cdots&A_{k-2}&A_{k-1}&A_{k}
\end{array}
\end{equation}

If $A_k=Y$ then Eq.\ref{eq:thm3proof2} does not hold. Therefore, there is no other neighbor of the $\bsv$ Pauli string $ZYXA_3\cdots A_k$, and a promising path is found for $ZXA_3\cdots A_{k-1}Y$.

If $A_k=X$, then Eq.\ref{eq:thm3proof2} is a nontrivial commutator if and only if $A_{k-1}=X$ or $Y$\footnote{If $A_{k-1}=Z$ or $I$, then there is no suitable $B_k$ that satisfies Eq.\ref{eq:thm3proof2}.}. In this case, $B_k=\overline{A_{k-1}}$, and we can find two $\rcv$ neighbors of the $\bsv$ vertex $ZYXA_3\cdots A_{k-1}X$: $ZXA_3\cdots A_{k-1}X$ and $ZYXA_3\cdots A_{k-2}\overline{A}_{k-1}$.

Next, consider a $\bsv$ neighbor $ZYXA_3\cdots A_{k-2}A_{k-1}XZ$ of $ZYXA_3\cdots A_{k-2}\overline{A_{k-1}}$, which is given by the following commutator relation:
\begin{equation}\label{eq:thm3proof3}
\begin{array}{rrrrrrrrrr}
 &Z&Y&X&A_3&\cdots&A_{k-2}&\overline{A_{k-1}}&&\\
 & & & &   &      &       &Z&X&Z \\\hline
 &Z&Y&X&A_3&\cdots&A_{k-2}&A_{k-1}&X&Z
\end{array}
\end{equation}
This is the only commutator relation that produces the length $k+2$ $bsv$ Pauli string $ZYXA_3\cdots A_{k-2}A_{k-1}XZ$. Therefore, we have found a promising path for $ZXA_3\cdots A_{k-1} X$.
\end{proof}

\begin{figure}[t!] 
\captionsetup{justification=centering}
\centering
  \begin{minipage}[b]{0.5\linewidth}
    \centering
    \subfloat[Theorem 1: Proof.]{\includegraphics[width=0.75\linewidth]{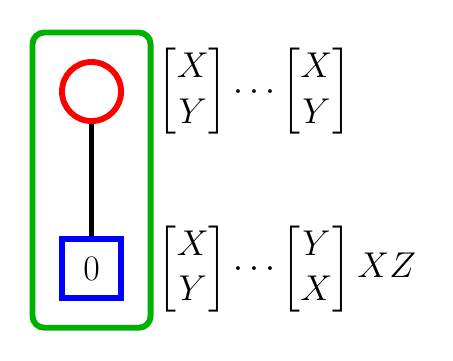}
    \label{fig:thm_1} }
    \vspace{4ex}
  \end{minipage}
  \begin{minipage}[b]{0.5\linewidth}
    \centering
    \subfloat[Theorem 2: Proof.]{\includegraphics[width=0.75\linewidth]{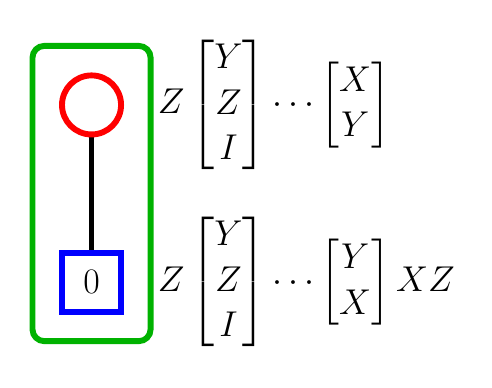}
    \label{fig:thm_2} }
    \vspace{4ex}
  \end{minipage} 
  \begin{minipage}[b]{1\linewidth}
    \centering
    \subfloat[Theorem 3: Proof.]{\includegraphics[width=0.75\linewidth]{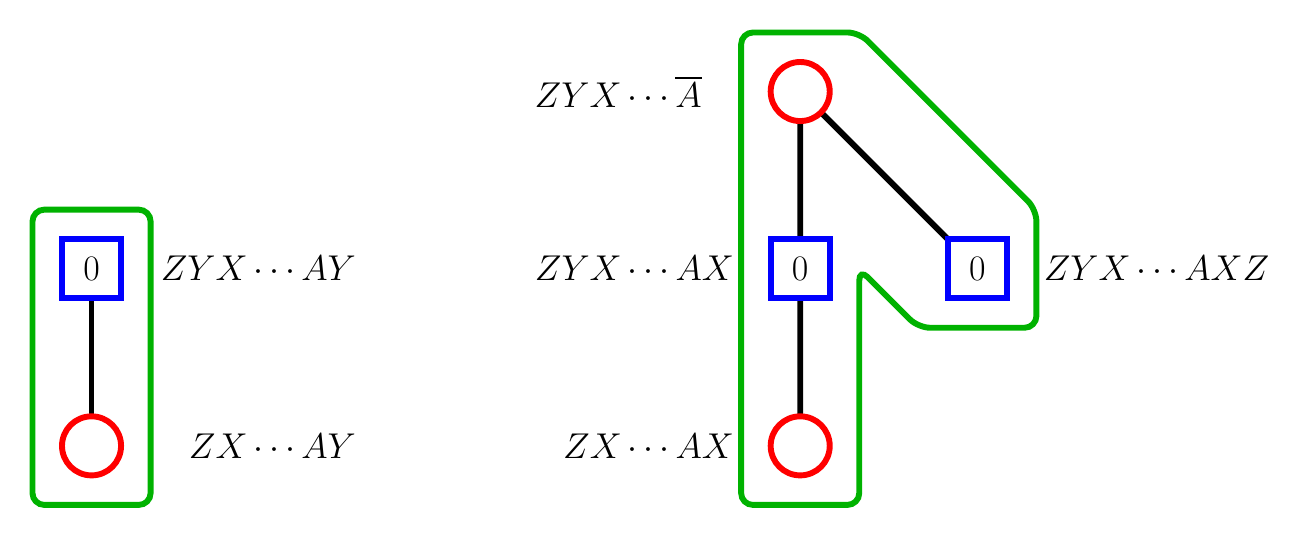}
    \label{fig:thm_3} }
  \end{minipage}
  \caption{The proofs of \ref{sub@fig:thm_1} Lemma \ref{thm:1}, \ref{sub@fig:thm_2} Lemma \ref{thm:2}, and \ref{sub@fig:thm_3} Lemma \ref{thm:3}. The green boxes represent promising paths.}
  \label{fig:proof_123} 
\end{figure}

Lemma \ref{thm:1}, \ref{thm:2}, and \ref{thm:3} demonstrate that every Pauli string of length $k$ that does not start or end with $Z$ has a promising path, and therefore has zero coefficient. This result not only reduces the number of Pauli string candidates related to a conserved quantity, if any exist, but also plays a crucial role in proving subsequent lemmas. It is important to note that the proofs of Lemmas \ref{thm:1}, \ref{thm:2}, and \ref{thm:3} can be easily visualized using graph representations, as illustrated in Fig. \ref{fig:proof_123}.

\textbf{Importance of the Edge.} --- Through the proofs of Lemmas \ref{thm:1}, \ref{thm:2}, and \ref{thm:3}, we observe a common pattern. In each case, we compute the commutator between $\rcv$ Pauli string and the Hamiltonian string, with the Hamiltonian string positioned on the ``edge" --- either the left or right edge of the Pauli string. This edge positioning is the only method by which we can obtain a $\bsv$ Pauli string of length greater than $k$ from an $\rcv$ Pauli strings of length $\leq k$.

Because this method restricts the number of neighboring $\rcv$ Pauli strings connected to the $\bsv$ Pauli string, it simplifies the search for a promising path. Indeed, in the proofs of Lemmas \ref{thm:1}, \ref{thm:2}, and \ref{thm:3}, this approach produces at most two neighboring $\rcv$ Pauli strings, which aligns perfectly with the process of identifying a promising path. If the two possible commutator representations of the $\bsv$ Pauli string arise from placing the Hamiltonian string on the left and right edges, respectively, we refer to them as ``expected" commutator representations.

For example, suppose we are trying to find a promising path starting from the Pauli string $ZZYXZ$ for $k=5$. The simplest approach is to commute the Hamiltonian string $XZ$ on the right edge, producing the Pauli string $ZZYXYZ$. Since we have already computed the commutator on the right edge, the next ``expected" commutator representation of its Pauli string should occur on the left side. Indeed, we have the following relation:
\begin{equation}\label{eq:expected_exmp}
\begin{array}{rrrrrrr|l}
 &Z&Z&Y&X&Z& &\textrm{Baseline}\\
 & & & & &X&Z&\\\hline
 &Z&Z&Y&X&Y&Z&\\\hline
 & &Y&Y&X&Y&Z&\textrm{Expected}\\
 &Z&X& & & & &\textrm{representation}\\
\end{array}
\end{equation}
This forms a possible route for the promising path of the Pauli string $ZZYXZ$. In fact, since the Pauli string $YYXYZ$ has already been shown to have a promising path (as demonstrated in Lemma \ref{thm:2}, we can conclude that the Pauli string $ZZYXZ$ also has a promising path.

\textbf{Putting and pulling method.} --- Eq.\ref{eq:expected_exmp} illustrates the process of ``putting" $XZ$ operator on the right(baseline) and ``pulling" $ZX$ operator from the left of the $\rcv$ Pauli string(expected representation). Since this is the fundamental strategy for finding the promising path of a given $\rcv$ Pauli string, we need to define it precisely.

``Putting" the Hamiltonian string on the right edge of the $\rcv$ Pauli string involves taking the commutator between the Hamiltonian string and the $\rcv$ Pauli string, where the Hamiltonian string positioned on the right edge, resulting in a $\bsv$ Pauli string longer than the original $\rcv$ one, as shown in the ``Baseline" commutator relation in Eq.\ref{eq:expected_exmp}.

``Pulling" the Hamiltonian string from the left edge of the Pauli string involves expressing the commutator relation where the Hamiltonian string is positioned on the left edge, as seen in the ``Expected" commutator relation in Eq.\ref{eq:expected_exmp}. For example, ``putting" $XZ$ on the right edge of $ZZYXZ$ gives $ZZYXYZ$, and ``pulling" $ZX$ from the left edge gives $YYXYZ$.

This ``putting and pulling" method generally yields at most two commutator representations. Thus, for any given Pauli string, we can attempt to find its promising path by repeatedly applying the putting and pulling method.

\textbf{``Unexpected" commutator representations.} --- When dealing with Pauli strings that start end with $Z$ operators, we occasionally encounter ``unexpected" commutators. These commutators are not anticipated by the putting and pulling method, and they result in more than two neighboring $\rcv$ Pauli strings connected to the $\bsv$ Pauli string, leading to cases classified as Exception 1.

These unexpected commutator representations can occur at various positions within the Pauli string. Depending on the position of the Hamiltonian string in unexpected commutator, the method to resolve the exceptional cases differs. Therefore, it is useful to classify Exception 1 cases into smaller subcategories based on the position of the Hamiltonian string for the unexpected commutator: right, left, or middle. This classification is general and can be applied to a wide range of spin-$1/2$ Hamiltonian systems.

\begin{figure}[t!] 
\captionsetup{justification=centering}
\centering
  \begin{minipage}[b]{0.5\linewidth}
    \centering
    \subfloat[Exception 1: Category 1.]{\includegraphics[width=0.9\linewidth]{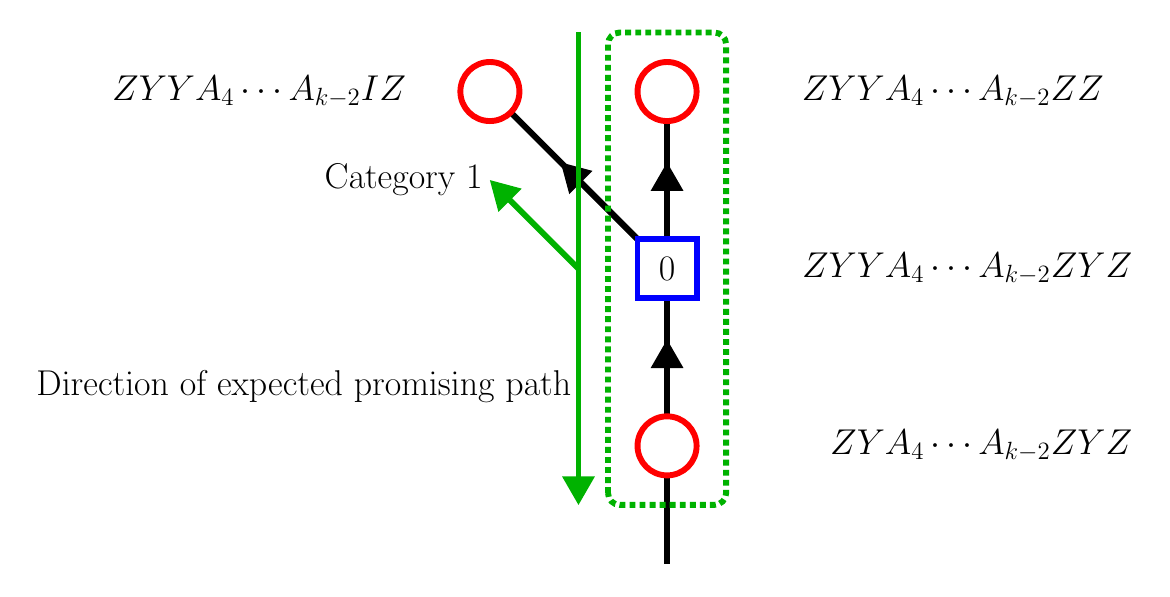}
    \label{fig:cat_1} }
    \vspace{4ex}
  \end{minipage}
    \begin{minipage}[b]{0.5\linewidth}
    \centering
    \subfloat[Avoiding Category 1.]{\includegraphics[width=0.9\linewidth]{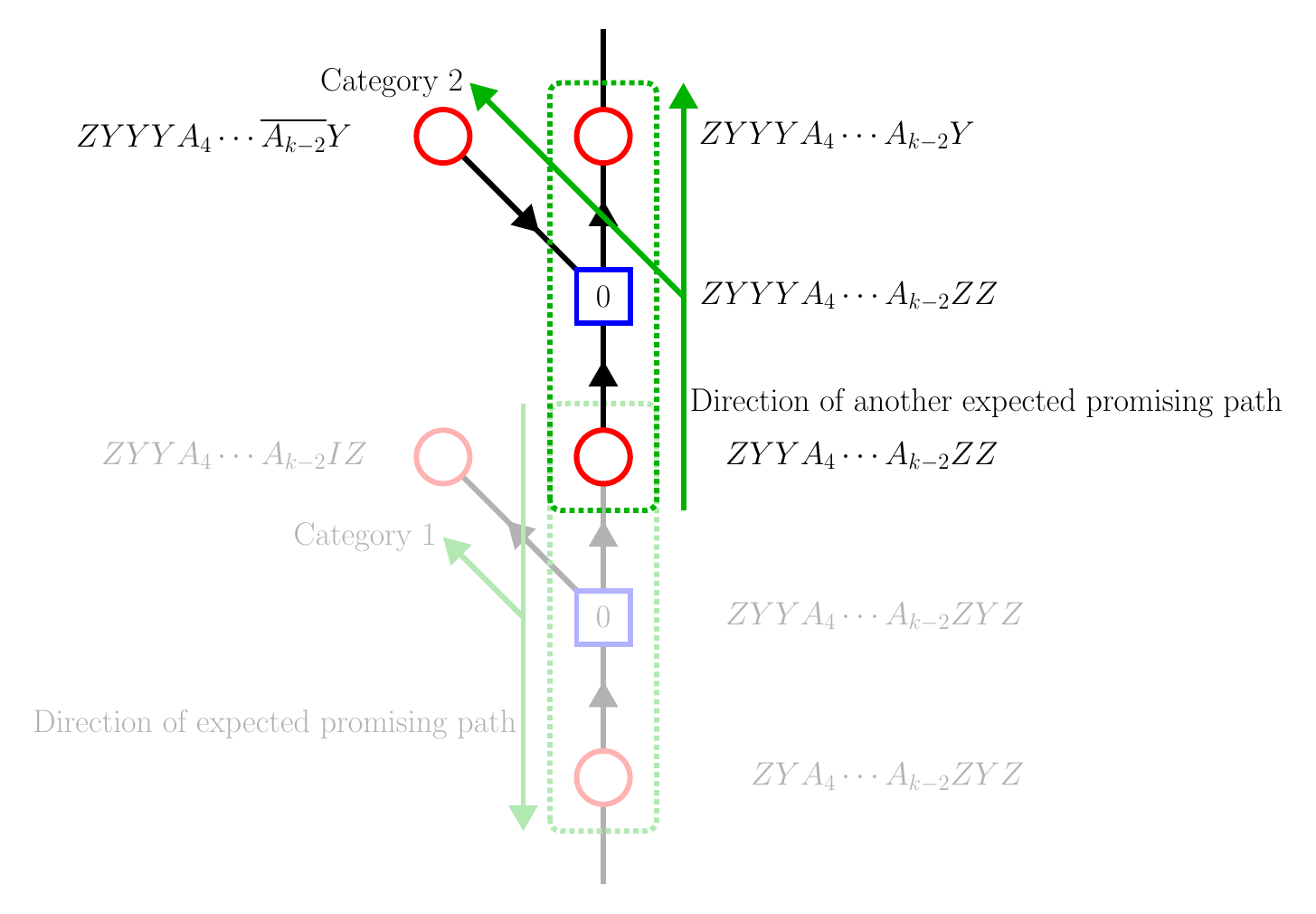}
    \label{fig:cat_1_avoid} }
        \vspace{4ex}
  \end{minipage}
    \begin{minipage}[b]{0.5\linewidth}
    \centering
    \subfloat[Impossible to avoid Category  1.]{\includegraphics[width=0.9\linewidth]{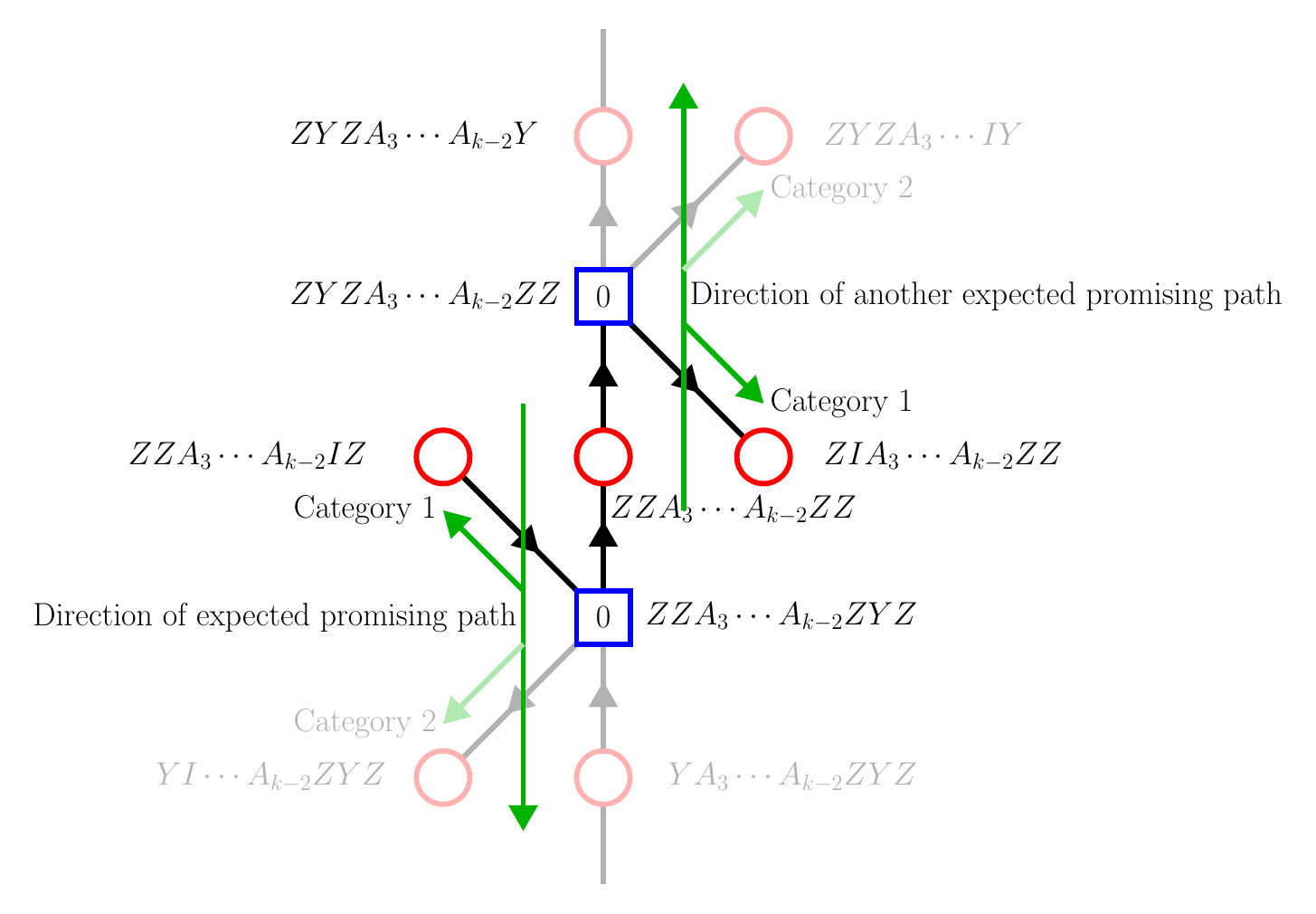}
    \label{fig:cat_1_avoid_exception} }
        \vspace{4ex}
  \end{minipage}
  \caption{\ref{sub@fig:thm_1} Diagrammatic representation of Category 1. \ref{sub@fig:thm_2} An example where the Category 1 case can be avoided. \ref{sub@fig:thm_3} An example where the Category 1 case cannot be avoided. Green boxes represent the expected Pauli strings, and green arrows indicate their directions.}
  \label{fig:category} 
\end{figure}

\textbf{Category 1.} --- Consider the $\rcv$ Pauli string that starts with $ZYY$ and ends with $ZZ$. Taking the commutator with the Hamiltonian string $XZ$ on the right edge results in a $\bsv$ Pauli string of length $k+1$. In this case, in addition to the ``expected" commutator with Hamiltonian string on the left edge, there is another commutator with Hamiltonian string on the \textit{right} edge. Specifically, we have the following relation:
\begin{equation}\label{eq:cat1eq1}
\begin{array}{rrrrrrrrrrr|l}
 &Z&Y&Y&A_4&\cdots& &A_{k-2}&Z&Z&&\textrm{Baseline}\\
 & & & &   &      & &       & &X&Z&\\\hline
 &Z&Y&Y&A_{4}&\cdots& &A_{k-2}&Z&Y&Z&\\\hline
 &\tcr{Z}&\tcr{Y}&\tcr{Y}&\tcr{A_{4}}&\tcr{\cdots}& &\tcr{A_{k-2}}&\tcr{I}&\tcr{Z}&&\textrm{Cat 1}\\
 & & & &   &      & &       &\tcr{Z}&\tcr{X}&\tcr{Z}&\\\hline
 & &Z&Y&A_{4}&\cdots& &A_{k-2}&Z&Y&Z&\textrm{Expected}\\
 &Z&X& &   &      & &       & & & &\textrm{representation}\\
\end{array}
\end{equation}
This relation introduces three neighboring $\rcv$ Pauli strings to a single length $k+1$ $\bsv$ Pauli string. The red-colored second commutator is an unexpected commutator, with the Hamiltonian string acting on the right side. See Fig.\ref{fig:cat_1} for a diagrammatic explanation. In the graph, while following the expected promising path by ``putting" the Hamiltonian string on the right and ``pulling" from the left, the promising path ``bounces back" due to the unexpected commutator on the right edge. This situation occurs for $\rcv$ Pauli strings that end with $ZZ$ or $IZ$.

If the Pauli string which ends with $ZZ$ or $IZ$ does not start with $ZZ$ or $ZI$ but instead starts with $ZX$ or $ZY$, such as the Baseline case in Eq.\ref{eq:cat1eq1}, we can avoid the Category 1 case by following a different direction for the expected promising path. Specifically, we now ``put" the Hamiltonian string $ZX$ on the \textit{left} edge and ``pull" the Hamiltonian string from the \textit{right} of the Pauli string, oppose to the original putting and pulling method. See Fig.\ref{fig:cat_1_avoid} for details. In this scenario, another type of unexpected representation arises (which we refer to as the Category 2 case, to be addressed later), which we will resolve soon.

On the other hand, if the Pauli string both ends with $ZZ$ or $IZ$ \textit{and} starts with $ZZ$ or $ZI$, it becomes impossible to avoid the Category 1 case. In Fig.\ref{fig:cat_1_avoid_exception}, we observe that in this situation the Category 1 case appears regardless of the direction taken for the expected promising path. Pauli strings in this category are strongly related to the trivial operators, as discussed in Theorem 1 of the main text.

Therefore, we conclude that each Pauli string in Category 1 can either be treated as a Pauli string in Category 2 or is related to the trivial operators. This implies that we do not need to further consider the Pauli strings in Category 1, or in other words, the Pauli strings that end with $ZZ$ or $IZ$.
 
\begin{figure}[t] 
\captionsetup{justification=centering}
    \centering
    \includegraphics[width=0.5\linewidth]{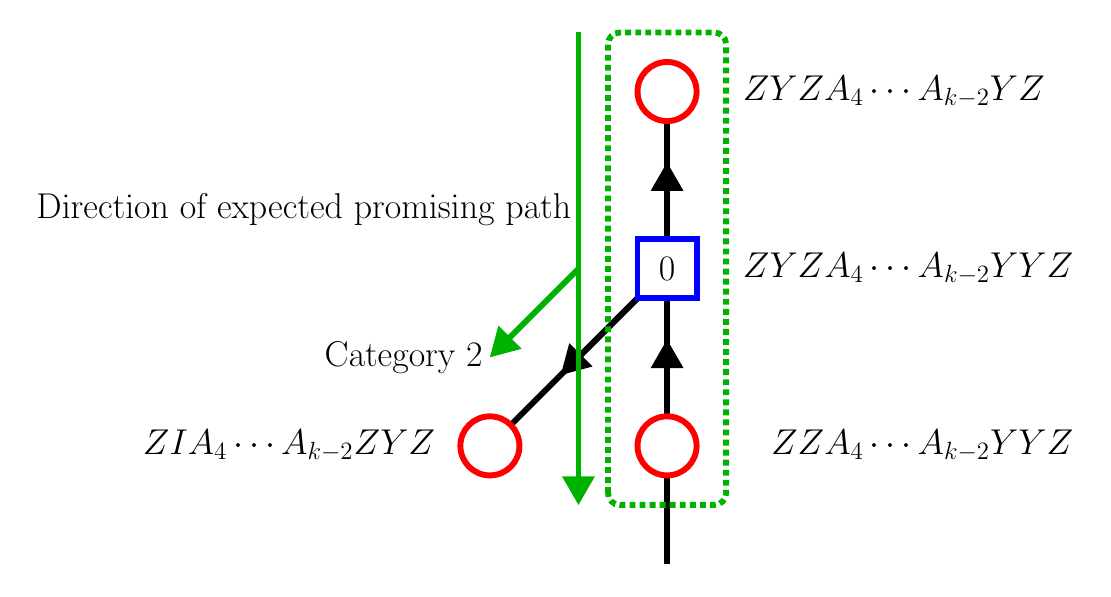}
\caption{The diagrammatic representation of Category 2.}
    \label{fig:cat_2} 
\end{figure}

\textbf{Category 2.} --- Consider the $\rcv$ Pauli string that starts with $ZYZ$ and ends with $Z$. Taking the commutator with the Hamiltonian string $XZ$ on the right edge results in a $\bsv$ Pauli string of length $k+1$. In this case, in addition to the ``expected" commutator with Hamiltonian string on the left edge, there is another commutator with Hamiltonian string on the \textit{left} edge. Specifically, we have the following relation:
\begin{equation}\label{eq:cat2eq1}
\begin{array}{rrrrrrrrrrr|l}
 &Z&Y&Z&A_4&\cdots& &A_{k-2}&Y&Z&&\textrm{Baseline}\\
 & & &   &      & &       & & &X&Z&\\\hline
 &Z&Y&Z&A_4&\cdots& &A_{k-2}&Y&Y&Z&\\\hline
 & &Z&Z&A_4&\cdots& &A_{k-2}&Y&Y&Z&\textrm{Expected}\\
 &Z&X&   &      & &       &&&&&\textrm{commutator}\\\hline
 & &\tcr{Z}&\tcr{I}&\tcr{A_4}&\tcr{\cdots}& &\tcr{A_{k-2}}&\tcr{Y}&\tcr{Y}&\tcr{Z}&\textrm{Cat 2}\\
 &\tcr{Z}&\tcr{X}&\tcr{Z}&      & &      & &&&&\\
\end{array}
\end{equation}
This relation introduces three neighboring $\rcv$ Pauli strings. The red-colored third commutator is an unexpected commutator, with the Hamiltonian string acting on the left edge. See Fig.\ref{fig:cat_2} for a diagrammatic explanation. In the graph description, we observe that while following the expected promising path by ``putting" the Hamiltonian string on the right and ``pulling" from the left, the promising path is ``branched" due to the unexpected commutator representation on the left edge. This situation occurs with Pauli strings that start with $ZYZ, ZYI, ZZZ,$ or $ZZI$ (or their reflected forms).

\begin{figure}[t] 
\captionsetup{justification=centering}
    \centering
    \includegraphics[width=0.5\linewidth]{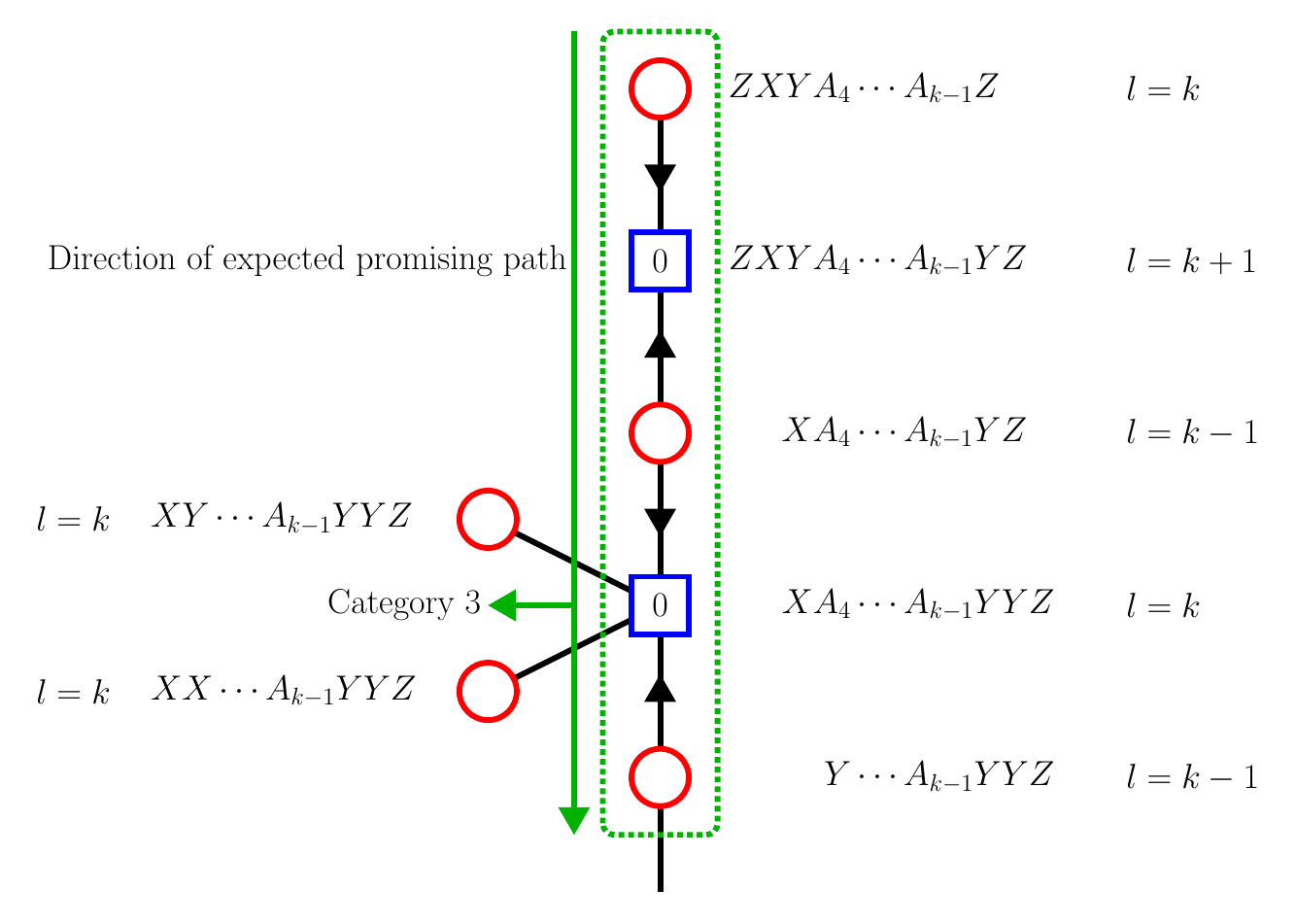}
\caption{The diagrammatic representation of Category 3.}
    \label{fig:cat_3} 
\end{figure}

\textbf{Category 3.} Consider the $\rcv$ Pauli string that starts with $ZXY$ and ends with $Z$. Take the commutator with Hamiltonian string $XZ$ on the right edge results in a $\bsv$ Pauli string of length $k+1$. This Pauli string has another neighboring $\rcv$ Pauli string of length $k-1$, as shown below.
\begin{equation}\label{eq:cat3eq2}
\begin{array}{rrrrrrrrrrr|l}
 &Z&X&Y&A_4&\cdots& &A_{k-1}&Z& & & l=k\\
 & & &   &   &      & &       &X&Z& & \\\hline
 &Z&X&Y&A_4&\cdots& &A_{k-1}&Y&Z& & l=k+1\\\hline
  & & &X&A_4&\cdots& &A_{k-1}&Y&Z& & l=k-1\\
 &Z&X&Z& &      & &       & & & & \\
\end{array}
\end{equation}

In this case there are only two commutators containing the $\bsv$ Pauli string, with no unexpected commutator. However, the problem arises when we  attempt the same process on the length $k-1$ Pauli string. At this point, there can be numerous neighboring $\rcv$ Pauli strings. For example,
\begin{equation}\label{eq:cat3eq1}
\begin{array}{rrrrrrrrrrrr|l}
 &X&Y&A_3&A_4&\cdots& &A_{k-2}&Y&Z& & & l=k-1\\
 & & &   &   &      & &       & &X&Z& & \textrm{Baseline}\\\hline
 &X&Y&A_3&A_4&\cdots& &A_{k-2}&Y&Y&Z& & l=k\\\hline
  & &X&?&?&\cdots& &A_{k-2}&Y&Y&Z& & l=k-1\\
 &X&Z& & &      & &       & & & & &\textrm{Exp. comm.} \\\hline
 &\tcr{X}&\tcr{Y}&\tcr{?}&\tcr{?}&\tcr{\cdots}& &\tcr{A_{k-2}}&\tcr{Y}&\tcr{Y}&\tcr{Z}& & l=k\\
 & & &\tcr{Z}&\tcr{X}&      & &       & & & & & \textrm{Cat 3}\\\hline
 &\tcr{X}&\tcr{X}&\tcr{?}&\tcr{?}&\tcr{\cdots}& &\tcr{A_{k-2}}&\tcr{Y}&\tcr{Y}&\tcr{Z}& & l=k\\
 & &\tcr{Z}&\tcr{X}&\tcr{Z}&      & &      & & & & & \textrm{Cat 3}\\\hline
 & &&&&\vdots      & &       &&& && \\
\end{array}
\end{equation}

Here, the red-colored third and fourth commutator representations (and many other possible representations not shown) are unexpected commutator representations, with the Hamiltonian string positioned in the middle. This occurs because the length of the $bsv$ Pauli string is $k$, and in the commutator that generates the $\bsv$ Pauli string, the Hamiltonian string does not necessarily need to be positioned on the edge. See Figure \ref{fig:cat_3} for the diagrammatic explanation. In the graph representation, we observe that while following the expected promising path by putting the Hamiltonian string on the right and pulling from the left, the promising path becomes ``dissipated" by various edges connected to the $\bsv$ Pauli string. This situation occurs with Pauli strings that start with $ZXY$ or $ZXX$.

\textbf{Summary in Exception 1.} --- In summary, we classified the Pauli strings in Exception 1 based on the position of the Hamiltonian string in the unexpected commutators. 

In \textbf{Category 1}, which involves $\rcv$ Pauli strings that end with $ZZ$ or $IZ$, the unexpected commutator occurs with the Hamiltonian string on the left edge of the $\rcv$ Pauli string. However, these Pauli strings can either be treated as belonging to Category 2 by using mirror symmetry of the Hamiltonian, or they are related to trivial operators. Therefore we can disregard Category 1 and instead focus on Pauli strings that end with $XZ$ or $YZ$.

In \textbf{Category 2}, which involves $\rcv$ Pauli strings that start with $ZYZ, ZYI, ZZZ,$ or $ZZI$, the unexpected commutator occurs with the Hamiltonian string on the right edge of the $\rcv$ Pauli string. This is a branching case, as mentioned in the main text, and will ultimately be shown to have a zero coefficient.

In \textbf{Category 3}, which involves $\rcv$ Pauli strings that start with $ZXY$ or $ZXX$, the unexpected commutator does not occurs directly. However, one can find a neighboring $\bsv$ Pauli string that has only two neighbors: one is the original $\rcv$ Pauli string, and the other is a $\rcv$ Pauli string with length $k-1$. All neighbors of this length $k-1$ $\rcv$ Pauli string has more than three neighbors, where the unexpected commutator occurs with Hamiltonian string positioned in the middle.

\textbf{Treating Category 3.} --- Now, we show that the exceptions in Category 3 can be effectively addressed. Specifically, for every $\rcv$ Pauli string in Category 3, we can identify a neighboring $\bsv$ Pauli string that has at most three neighboring $\rcv$ Pauli strings. In fact, most commutators with the Hamiltonian string positioned in the middle do not contribute to the $\bsv$ Pauli string. The proof relies on \textbf{Lemmas 1}, \textbf{2}, and \textbf{3}, which eliminated a number of Pauli strings with zero coefficient by identifying a promising path.

\begin{theorem}\label{thm:4}
In the commutator graph of the PXP model for quantities of length $k$, consider a $\rcv$ Pauli string of length $k-1$ that ends with $Z$ and does not start with $Z$. Then we can always find a neighboring $\bsv$ Pauli string by putting the $XZ$ Hamiltonian string on the right edge of the Pauli string, such that all its neighboring $\rcv$ have zero coefficients, except for at most three: the two Pauli strings which gives the expected commutators, and a Category 2-type unexpected commutator (if exists).
\end{theorem}

\begin{proof}
First, consider the Pauli string $XA_2\cdots A_{k-2}Z$, and compute the following commutator relation:
\begin{equation}\label{eq:thm4proof1}
\begin{array}{rrrrrrrrr|l}
 &X&A_2&\cdots&A_{k-2}&Z&& & & l=k-1\\
 & &   &      & &X&Z& & & \\\hline
 &X&A_2&\cdots&A_{k-2}&Y&Z& & & l=k
\end{array}
\end{equation}
Now suppose that there exists another commutator which gives $XA_3\cdots A_{k-2} YZ$, where the Hamiltonian string is positioned in the middle of the Pauli string. In that case, the leftmost character $X$ will not be altered, resulting in the following commutator relation:
\begin{equation}\label{eq:thm4proof2}
\begin{array}{rrrrrrrrrrrrr|l}
 &X&A_2&\cdots&?&?&?&\cdots&A_{k-2}&Y&Z& & &l=k\\
 & &   &      &Z&X&Z&& & & & & & \\\hline
 &X&A_2&\cdots&?&?&?&\cdots&A_{k-2}&Y&Z& & & l=k
\end{array}
\end{equation}
However, \textbf{Lemma 2} states that the coefficient of $XA_2\cdots A_{k-2}YZ$ is zero. Since this holds unless the Hamiltonian string changes the leftmost character $X$ to $Z$ or $I$, the only possible commutator relation is:
\begin{equation}\label{eq:thm4proof3}
\begin{array}{rrrrrrrrr|l}
 &X&A_2&\cdots&A_{k-2}&Y&Z& & & l=k\\
 &X&  Z&      & & & & & & \\\hline
 & &\overline{A_2}&\cdots&A_{k-2}&Y&Z& & & l=k-1
\end{array}
\end{equation}
This shows that for the length $k$ $\bsv$ Pauli string $XA_2\cdots A_{k-2}YZ$, there are at most two neighboring $\rcv$ Pauli strings: these are the expected commutators.

For the Pauli string $YA_2\cdots A_{k-2}Z$, the logic is similar, but there may be two additional commutators that give the same $\bsv$ Pauli string:
\begin{equation}\label{eq:thm4proof4}
\begin{array}{rrrrrrrrr|l}
 &Y&A_2&\cdots&A_{k-2}&Z&& & & l=k-1\\
 & &   &      & &X&Z& & & \\\hline
 &Y&A_2&\cdots&A_{k-2}&Y&Z& & & l=k\\\hline
 &Z&\overline{A_2}&\cdots&A_{k-2}&Y&Z& & & l=k\\
 &X&  Z&      & & & & & & \\\hline
 &Z&A_2&\cdots&A_{k-2}&Y&Z& & & l=k\\
 &X&&      & & & & & & \\
\end{array}
\end{equation}

This shows that for the length $k$ $\bsv$ Pauli string $YA_2\cdots A_{k-2}YZ$, there are at most three neighboring $\rcv$ Pauli strings: the last one gives the Category 2-type unexpected commutator. This completes the proof.
\end{proof}

From the argument regarding Category 1 and \textbf{Lemma 4}, we can conclude that the unexpected commutators in Category 1 and 3 do not need to be considered. This leaves us with only the Category 2 case, which includes a variety of Pauli string scenarios that require careful treatment.

\textbf{Treating Category 2.} --- Now we analyze the unexpected commutators in the Category 2 case. In Fig.\ref{fig:cat_2}, the expected promising path branches due to an unexpected commutator. However, we can still attempt to find the promising path along each branch by putting the $XZ$ string on the right and pulling the Hamiltonian string from the left.

The key point is that this repetitive application of the putting and pulling method only generates unexpected commutators in Category 2. This is because (i) we have already determined that all unexpected commutators in Category 3 can be disregarded, and (ii) putting the $XZ$ string on the right changes the right edge into $\cdots YZ$, which prevents the creation of an unexpected commutator in Category 1, as those only occur when the Pauli string ends with $\cdots ZZ$ or $\cdots IZ$.

Thus, we are left with two possibilities for a $\rcv$ Pauli string: either every branch caused by the unexpected commutators in Category 2 becomes part of a promising path, or the $\rcv$ Pauli string is part of a loop.

To examine the case where every branch becomes part of a promising path, recall Eq.\ref{eq:cat2eq1} and the graph representation in Fig.\ref{fig:cat_2}. The $\bsv$ vertex $ZYZA_4\cdots A_{k-2}YYZ$ has three neighboring $\rcv$ vertices: the original $\rcv$ vertex, the $\rcv$ vertex $ZZA_4\cdots A_{k-2}YYZ$ given by the expected commutator, and the $\rcv$ vertex $ZIA_4\cdots A_{k-2}ZYZ$ given by the unexpected commutator.

First, focus on the $\rcv$ vertex resulting from the expected commutator. By putting $XZ$ on the right edge and pulling the Hamiltonian strings from the left, we get the following:
\begin{equation}\label{eq:midthm1}
\begin{array}{rrrrrrrrrrrr}
 &Z&Z&A_4&\cdots& &A_{k-2}&Y&Y&Z&\\
 & & &   &      & &       & & &X&Z\\\hline
 &Z&Z&A_4&\cdots& &A_{k-2}&Y&Y&Y&Z\\\hline
 & &Y&A_4&\cdots& &A_{k-2}&Y&Y&Y&Z\\
 &Z&X&   &      & &       &&&&\\\hline
 & &Y&\overline{A_4}&\cdots& &A_{k-2}&Y&Y&Y&Z\\
 &Z&X&Z&      & &       &&&&\\\hline
\end{array}
\end{equation}
Note that the third commutator gives a nontrivial result only when $A_4=I$ or $Z$; here, we use the notation $\overline{I}=Z$ and $\overline{Z}=I$. This shows that from the $\rcv$ vertex $ZZA_4\cdots A_{k-2}YYZ$, we encounter another unexpected commutator in Category 2 when $A_4=I$ or $Z$, resulting in an additional branch.

One surprising point is that, due to \textbf{Lemmas 2} and \textbf{3}, each $\rcv$ Pauli string in the second and third commutator of Eq.\ref{eq:midthm1} is part of a promising path. Therefore, each branch of the $\rcv$ vertex $ZZA_4\cdots A_{k-2} YYZ$ is part of a promising path.

\begin{figure}[h!]
\captionsetup{justification=centering}
\centering
 \subfloat[]{\label{fig:flowchart_pre_a}\includegraphics[height=0.4\textwidth]{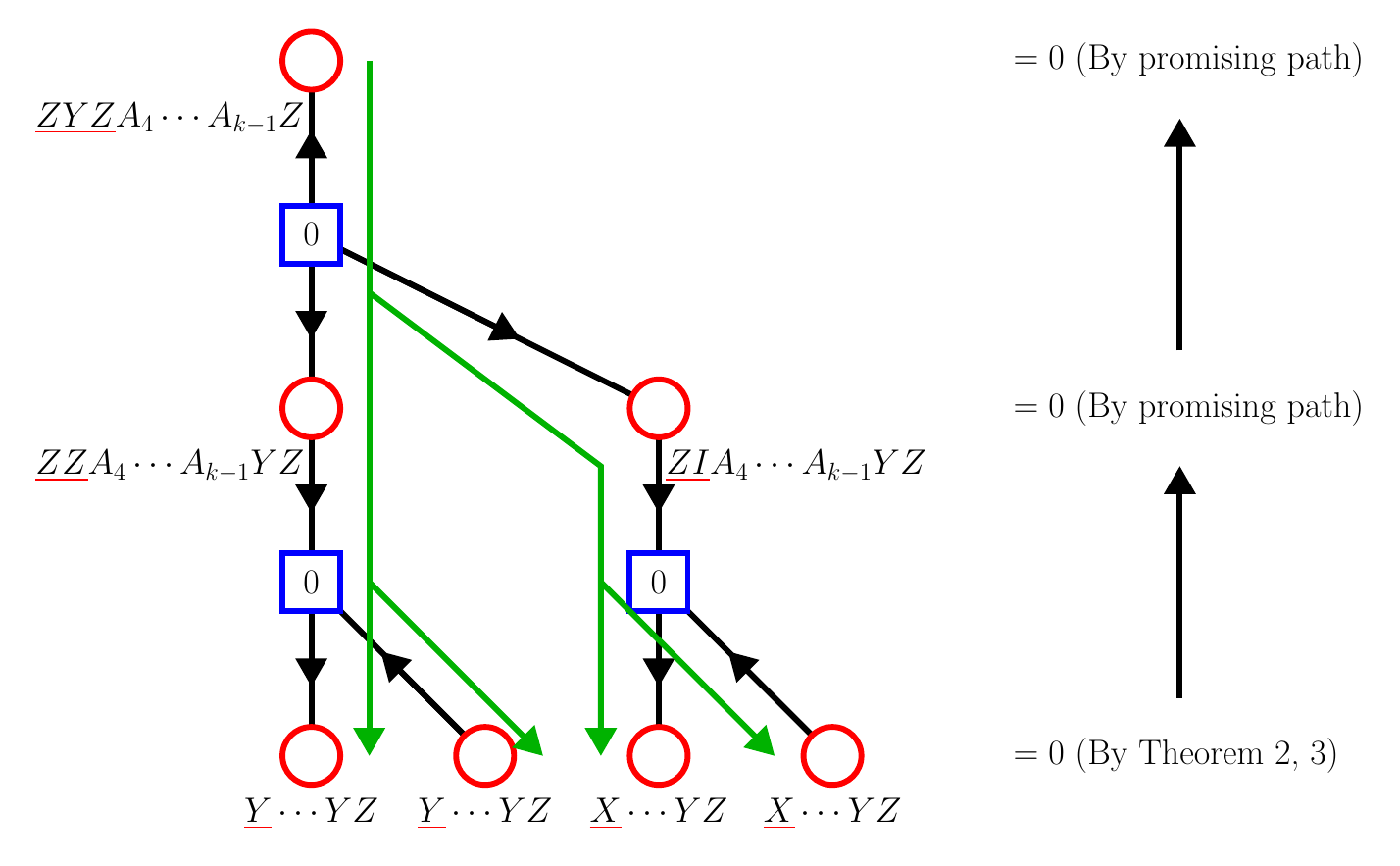}}
 \qquad
 \subfloat[]{\label{fig:flowchart_pre_b}\includegraphics[height=0.4\textwidth]{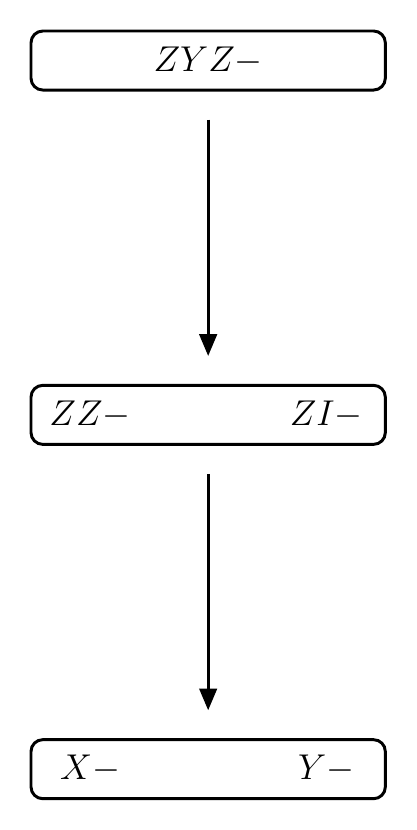}}
  \caption{\ref{sub@fig:flowchart_pre_a} Graph representation showing that the coefficient of $ZYZA_4\cdots A_{k-1}Z$ vanishes. The thick arrows on the right indicate the logical progression we follow. \ref{sub@fig:flowchart_pre_b} Flowchart of the left-edge Pauli substring. Each box represents a set of Pauli strings whose left edge substring, highlighted by red underscore, matches the string inside the box. Following the arrows, we observe how the left-edge Pauli substring changes as we repeatedly apply the process of putting $XZ$ string on the right and pulling the Hamiltonian strings from the left.}
  \label{fig:flowchart_pre}
\end{figure}

If we apply the same process to the $\rcv$ vertex $ZIA_4\cdots A_{k-2}ZYZ$, we again find a branch where each part is included in a promising path. Fig.\ref{fig:flowchart_pre_a} illustrates the entire process, which is followed while finding the promising path that includes the top $\rcv$ vertex, $ZYZA_4\cdots A_{k-1}Z$. Since each of the $\rcv$ vertices at the bottom is part of a promising path, we can ignore them, leaving the lowest $bsv$ vertices with only one $\rcv$ neighbor that has a nonzero coefficient. This directly shows that each $\rcv$ vertex in the middle is part of a promising path, meaning that the top $\bsv$ vertex has only have one $\rcv$ neighbor that has a nonzero coefficient, thus confirming a promising path to the top $\rcv$ vertex.

A crucial point here is that, at each step, the details of the middle and right edge of the $\rcv$ Pauli string are not very important (as long as it does not end with $\cdots ZZ$ or $\cdots IZ$, which do not occurs since we already removed every Pauli string occuring unexpected commutators in Category 1). The left-edge Pauli substrings are the key focus. 

Fig.\ref{fig:flowchart_pre_b} presents a simplified diagram, showing only the left edge Pauli substrings of the Pauli strings in Fig.\ref{fig:flowchart_pre_a}. We begin with a length $k$ $\rcv$ Pauli string that starts with $ZYZ$, then move to length $k$ $\rcv$ Pauli strings that start with $ZZ-$ or $ZI-$, followed by length $k$ $\rcv$ $X-$ and $Y-$ Pauli strings. We can stop here, as we know that each of these types of Pauli strings is part of a promising path, as shown in \textbf{Lemmas 2} and \textbf{3}, meaning our task is complete.

As we have demonstrated above, determining whether a Pauli string is part of a promising path or is included in a loop strongly depends on understanding how the left edge of the Pauli string changes during the putting and pulling process. The following lemma and its proof explain how this change in the left edge occurs and how it can be used to classify Pauli strings.

\begin{theorem}\label{thm:5}
For an $\rcv$ Pauli string that does not end with $IZ$ or $ZZ$, the following statements hold:
\begin{enumerate}
\item If the $\rcv$ Pauli string starts and ends with $Z$ and contains only $X$ or $Y$ operators between these $Z$'s, then it falls under the Type 2 Exception. Specifically, the collection of such Pauli strings with an even number of $X$ operators forms one loop $L_e$, while the collection of such Pauli strings with an odd number of $X$ operators forms a different loop $L_o$.
\item If the above condition is not met, then the $\rcv$ Pauli string is part of a promising path.
\end{enumerate}
\end{theorem}

\begin{figure}[h!]
\captionsetup{justification=centering}
\centering
  \includegraphics[width=0.9\textwidth]{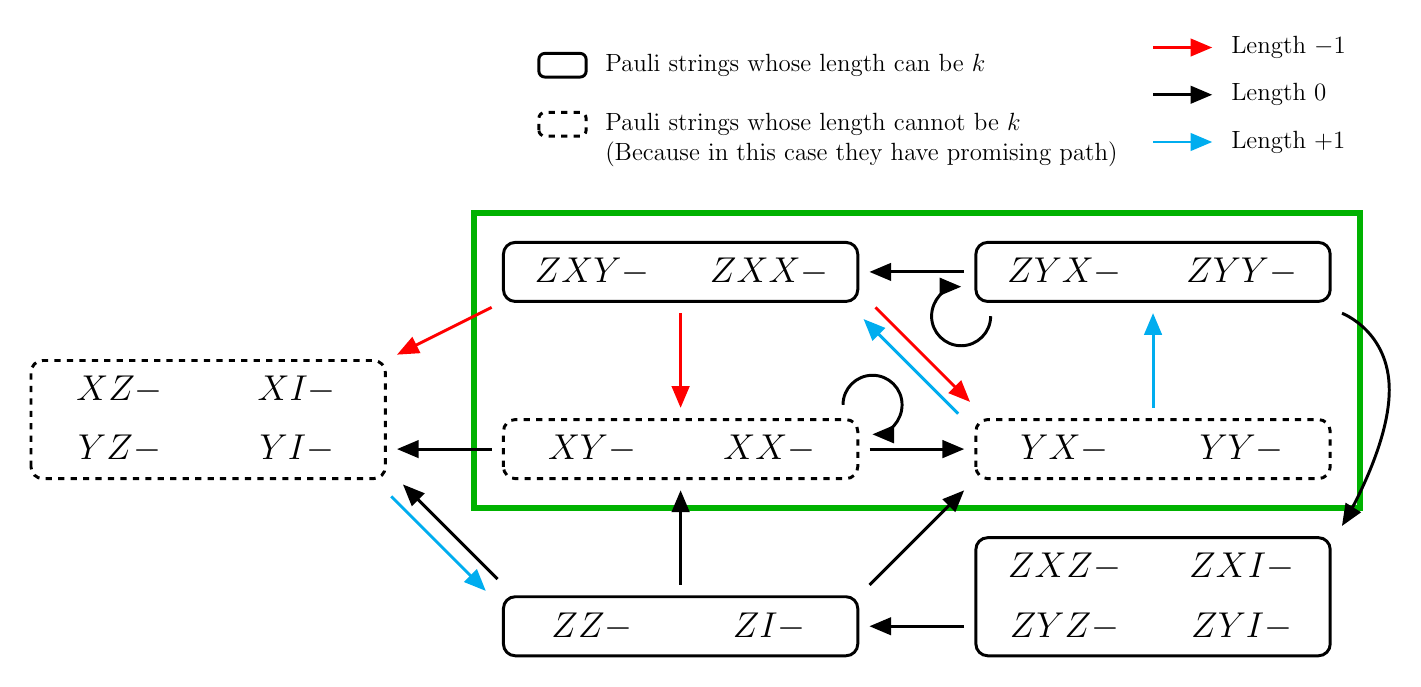}
  \caption{Flowchart of the left-edge substring of the Pauli string during the putting and pulling process. Each black box represents a set of Pauli strings that share the same left-edge substring as the Pauli string shown in the box. Each arrow represents how the left-edge substring changes after a single putting and pulling process. The red arrow indicates a decrease in the length of the Pauli string by $1$, the blue arrow indicates an increase in length by $1$, and the black arrow indicates that the length of the Pauli string remains unchanged. For more details, see the proof of \textbf{Lemma 5}.}
  \label{fig:flowchart}
\end{figure}

\begin{proof}
Fig.\ref{fig:flowchart} illustrates every possible change to the left-edge Pauli substring during the putting and pulling process. The figure is read as follows:

Consider a length $k$ $\rcv$ Pauli string $P$. Find the box that includes the left-edge Pauli substring of $P$, i.e. the box containing a Pauli substring that mathes the left-edge of $P$. Track the arrows from that box and collect all the left-edge Pauli substrings found at the end of the arrows. If the arrow is red (blue), it indicates that the length of the $\rcv$ Pauli string increases (decreases) by $1$ compared to $P$. This collection of the left-edge Pauli substrings provides the possible range of Pauli strings obtained after a single putting and pulling process on $P$.

Here are some examples:
\begin{enumerate}
\item If $P$ is a length $k$ Pauli string starting with $ZXZ-$, then after a single putting and pulling process, we get length $k$ Pauli strings starting with $ZZ-$ or $ZI-$.
\item If $P$ is a length $k-1$ Pauli string starting with $YX-$, then we can find its neighboring $\bsv$ vertex whose other $\rcv$ neighbors are represented by length $k$ Pauli strings starting with $ZXY-, ZXX-, ZYX-,$ or $ZYY-$.
\end{enumerate}

Notice that the number of neighbors is not restricted; there may be none or more than one. The \textit{actual} $\rcv$ Pauli strings obtained from a single putting and pulling process on $P$ depend on the operators immediately following the left-edge Pauli substring. For example, after a single putting and pulling process, a length $k-1$ Pauli string $P$ starting with $YX-$ becomes a length $k$ Pauli string starting with $ZXY-$ when the operator following $X$ in $P$ is $Y$, and becomes a length $k$ Pauli string starting with $ZXX-$ when the operator following $X$ in $P$ is $X$.

From the definition of the flowchart, we can deduce the following: If we start from a particular box and follow the arrows, eventually reaching the dashed boxes with length $k$, then since those Pauli strings have promising paths, each Pauli string in the initial box has also has a promising path. This follows by similar reasoning as shown in Fig.\ref{fig:flowchart_pre}.

Now we examine all the possibilities for the $\rcv$ Pauli string.

Suppose we have a length $k$ Pauli string that starts with $ZZ-$ or $ZI-$. Following the arrows, we obtain length $k$ Pauli strings that start with $X-$ or $Y-$, both of which have a promising path. Therefore, each Pauli string starting with $ZZ-$ and $ZI-$ also has a promising path.

Next, Suppose we have a length $k$ Pauli string starting with $ZXZ-$, $ZXI-$, $ZYZ-$, or $ZYI-$. Following the arrows, we reach length $k$ Pauli strings starting with $ZZ-$ or $ZI-$, both of which, as shown earlier, have a promising path. Therefore, each Pauli string starting with $ZXZ-$, $ZXI-$, $ZYZ-$, or $ZYI-$ also has a promising path.

Now, consider a length $k$ Pauli string starting with $ZYX-$ or $ZYY-$. Suppose that, possibly after some self-loop, we follow the arrow toward the box containing $ZXZ-$, with length $k$. Since each Pauli strings in this box has a promising path, as shown earlier, the original length $k$ Pauli string is also part of a promising path.

Similarly, consider a length $k$ Pauli string starting with $ZXY-$ or $ZXX-$. If we follow the arrow toward the box containing $XZ-$, with length $k-1$, and then follow the arrows again, we get length $k$ Pauli strings starting with $ZZ-$ or $ZI-$, for which we have already shown that these Pauli strings are always included in a promising path. Hence, the original length $k$ Pauli string is also part of a promising path.

Finally, consider a length $k$ Pauli string starting with $ZXY-$ or $ZXX-$, and follow the red arrow toward the box containing $XY-$ and $XX-$. Suppose that, possibly after some self-loop, we follow the arrow toward the box containing $XZ-$, with length $k-1$. We have already shown that each Pauli string in this box has a promising path. Therefore, the original length $k$ Pauli string is also part of a promising path.

The key point of these arguments is as follows: if we leave the green box area, we can always find a promising path. The only situation in which we cannot find a promising path is when we remain entirely within the green box area. Now, if a $Z$ or $I$ operator exists between the two $Z$ operators on the edge of the original Pauli string, then by repeatedly applying the putting and pulling process, this $Z$ or $I$ operator is pushed out to the left edge of the Pauli string, resulting in a Pauli string outside the green box\footnote{Notice that in the green box, there are no $Z$ or $I$ operators except the leftmost $Z$ operator, while outside the green box, every Pauli string contains $Z$ or $I$ operators on the left edge in addition to the leftmost $Z$ operator (if exists).}. If there is no such $Z$ or $I$ operator, then during the putting and pulling process, the string never exits the green box.

This argument shows that if a Pauli string starts with $Z-$ and ends with $-XZ$ or $-YZ$, and contains a $Z$ or $I$ operator in the middle, one can always find a promising path that includes it. This proves the second part of \textbf{Lemma 5}.

\begin{figure}[h!]
\captionsetup{justification=centering}
\centering
  \includegraphics[width=0.9\textwidth]{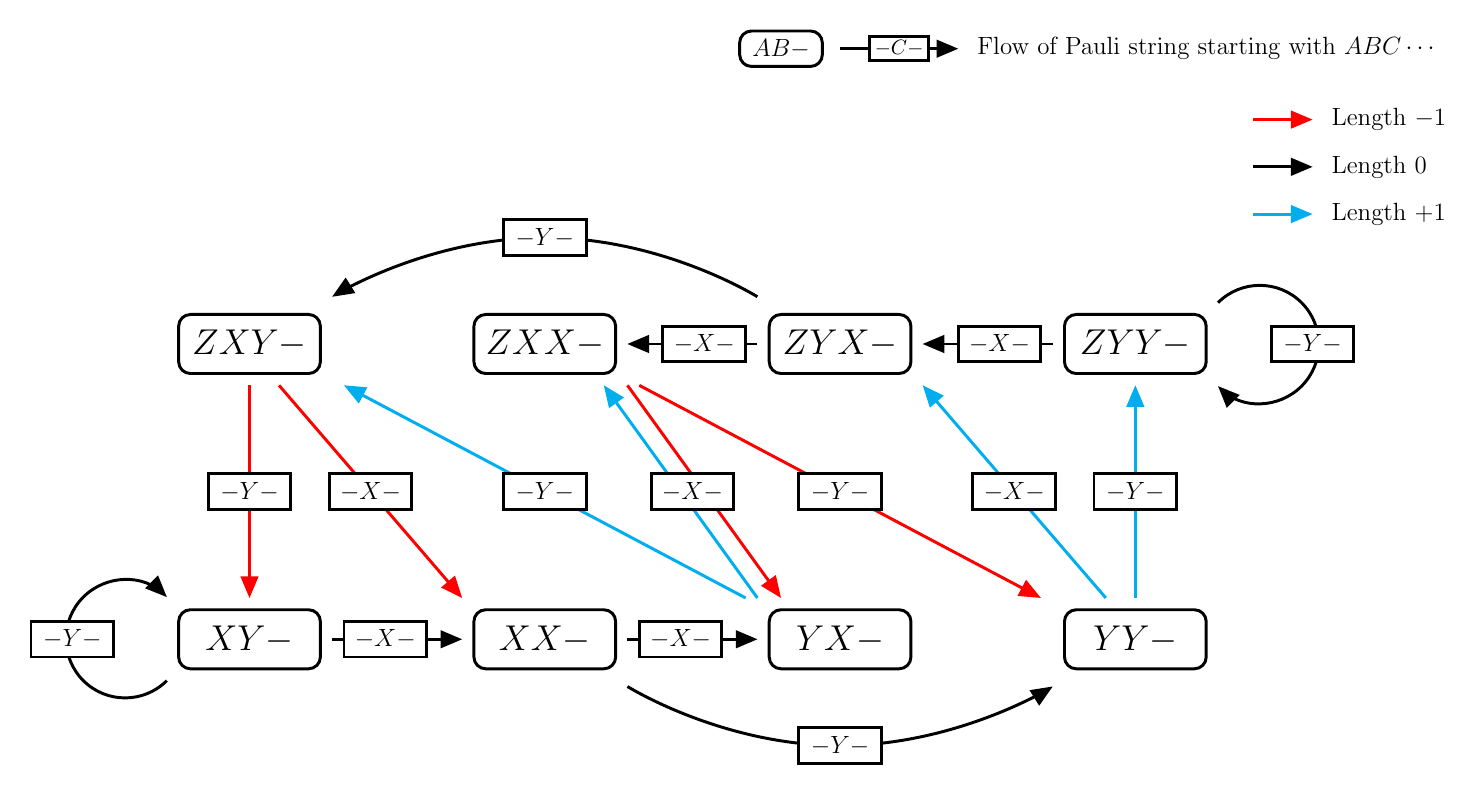}
  \caption{Detailed flowchart inside the green box of Fig.\ref{fig:flowchart}. The symbols on the arrows represent the Pauli operators immediately following the Pauli substrings at the tail of the arrow. Refer to the Proof of \textbf{Lemma 5} for more details.}
  \label{fig:flowchart_2}
\end{figure}

To prove the first part of \textbf{Lemma 5}, we need to describe the flowchart within the green box in detail. In Fig.\ref{fig:flowchart_2}, we present the flowchart for every possible left edge Pauli substring in the green box from Figure \ref{fig:flowchart}. We have omitted the arrows leading to or from the outside the green box, as we already know that such flows always result in a promising path. The characters on the arrows represents the operator placed immediately to the right of the left edge substring. For example, if we want to track the change of the left edge substring for a Pauli string starting with $ZXXY-$, we follow the red arrow originating from $ZXX-$ with the symbol $-Y-$ on it. 

Since putting $XZ$ operator on the right side of a Pauli string that ends with $Z$ always produces a Pauli string ending with $YZ$, e.g.
\begin{equation}
\begin{array}{rrrrrrr}
 &Z&A_2&\cdots&A_{k-1}&Z& \\
 & & & & &X&Z\\\hline
 &Z&A_2&\cdots&A_{k-1}&Y&Z\\
\end{array}
\end{equation}
we can conclude that starting from a length $k$ Pauli string that starts and ends with $Z$, and contains only $X$ and $Y$ operators between these $Z$ operators, one can endlessly follow the flow in Fig.\ref{fig:flowchart_2},forming a loop.

Moreover, each flow in Fig.\ref{fig:flowchart_2} neither increase the number of $X$ operators nor changes the parity of the number of $X$ operators. These observations demonstrate properties (A) and (B) in Theorem 2 of the main text.

Finally, by starting from the Pauli string $ZYYY\cdots YXZ$ and following the arrows in Fig.\ref{fig:flowchart_2}, we can confirm property (C) in Theorem 2 of the main text: all Pauli strings $ZXYY\cdots YYZ$, $ZYXY\cdots YYZ$, $\cdots$, $ZYYY\cdots YXZ$ form a single loop.

Having demonstrated properties (A), (B), and (C), we can now prove the first part of \textbf{Lemma 5} using the same reasoning as in Theorem 1 of the main text.
\end{proof}

\textbf{Theorem 1 in the main text.} --- In the main text, we presented Theorem 1 as a classification of every length-$k$ Pauli string. The detailed proof of this Theorem can be constructed using \textbf{Lemma 1} through \textbf{5}. In summary, we demonstrated that every length-$k$ Pauli string has a zero coefficient, except for the following cases. We use the notation $(A)^{(n)}$ to represent $n$ repetition of operator $A$. 

\begin{enumerate}
\item A Pauli string $Z(Y)^{(k-2)}Z$, or other Pauli strings in the loop $L_e$.
\item A Pauli string $ZX(Y)^{(k-3)}Z$, or other Pauli strings in the loop $L_o$.
\item Pauli strings that start with $ZZ$ or $ZI$ and end with $IZ$ or $ZZ$, which are related to the trivial operators.
\end{enumerate}

\section{Translationally Non-Invariant Conserved Quantities}

Before proceeding further, it is a good time to introduce how we can eliminate the possibility of translationally non-invariant conserved quantities in the PXP model. This argument follows the work in \cite{shiraishi2019proof}.

Suppose there exists a translationally non-invariant local quantity $C$, where $C$ is a length $k$ quantity with $k>3$. Let $T^{(j)}$ represent the translational shifting operator by distance $j$. We can define:
\begin{equation}
C_0:= \sum_{j=1}^{L} T^{(-j)} C T^{(j)}.
\end{equation}
It is straightforward to see that $C_0$ is a translational invariant local quantity. However, this does not immediately imply that we can restrict our focus to translationally invariant local quantities, because there is a chance that $C_0$ could become a conserved quantity with length $\leq 3$, such as $H, I, 0$, or a trivial operator in the Hilbert space under consideration, even if $C$ is not one of these. Thus, we need to rule out this possibility.

Consider the following quantities:
\begin{equation}
C_a = \sum_{j=1}^m e^{2\pi i a j/m }T^{(-j)}C T^{(j)}
\end{equation}
where $a=1,2,\cdots, m-1$, and $m$ is the smallest positive integer such that $T^{-m} C T^{m}=C$, which always exists for finite $L$. Note that $[C,H]=0$ implies $[C_a,H]=0$ for all $a$. Since $C$ is a length $k$ operator and $mC=C_0+\sum_{a=1}^{m-1} C_a$, at least one of the $C_a$ operators must be a length $k$ operator, denoted as $\overline{C}$.

Our goal is to show that $[C,H]=0$ implies $\overline{C}$ is a trivial operator, which we have already demonstrated for the case $\overline{C}=C_0$ in the main text. The key point is that, regardless of the value of $a$ for which $C_a=\overline{C}$, all the arguments used in the proof of \textbf{Theorem 1} in the main text and \textbf{Lemmas 1} through \textbf{5} remain applicable. This can be understood as follows.

Consider length $k=4$ quantity $C_a$. From the commutators 
\begin{equation*}
[\{ZYXZ\}_j , \{XZ\}_{j+3}] = 2i\{ZYXYZ\}_j = [\{ZXYZ\}_{j+1},\{ZX\}_j],
\end{equation*}
we have $q(\{ZYXZ\}_j) + q(\{ZXYZ\}_{j+1})=0$. Due to the definition of $C_a$, we obtain 
\begin{equation*}
q(\{ZYXZ\}_1) e^{2\pi i a (j-1)/m} + q(\{ZXYZ\}_1) e^{2\pi i a j/m}=0,
\end{equation*}
or equivalently, 
\begin{equation*}
q(\{ZYXZ\}_1) + q(\{ZXYZ\}_1) e^{2\pi i a /m}=0.
\end{equation*}
The important point here is that changing $a$ in $C_a$ only introduces a nonzero factor like $e^{2\pi i a/m}$ to the coefficient, without altering the graph structure. Since the promising path argument and the trivial operator argument do not depend on this coefficient scaling, we can directly apply \textbf{Theorem 1} to $C_a=\overline{C}$, showing that all length $k$ Pauli strings, except those that start and end with $Z$ and have only $X$ or $Y$ operators in between, have zero coefficients in $\overline{C}$.

However, addressing the \textbf{Exception 2} type Pauli strings, which form loops, presents a more complicated situation. We will address these cases in the following lemmas.

\section{Theorem 2 in the Main Text}

In Theorem 2 in the main text, we established that every Pauli string in the loop $L_e$ must have a zero coefficient, and we provided a proof for this statement. While this proof holds for translationally invariant quantity $C_0$, extending it to general $C_a$, including non-translationally invariant quantities, becomes more complex. Here, we will prove this statement for general $a$, encompassing non-translationally invariant quantities as well.

\begin{theorem}\label{thm:6}
The coefficient of the Pauli string $Z(Y)^{(k-2)}Z$ in $C_a$  is zero.
\end{theorem}

\begin{proof}
We consider the case where $k\geq 6$; the cases for $k=4$ and $k=5$ can be handled in a straightforward manner. Since
\begin{equation}
[\{Z(Y)^{(k-2)}Z\}_1,\{XZ\}_{k}]= [\{Z(Y)^{(k-2)}Z\}_2,\{ZX\}_1]
\end{equation}
and these are the only possible commutators that generate $\{Z(Y)^{(k-1)}Z\}_{1}$, we have the following relation:
\begin{align*}
&q(\{Z(Y)^{(k-2)}Z\}_1)+q(\{Z(Y)^{(k-2)}Z\}_2)\\
&=q(\{Z(Y)^{(k-2)}Z\}_1)+e^{2\pi i a/m} q(\{Z(Y)^{(k-2)}Z\}_1)\\
&=0.
\end{align*}
Thus, we conclude that $q_{Z(Y)^{(k-2)}Z}=0$ when $a/m\neq 1/2$. This indicates that we need to handle the case for the operator $C_{m/2}$ separately, as it has an eigenvalue $-1$ under translation and contains the Pauli string $Z(Y)^{(k-2)}Z$.

Before proceeding further, we first note that since we are focusing on the operator $C_{m/2}$ which has a translation eigenvalue $-1$, if $\left\{Z(Y)^{(k-2)}Z\right\}_1$ is present in $C_{m/2}$ with coefficient $q$, then $\left\{Z(Y)^{(k-2)}Z\right\}_2$ must have a coefficient $-q$. To distinguish between these two coefficients, we will use subscripts $1$ or $2$ to the right of the Pauli string, indicating that the leftmost Pauli matrix acts on the odd or even site of the chain, respectively. 

Consider the commutators that generate the Pauli string $Z(Y)^{(s)}Z(Y)^{(k-s-3)}Z_1$, where $s=1,2,\cdots, k-4$. For $s=2,\cdots, k-5$, the following relation holds:
\begin{align*}\label{eq:ZYZ1}
&q(Z(Y)^{(s-1)}XYX(Y)^{(k-s-4)}Z_1) - q(Z(Y)^{(k-2)}Z_1)\\
& - q(Z(Y)^{(s-1)}Z(Y)^{(k-s-3)}Z_2) - q(Z(Y)^{(s)}Z(Y)^{(k-s-4)}Z_1)= 0.\numberthis
\end{align*}
For $s=1$, we have:
\begin{align*}\label{eq:ZYZ2}
&q(ZXYX(Y)^{(k-5)}Z_1)- q(Z(Y)^{(k-2)}Z_1)\\
& - q(ZZ(Y)^{(k-4)}Z_2)  - q(ZYZ(Y)^{(k-5)}Z_1) + q(ZI(Y)^{(k-4)}Z_2) = 0.\numberthis
\end{align*}
For $s=k-4$, we have:
\begin{align*}\label{eq:ZYZ3}
&q(Z(Y)^{(k-5)}XYXZ_1)  - q(Z(Y)^{(k-2)}Z_1)\\
& - q(Z(Y)^{(k-5)}ZYZ_2) - q(Z(Y)^{(k-4)}ZZ_1) + q(Z(Y)^{(k-4)}IZ_1) = 0.\numberthis
\end{align*}
Now, by following the loop $L_e$, we can show the following. Let 
\begin{equation}\label{eq:ZYZ4}
q_0\coloneqq q(ZXYX(Y)^{(k-5)}Z_1).
\end{equation}
Then for all $s=1,\cdots, k-5$, we have:
\begin{equation}\label{eq:ZYZ5}
q(Z(Y)^{(s)}XYX(Y)^{(k-s-5)}Z_1) = (-1)^s q_0.
\end{equation}
Additionally, we can show that:
\begin{align*}\label{eq:ZYZ6}
q_0&= -q(XX(Y)^{(k-4)}Z_1)\\
&= q((Y)^{(k-2)}Z_2)\\
&= q(Z(Y)^{(k-2)}Z_2)\\
&= -q(Z(Y)^{(k-2)}Z_1).\numberthis
\end{align*}
Using Eqs.\ref{eq:ZYZ4}, \ref{eq:ZYZ5}, and \ref{eq:ZYZ6}, we can transform Eqs.\ref{eq:ZYZ1}, \ref{eq:ZYZ2}, and \ref{eq:ZYZ3} into the followings.
\begin{equation}\label{eq:ZYZ7}
((-1)^{s-1}+1)q_0 - q(Z(Y)^{(s-1)}Z(Y)^{(k-s-3)}Z_2) - q(Z(Y)^{(s)}Z(Y)^{(k-s-4)}Z_1)= 0
\end{equation}
\begin{equation}\label{eq:ZYZ8}
2q_0 - q(ZZ(Y)^{(k-4)}Z_2)  - q(ZYZ(Y)^{(k-5)}Z_1) + q(ZI(Y)^{(k-4)}Z_2) = 0
\end{equation}
\begin{equation}\label{eq:ZYZ9}
((-1)^{k-5}+1)q_0 - q(Z(Y)^{(k-5)}ZYZ_2) - q(Z(Y)^{(k-4)}ZZ_1) + q(Z(Y)^{(k-4)}IZ_1) = 0
\end{equation}
From these, we derive the following equations;
\begin{align*}
2q_0 - q(ZZ(Y)^{(k-4)}Z_2)  - q(Z(Y)^{(1)}Z(Y)^{(k-5)}Z_1) + q(ZI(Y)^{(k-4)}Z_2) &= 0\\
  -q(Z(Y)^{(1)}Z(Y)^{(k-5)}Z_2) - q(Z(Y)^{(2)}Z(Y)^{(k-6)}Z_1)&= 0\\
2q_0 - q(Z(Y)^{(2)}Z(Y)^{(k-6)}Z_2) - q(Z(Y)^{(3)}Z(Y)^{(k-7)}Z_1)&= 0\\
\vdots\\
((-1)^{k-5}+1)q_0&\\
-q(Z(Y)^{(k-5)}ZYZ_2) - q(Z(Y)^{(k-4)}ZZ_1) + q(Z(Y)^{(k-4)}IZ_1)& = 0
\end{align*}
Summing over these equations, we get:
\begin{align*}
 2\left\lfloor\frac{k-3}{2}\right\rfloor q_0  - q(ZZ(Y)^{(k-4)}Z_2)  + q(ZI(Y)^{(k-4)}Z_2)&\\
  - q(Z(Y)^{(k-4)}ZZ_1) + q(Z(Y)^{(k-4)}IZ_1) &=0\numberthis\label{eq:ZYZfirst}
\end{align*}

Similarly, consider the commutators that generate the Pauli string $Z(Y)^{(s)}I(Y)^{(k-s-3)}Z_1$. For $s=2,\cdots, k-5$, we have:
\begin{align*}\label{eq:ZYZ10}
&q(Z(Y)^{(s-1)}XX(Y)^{(k-s-3)}Z_1)+ q(Z(Y)^{(s)}XX(Y)^{(k-s-4)}Z_1)\\
& -q(Z(Y)^{(s-1)}I(Y)^{(k-s-3)}Z_2) - q(Z(Y)^{(s)}Z(Y)^{(k-s-4)}Z_1)= 0.\numberthis
\end{align*}
For $s=1$, we have:
\begin{align*}\label{eq:ZYZ11}
&q(ZXX(Y)^{(k-4)}Z_1)+ q(ZYXX(Y)^{(k-5)}Z_1)\\
& + q(ZZ(Y)^{(k-4)}Z_2) - q(ZI(Y)^{(k-4)}Z_2)- q(ZYI(Y)^{(k-5)}Z_1) = 0.\numberthis
\end{align*}
For $s=k-4$, we have:
\begin{align*}\label{eq:ZYZ12}
&q(Z(Y)^{(k-4)}XXZ_1)  + q(Z(Y)^{(k-5)}XXYZ_1)\\
&  + q(Z(Y)^{(k-4)}ZZ_1) - q(Z(Y)^{(k-4)}IZ_1) - q(Z(Y)^{(k-5)}IYZ_2) = 0.\numberthis
\end{align*}
Following the loop $L_e$, we find the following. Let
\begin{equation}\label{eq:ZYZ14}
q_1\coloneqq q(ZXX(Y)^{(k-4)}Z_1).
\end{equation}
For all $s=1,\cdots, k-5$, we have:
\begin{equation}\label{eq:ZYZ15}
q(Z(Y)^{(s)}XX(Y)^{(k-s-4)}Z_1) = (-1)^s q_1.
\end{equation}
By using Eqs.\ref{eq:ZYZ14} and \ref{eq:ZYZ15}, the first lines of Eqs.\ref{eq:ZYZ10}, \ref{eq:ZYZ11}, and \ref{eq:ZYZ12} become zero. This leads to the following equations:
\begin{align*}
q(ZZ(Y)^{(k-4)}Z_2) - q(ZI(Y)^{(k-4)}Z_2)- q(Z(Y)^{(1)}I(Y)^{(k-5)}Z_1) &= 0\\
-q(Z(Y)^{(1)}I(Y)^{(k-5)}Z_2) - q(Z(Y)^{(2)}Z(Y)^{(k-6)}Z_1)&= 0\\
-q(Z(Y)^{(2)}I(Y)^{(k-6)}Z_2) - q(Z(Y)^{(3)}Z(Y)^{(k-7)}Z_1)&= 0\\
&\vdots\\
q(Z(Y)^{(k-4)}ZZ_1) - q(Z(Y)^{(k-4)}IZ_1) - q(Z(Y)^{(k-5)}IYZ_2) &= 0
\end{align*}
Summing over these equations, we get:
\begin{align*}
q(ZZ(Y)^{(k-4)}Z_2) - q(ZI(Y)^{(k-4)}Z_2)&\\
+q(Z(Y)^{(k-4)}ZZ_1) - q(Z(Y)^{(k-4)}IZ_1)&=0\numberthis\label{eq:ZYZsecond}
\end{align*}

Adding Eq.\ref{eq:ZYZfirst} to Eq.\ref{eq:ZYZsecond} gives:
\begin{equation}
2\left\lfloor \frac{k-3}{2}\right\rfloor q_0 = 0.
\end{equation}
Since $k\geq 6$, we conclude that $q_0=0$, and from Eq.\ref{eq:ZYZ6}, this shows that $q(Z(Y)^{(k-2)}Z_1)=0$.
\end{proof}

\section{Theorem 3 in the Main Text}
In Theorem 3 of the main text, we discussed that every Pauli string in the loop $L_o$ must have a zero coefficient. A brief proof using the concept of the quasi-promising path was presented for translationally invariant quantity. Here we prove this statement for general $a$, and discuss the proof for the translationally invariant case in detail.

\begin{theorem}\label{thm:7}
The coefficient of $ZX(Y)^{(k-3)}Z$ in $C_a$ is zero.
\end{theorem}

\begin{proof}
We consider the case where $k\geq 7$; the cases for $k=4,5,$ and $6$ can be handled in a straghtforward manner. Since
\begin{align*}
[\{ZX(Y)^{(k-3)}Z\}_1&,\{XZ\}_{k}] \\
&= -[\{X(Y)^{(k-3)}Z\}_3,\{ZXZ\}_1]\\
[\{X(Y)^{(k-3)}Z\}_3&,\{XZ\}_{k+1}]\\
 &= -[\{X(Y)^{(k-3)}Z\}_4, \{XZ\}_3]\\
[\{ZXY\cdots YYZ\}_2&,\{XZ\}_{k+1}]\\
 &= -[\{X(Y)^{(k-3)}Z\}_4,\{ZXZ\}_2],\\
\end{align*}
and these are all the possible commutators that give the same result in each row, we have the following relation: 
\begin{align*}
&q(\{ZX(Y)^{(k-3)}Z\}_1)-q(\{ZX(Y)^{(k-3)}Z\}_2)\\
&=q(\{ZX(Y)^{(k-3)}Z\}_1)-e^{2\pi i a/m} q(\{ZX(Y)^{(k-3)}Z\}_1)\\
&=0.
\end{align*}
From this equation, we conclude that $q_{ZX(Y)^{(k-3)}Z}=0$ when $a/m\neq 0$. This indicates that the case for the operator $C_0$, which has an eigenvalue $1$ under the translation operator and contains $ZX(Y)^{(k-3)}Z$ operator, must be treated separately.

Before proceeding further, it is useful to define following lemmas, which will be helpful in proving Lemma \ref{thm:7}.

\textbf{Lemma 7.1.} \textit{The coefficients of length $k$ Pauli strings in the loop $L_o$ can be determined in the following way. Define $X^+\coloneqq 2QX = X+iY$ and $X^-\coloneqq 2PX=X-iY$. Then, the sum of length-$k$ Pauli strings in the loop $L_o$ is proportional to: 
\begin{itemize}
\item The real part of $Z(X^+X^-)^{(k-3)/2} X^+ Z$ if $k$ is odd.
\item The imaginary part of $Z(X^+X^-)^{(k-2)/2}Z$ if $k$ is even.
\end{itemize}
}

For example, take $k=6$. The imaginary part of $Z(X^+X^-)^{(2)}Z$ is
\begin{align*}
\Im[Z(X^+X^-)^{2}Z]&= -ZXYYYZ+ZYXYYZ-ZYYXYZ+ZYYYXZ\\
&-ZXXXYZ+ZXXYXZ-ZXYXXZ+ZYXXXZ,
\end{align*}
which gives the relationships between the coefficients of length $k=6$ Pauli strings. For example, this gives $q(ZXYYYZ)=-q(ZYXYYZ)=q(ZXYXXZ)$.

\textbf{Lemma 7.2.} \textit{Consider a length $k$ Pauli string in $L_o$, and a length $k-1$ Pauli string obtained by removing the $Z$ operator on either the left or right edge. Then, the sum of the coefficients of these two Pauli strings is zero.}

For example, take $k=6$ and choose $ZXYYYZ$. Then, one can show that $q(ZXYYYZ)+q(XYYYZ)=0$ and $q(ZXYYYZ)+q(ZXYYY)=0$.

\textbf{Proof of Lemma 7.1 and Lemma 7.2.} --- The proof of \textbf{Lemma 7.1} and \textbf{7.2} can be directly obtained by following the flowchart in \ref{fig:flowchart_2}, with precise calculations of the coefficients. These results can also be understood in the following way.

Let $k$ odd and consider the operator:
\begin{equation}
V_k \coloneqq \sum_j \{P(X^+X^-)^{(k-3)/2}X^+P\}_j + \{P(X^-X^+)^{(k-3)/2}X^-P\}_j.
\end{equation}
Now, consider the commutator $[V_k,H]$, focusing on the cases where the $PXP$ operators in $H$ are positioned on the edges of the operator $V_k$. We get the following relations:
\begin{equation*}
\begin{array}{cccccccccc}
 &P&X^+&X^-&\cdots&X^-&X^+&P  & \\
 & &   &   &      &   &P  &X  &P\\\hline
 &P&X^+&X^-&\cdots&X^-&X^+&X^-&P
\end{array}
\end{equation*}
\begin{equation*}
\begin{array}{cccccccccc}
 &&P&X^+&X^-&\cdots&X^-&X^+&P   \\
 &P&X&P   &   &      &   &  &  \\\hline
 &P&X^-&X^+&X^-&\cdots&X^-&X^+&P
\end{array}
\end{equation*}
\begin{equation*}
\begin{array}{cccccccccc}
 &P&X^-&X^+&\cdots&X^+&X^-&P  & \\
 & &   &   &      &   &P  &X  &P\\\hline
 -&P&X^-&X^+&\cdots&X^+&X^-&X^+&P
\end{array}
\end{equation*}
\begin{equation*}
\begin{array}{cccccccccc}
 &&P&X^-&X^+&\cdots&X^+&X^-&P   \\
 &P&X&P   &   &      &   &  &  \\\hline
 -&P&X^+&X^-&X^+&\cdots&X^+&X^-&P
\end{array}
\end{equation*}
This equations show that the commutator $[V_k,H]$ does not contain any Pauli string with length $\geq k$, except for those starting and ending with $Z$. Since this is the condition used to demonstrate the loop structure $L_o$, we conclude that the coefficients of Pauli strings in $V_k$ match appropriate coefficients in the $L_o$ subgraph. 

Since the coefficients in the $L_o$ subgraph are uniquely determined up to scale, we conclude that $V_k$ determines the coefficients of Pauli strings in $L_o$. Calculating the coefficients in $V_k$ then yields the desired result in \textbf{Lemma 7.1} and \textbf{Lemma 7.2}. The same can be shown for even $k$ in a similar manner.

\textbf{Lemma 7.3.} \textit{Every length-$k-1$ Pauli strings that starts with $X$, ends with $Z$, and contains $I$ or $Z$ in the middle has a zero coefficient.}

For example, for $k=5$, $q(XIYZ)=0$. This can be directly shown by following the flowchart in Fig.\ref{fig:flowchart}, using the same argument applied to length-$k$ Pauli strings.

\textbf{Lemma 7.4.} \textit{The coefficients of the length-$k$ Pauli strings $ZZ\cdots ZZ$, $ZZ\cdots IZ$, $ZI\cdots ZZ$, and $ZI\cdots IZ$, which differ only by the second operator counting from the left or right, are equal.} 

For example, for $k=5$, we have $q(ZZXZZ)=q(ZIXZZ)=q(ZZXIZ)=q(ZIXIZ)$. This can be proven by considering the following commutators:
\begin{equation*}
\begin{array}{cccccccccc}
 &Z&Z&A_3&\cdots&A_{k-2}&Z&Z&   \\
 & & &   &      &       & &X&Z  \\\hline
 &Z&Z&A_3&\cdots&A_{k-2}&Z&Y&Z \\\hline
 &Z&Z&A_3&\cdots&A_{k-2}&I&Z&   \\
 & & &   &      &       &Z&X&Z  \\\hline
 & &Y&A_3&\cdots&A_{k-2}&I&Z&   \\
 &Z&X&   &      &       & & &   \\\hline
 & &Y&\overline{A_3}&\cdots&A_{k-2}&I&Z&   \\
 &Z&Z&Z  &      &       & & &   \\\hline
\end{array}
\end{equation*}
Since the coefficient of the Pauli strings in third and fourth commutators are zero, they do not contribute to the $ZZ\cdots ZYZ$ Pauli string. Therefore, we obtain:
\begin{equation*}
q(ZZ\cdots ZZ) = q(ZZ\cdots IZ).
\end{equation*}
A similar argument applies to the $ZI\cdots ZZ$ and $ZI\cdots IZ$ Pauli strings.

\textbf{Proof of Lemma 7.} --- Let $q_0\coloneqq q(ZX(Y)^{(k-3)}Z)$. Now, consider the commutators that contribute to the Pauli string $X(Y)^{(s)}Z(Y)^{(k-s-4)}Z$. For all $s=1,2,\cdots, k-6$, we have:
\begin{align*}
q(X(Y)^{(k-3)}Z) - q(X(Y)^{(s-1)}XYX(Y)^{(k-s-5)}Z)&\\
-q(X(Y)^{(s-1)}Z(Y)^{(k-s-4)}Z)+q(X(Y)^{(s)}Z(Y)^{(k-s-5)}Z)&=0.
\end{align*}
Here, all Pauli strings with zero coefficient, as stated in \textbf{Lemma 7.3}, are omitted. Using \textbf{Lemma 7.1} and \textbf{Lemma 7.2}, which provide:
\begin{align*}
q(X(Y)^{(k-3)}Z)&= -q_0\\
q(X(Y)^{(s)}XYX(Y)^{(k-s-5)}Z) &= q_0,
\end{align*}
we obtain:
\begin{equation}\label{eq:ZXY1}
-q(X(Y)^{(s-1)}Z(Y)^{(k-s-4)}Z)+q(X(Y)^{(s)}Z(Y)^{(k-s-5)}Z)=2q_0.
\end{equation}
Summing Eq.\ref{eq:ZXY1} telescopically for all $s=1,2,\cdots, k-6$, we arrive at:
\begin{equation}\label{eq:ZXY2}
-q(XZ(Y)^{(k-5)}Z)+q(X(Y)^{(k-6)}ZYZ)=2(k-6)q_0.
\end{equation}

Next, consider the commutators giving $XZ(Y)^{(k-4)}Z$. With the aid of \textbf{Lemma 7.3}, we find:
\begin{equation*}
q(YYX(Y)^{(k-5)}Z) + q(X(Y)^{(k-3)}Z)+q(XZ(Y)^{(k-5)}Z) = 0.
\end{equation*}
Using \textbf{Lemma 7.1} and \textbf{7.2}, we obtain
\begin{equation}\label{eq:ZXY3}
q(XZ(Y)^{(k-5)}Z) = 2q_0.
\end{equation}
Combining Eqs.\ref{eq:ZXY2} and \ref{eq:ZXY3}, we get:
\begin{equation}\label{eq:ZXY4}
q(X(Y)^{(k-6)}ZYZ)=2(k-5)q_0.
\end{equation}

Now, consider the commutators giving $ZYX(Y)^{k-5}ZZ$ and $ZYX(Y)^{k-5}IZ$. These yield the following equations:
\begin{align*}
q(ZYX(Y)^{(k-4)}Z) - q(ZZX(Y)^{(k-5)}ZZ)&\\
+q(ZX(Y)^{(k-5)}ZZ) - q(ZYX(Y)^{(k-4)})&=0,\\
-q(ZYX(Y)^{(k-6)}XXZ)-q(ZZX(Y)^{(k-5)}IZ)&\\
+q(ZX(Y)^{(k-5)}IZ) + q(ZYX(Y)^{(k-6)}XX)&=0.
\end{align*}
Using \textbf{Lemma 7.1} and \textbf{7.2}, we get
\begin{align*}
- q(ZZX(Y)^{(k-5)}ZZ)+q(ZX(Y)^{(k-5)}ZZ) &=2q_0,\\
-q(ZZX(Y)^{(k-5)}IZ)+q(ZX(Y)^{(k-5)}IZ) &=-2q_0.
\end{align*}
Subtracting these two equations and applying \textbf{Lemma 7.4} we get:
\begin{equation}\label{eq:ZXY5}
q(ZX(Y)^{(k-5)}ZZ)-q(ZX(Y)^{(k-5)}IZ)=4q_0.
\end{equation}

Finally, consider the commutators giving $ZX(Y)^{(k-5)}ZYZ$, which yield the following equation:
\begin{align*}
q(ZX(Y)^{(k-3)}Z)-q(ZX(Y)^{(k-6)}XYXZ)&\\
+q(ZX(Y)^{(k-5)}ZZ)-q(ZX(Y)^{(k-5)}IZ)&\\
+q(X(Y)^{(k-6)}ZYZ)&=0.
\end{align*}
Using Eq.\ref{eq:ZXY5} and \textbf{Lemma 7.1} and \textbf{Lemma 7.2}, we obtain:
\begin{equation}\label{eq:ZXY6}
q(X(Y)^{(k-6)}ZYZ)=-6q_0.
\end{equation}
Combining Eqs.\ref{eq:ZXY4} and \ref{eq:ZXY6}, we arrive at the final result:
\begin{equation}
2(k-2)q_0=0.
\end{equation}
Thus, $q_0=0$ for $k\geq 7$, proving the statement.
\end{proof}

\section{Graph Theoretical Representation of the Proof of Lemma 7}

\begin{figure}[h]
\captionsetup{justification=centering}
\centering
\includegraphics[width=1\textwidth]{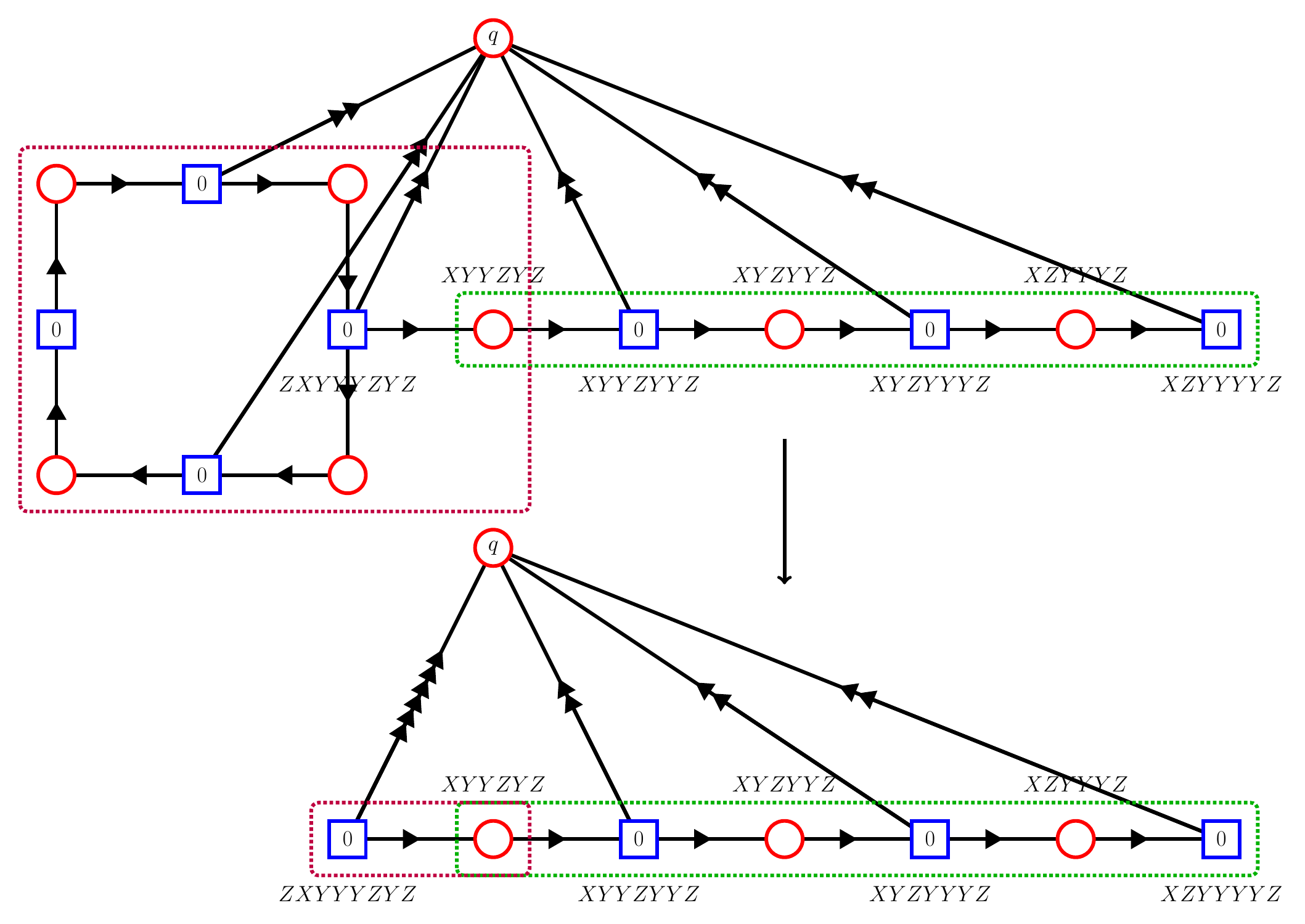}
  \caption{(Top Panel) Graph theoretical representation of the proof of Lemma \ref{thm:7} for $k=8$. The right side highlights a quasi-promising path. The left box highlights a graph structure which, after some modifications, can be transformed into a quasi-promising path. (Bottom Panel) After the modification, the graph structure shows the existence of two quasi-promising paths, supporting the conclusion of Theorem 3 in the main text.}
  \label{fig:appendix_qpp_qloop}
\end{figure}

In the main text, we introduced the concept of a quasi-promising path and asserted that it can be used to demonstrate the nonintegrability of a given system. In this section, we explore this demonstration in greater detail, explaining how quasi-promising paths emerge in our proof of Lemma \ref{thm:7}.

The top panel of Fig.\ref{fig:appendix_qpp_qloop} presents the graph representation of our proof of Lemma \ref{thm:7} for $k=8$. The structure inside the box on the right is a quasi-promising path with a disturbing vertex $q$, while the structure inside the box on the left does not initially appear to be a quasi-promising path. In fact, after preforming some modifications to the graph --- modifications justified by basic algebraic manipulations of the linear equations --- we can transform the subgraph in the left box of the top panel into a quasi-promising path, as shown in the bottom panel of Fig.\ref{fig:appendix_qpp_qloop}.

\begin{figure}[h]
\captionsetup{justification=centering}
\centering
\includegraphics[width=0.6\textwidth]{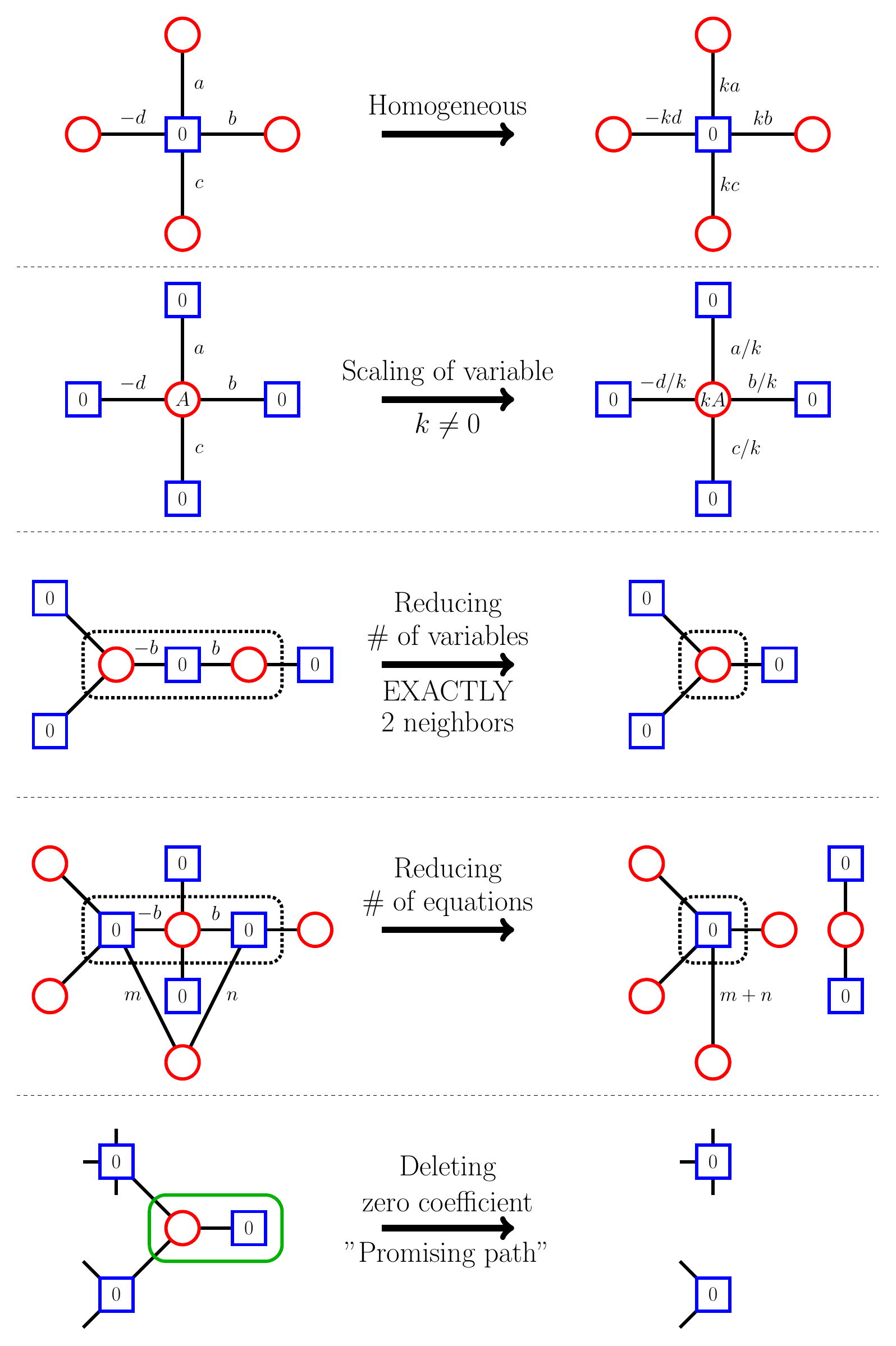}

  \caption{Graph modifications applicable to the commutator graph. The first and second diagrams adjust the coefficients of the equations, while the third, fourth, and fifth diagrams simplify the system by reducing the number of parameters or equations.}
  \label{fig:graph_mod}
\end{figure}

Figure \ref{fig:graph_mod} illustrates the modifications that can be applied to the commutator graph, generating a new graph that preserves the same set of linear equations. These modifications are derived from basic algebraic operations on linear equations, as explained below.

The first diagram corresponds to the transformation of the equation $ax+by+cz-dw=0$ into $-2ax-2by-2cz+2dw=0$, achieved by multiplying both sides by $2$. The second diagram reflects the situation where a parameter $A$ is changed to $2A$. In the third diagram, the $\bsv$ vertex in the green box must have only two neighboring red circles, corresponding to the equation $-bx+by=0$, which simplifies to $x=y$ allowing us to identify the two red circles. The fourth diagram represents the reduction of two equations, $ax-by+cz+mt=0$ and $by+dw+nt=0$, into a single equation $ax+cz+dw+(m+n)t=0$. The fifth diagram removes a zero coefficient, which is equivalent to the argument about the promising path discussed in the main text. 

Returning to Fig.\ref{fig:appendix_qpp_qloop}, the subgraph in the top panel can be transformed into the subgraph in the bottom panel by repeatedly applying the third and fourth modifications from Fig.\ref{fig:graph_mod}. Therefore, we can conclude that the proof of \textbf{Lemma 7} is also based on the quasi-promising path method.

\section{Trivial Operators}

Theorem 1 in the main text discusses trivial operators, i.e. those composed of operator strings containing $QQ$, and argues that ignoring Pauli strings of the form $ZZ\cdots ZZ$, $ZI\cdots ZZ$, $ZZ\cdots IZ$, and $ZI\cdots IZ$ is valid. In this section, we provide a more detailed argument for this claim.  

\begin{theorem}\label{thm:8} Let $C$ be a length $k$ conserved quantity in the PXP model, containing at least one length $k$ Pauli string that neither starts with $ZZ\cdots$ or $ZI\cdots$ nor ends with $\cdots ZZ$ or $\cdots IZ$. Then there exists a length $k$ conserved quantity $C'$ such that $C'$ contains no Pauli strings of the form $ZZ\cdots ZZ$, $ZI\cdots ZZ$, $ZZ\cdots IZ$, or $ZI\cdots IZ$, and $C-C'$ is a trivial operator.
\end{theorem}

\begin{proof} Suppose $C$ contains one of the Pauli strings $ZZA_3\cdots A_{k-2}ZZ$, $ZIA_3\cdots A_{k-2}ZZ$, $ZZA_3\cdots A_{k-2} IZ$, or $ZIA_3 \cdots A_{k-2} IZ$. Using \textbf{Lemma 7.4}, we observe that the coefficients of these Pauli strings in $C$ must always be equal. Let $q_0\coloneqq q_{ZZA_3\cdots A_{k-2}ZZ}$. Then, we have:
\begin{align*}
&q(ZZA_3\cdots A_{k-2}ZZ)ZZA_3\cdots A_{k-2}ZZ\\
&+q(ZIA_3\cdots A_{k-2}ZZ)ZIA_3\cdots A_{k-2}ZZ\\
&+q(ZZA_3\cdots A_{k-2}IZ)ZZA_3\cdots A_{k-2}IZ\\
&+q(ZIA_3\cdots A_{k-2}IZ)ZZA_3\cdots A_{k-2}IZ\\
=&q_0(ZZA_3\cdots A_{k-2}ZZ + ZIA_3\cdots A_{k-2}ZZ\\
&+ZZA_3\cdots A_{k-2}IZ+ZIA_3\cdots A_{k-2}IZ)\\
=&4q_0ZQA_3\cdots A_{k-2}QZ.
\end{align*}
Since $Z=2Q-I$, we have:
\begin{align*}
ZQ\cdots QZ &= 4QQ\cdots QQ-2QQ\cdots QI\\
&-2IQ\cdots QQ+IQ\cdots QI.\numberthis
\end{align*}

Now, define:
\begin{equation}
C'=C-16 q_0 QQA_3\cdots A_{k-2}QQ.
\end{equation}
By definition, $C'$ does not contain any of the previously mentioned Pauli strings. Since removing $QQ\cdots QQ$ from $C$ does not eliminate all the length $k$ Pauli strings in $C$ (due to the assumption about $C$), $C'$ remains a length $k$ conserved quantity, and $C-C'$ is a trivial operator.

If $C$ still contains other Pauli strings like $ZZB_3\cdots B_{k-2}ZZ$, we can apply the same process and define $C''=C-16 q'_0 QQB_3\cdots B_{k-2}QQ$. Repeating this process until no Pauli strins of the form $ZZ\cdots ZZ$, $ZI\cdots ZZ$, $ZZ\cdots IZ$, or $ZI\cdots IZ$ remain, we obtain the desired quantity.
\end{proof}

\section{Demonstrating Nonintegrability in Other Spin-$1/2$ Models}
In the main text and the suppliment material, we demonstrated that the PXP model has no nontrivial conserved quantities using a graph theoretical approach. This method is both simple and robust, making it applicable to a variety of spin-$1/2$ models. In this appendix, we illustrate how the graph theoretical approach can also be used to demonstrate the nonintegrability of the $XYZ$ model with a magnetic field and the mixed-field Ising chain model, both of which have been previously shown to be nonintegrable through direct calculation.

\begin{figure}[h]
\centering
\includegraphics[width=0.6\textwidth]{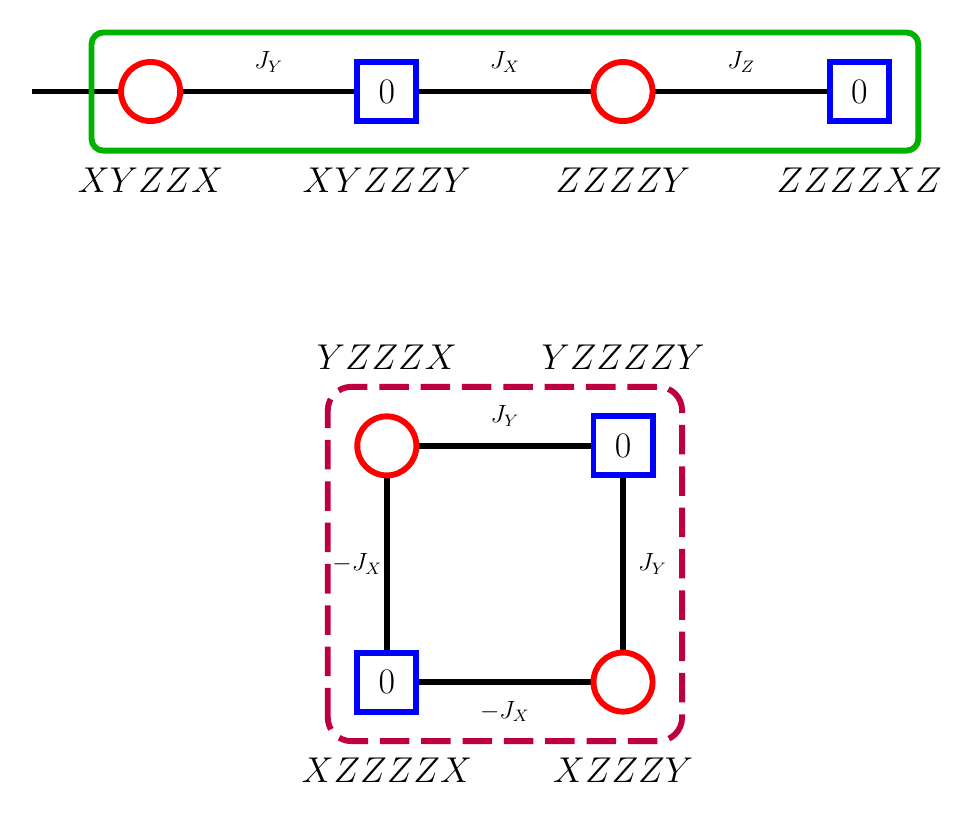}
  \caption{Subgraphs of the commutator graph for length $k=5$ quantities. The two-headed or six-headed arrows indicate that the weights are multiplied by two or six, respectively, compared to the single arrows. (Top) The graph visualisation shows that a length $5$ Pauli string, which is not a doubling-product operator, is part of a promising path and therefore has a zero coefficient. (Bottom) The doubling-product operator falls under the Exception 2 case discussed in the main text and is part of a loop.}
  \label{fig:appendix_xyz_1}
\end{figure}

For the $XYZ$ model with a magnetic field\cite{shiraishi2019proof}, the proof is divided into two parts. First, it asserts that all length $k$ Pauli strings that are not doubling-product operators have zero coefficients. Second, it claims that all doubling-product operators, which have linearly related coefficients, also have zero coefficients. Fig.\ref{fig:appendix_xyz_1} illustrates this fact using the subgraph of the commutator graph for length $k=5$ quantities. The top panel shows a promising path including a Pauli string $XYZZX$, which is not a doubling-product operator. The small coefficients $J_X, J_Y,$ and $J_Z$ above the edges represent the weights of the edges. The lower panel shows the loop involving the Pauli strings $YZZZX=\overline{YXYX}$ and $XZZZY=\overline{XYXY}$, both of which are doubling-product operators.

\begin{figure}[h]
\captionsetup{justification=centering}
\centering
\includegraphics[width=1\textwidth]{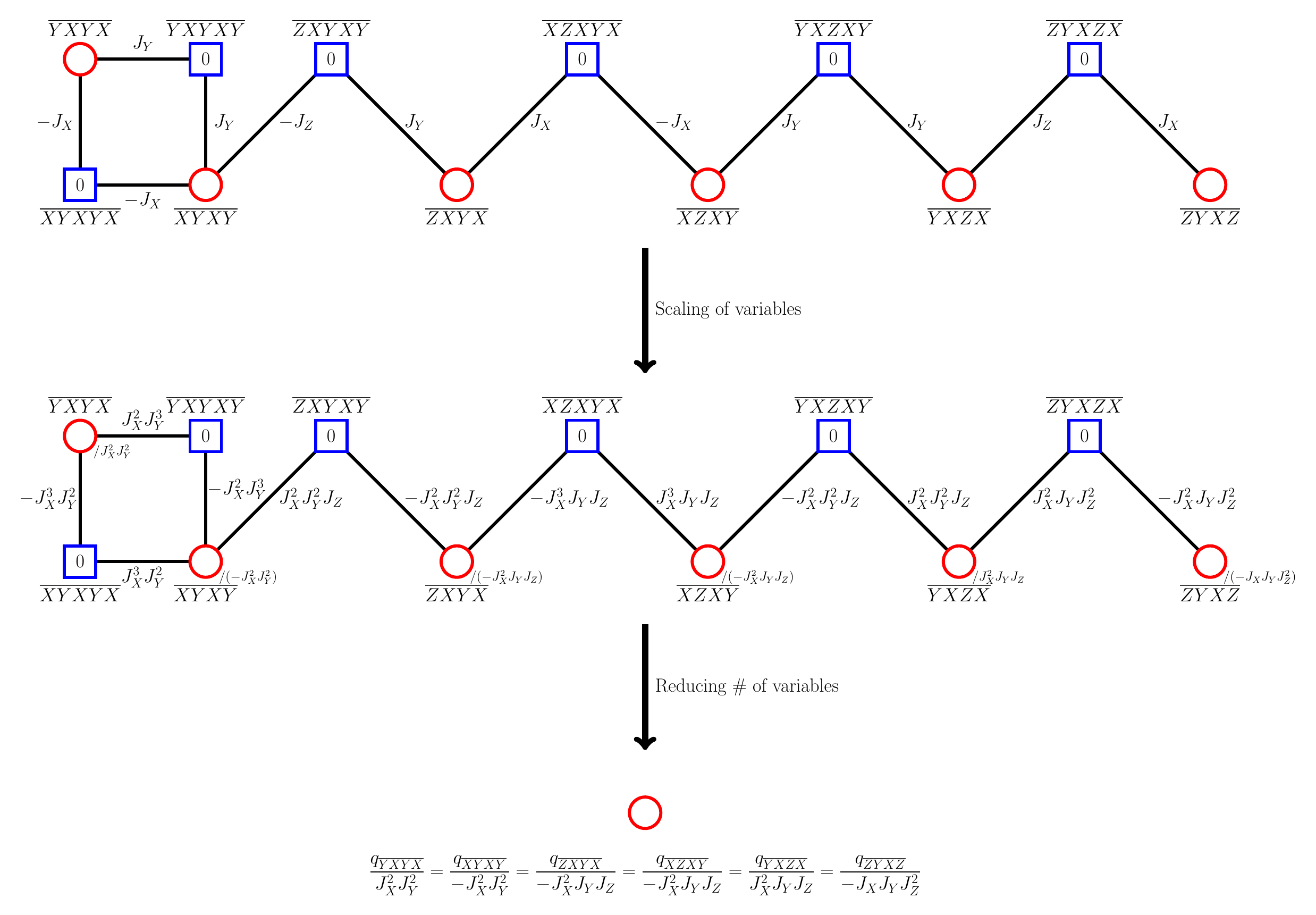}
  \caption{Demonstrating that all doubling-product operators have the same coefficient, with appropriate scaling. In the first step, the modification shown in the second row of \ref{fig:graph_mod} is applied. In the second step, the modification shown in the third row of \ref{fig:graph_mod} is used.}
  \label{fig:appendix_xyz_4}
\end{figure}

Using the method suggested in Fig.\ref{fig:graph_mod}, we can demonstrate the linear relationships between the coefficients of doubling-product operators by modifying the graph. Fig.\ref{fig:appendix_xyz_4} shows the result. First, by scaling the doubling-product operators with appropriate coefficients, we adjust the edge coefficients such that every two edges connected to a $\bsv$ vertex have the same absolute values but opposite signs. More specifically, scale the doubling-product operators as follows: if $X, Y,$ and $Z$ appears $n_X, n_Y,$ and $n_Z$ times respectively in the doubling-product form respectively, scale by dividing by $J_X^{n_X}J_Y^{n_Y}J_Z^{n_Z}$. If there is an odd number of $XZ$, $ZY$, or $YX$ pairs in the doubling-product form, multiply $-1$. The graph modification according to this scaling of coefficients allows every $\bsv$ vertex in Figure \ref{fig:appendix_xyz_4} to be eliminated using the transformation shown in the third row of Fig.\ref{fig:graph_mod}, leaving a single $\rcv$ vertex.

\begin{figure}[h]
\captionsetup{justification=centering}
\centering
\includegraphics[width=0.6\textwidth]{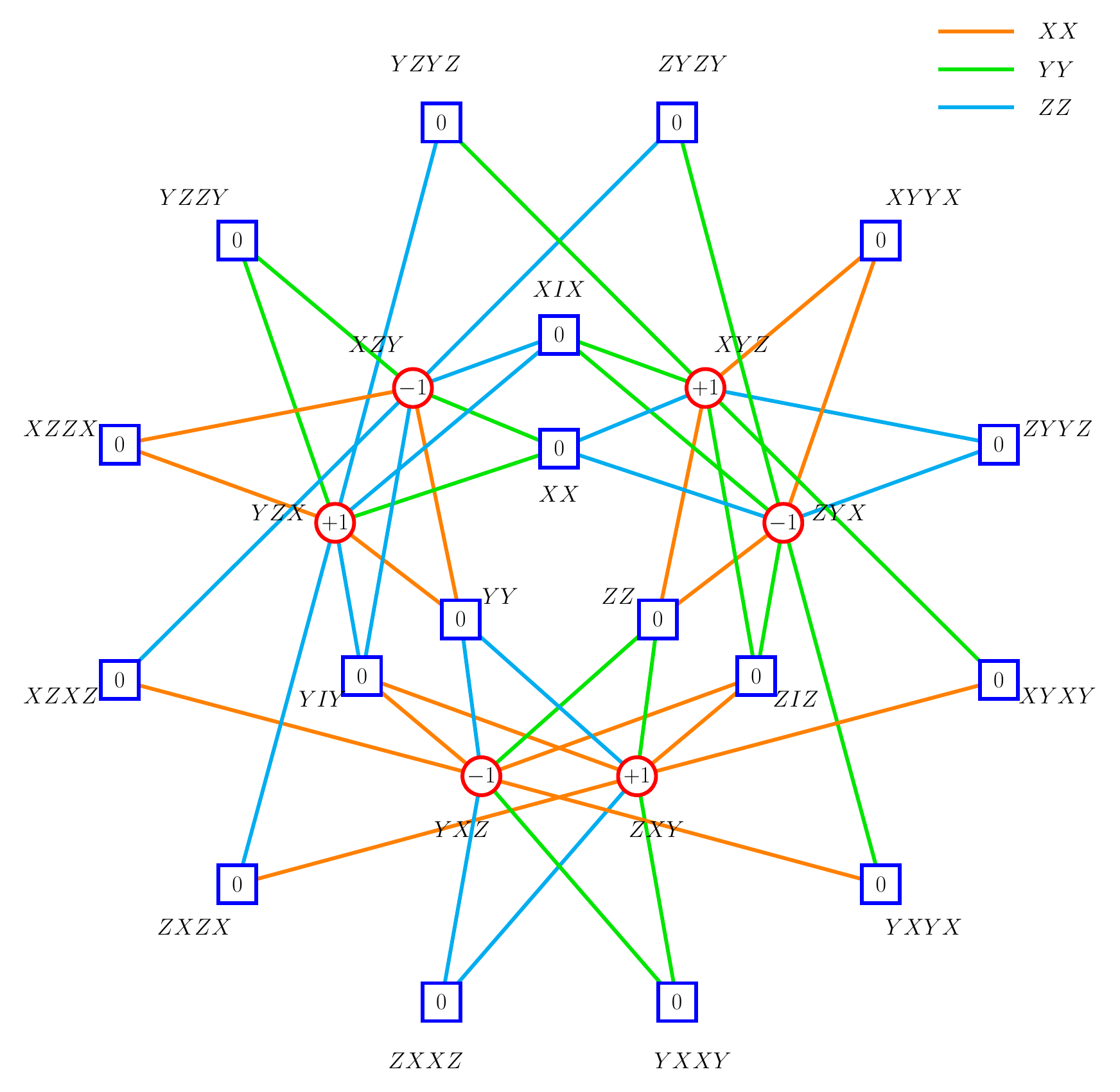}
  \caption{A connected component of the commutator graph for length $k=3$ quantities in $XYZ$ model, with $J_{XX}=J_{YY}=J_{ZZ}$. No vertices are omitted in this graph. The orange, green, and cyan lines represent the $J_{XX}, J_{YY},$ and $J_{ZZ}$ coefficients, respectively. Signs are ignored for clarity.}
  \label{fig:consquan_xyz}
\end{figure}

There are three notable points here. First, the doubling-product operators form a loop in the commutator graph, corresponding to the Exception 2 case discussed in the main text, and there are no other independent loop structures in the commutator graph. Second, in this case, the total degrees of freedom of coefficients we need to consider (i.e., the number of length-$k$ Pauli strings with independent coefficients) is $1$, which is constant $\mathcal{O}(1)$ and independent of the length of the quantity $k$. Third, if we consider the Hamiltonian with no magnetic field, i.e. $h=0$, this loop structure results in a conserved quantity. See Fig.\ref{fig:consquan_xyz} for an example of length-$3$ conserved quantity in the $XYZ$ model with $J_{XX}=J_{YY}=J_{ZZ}$, which exhibits this loop structure.

\begin{figure}[h]
\captionsetup{justification=centering}
\centering
\includegraphics[width=1\textwidth]{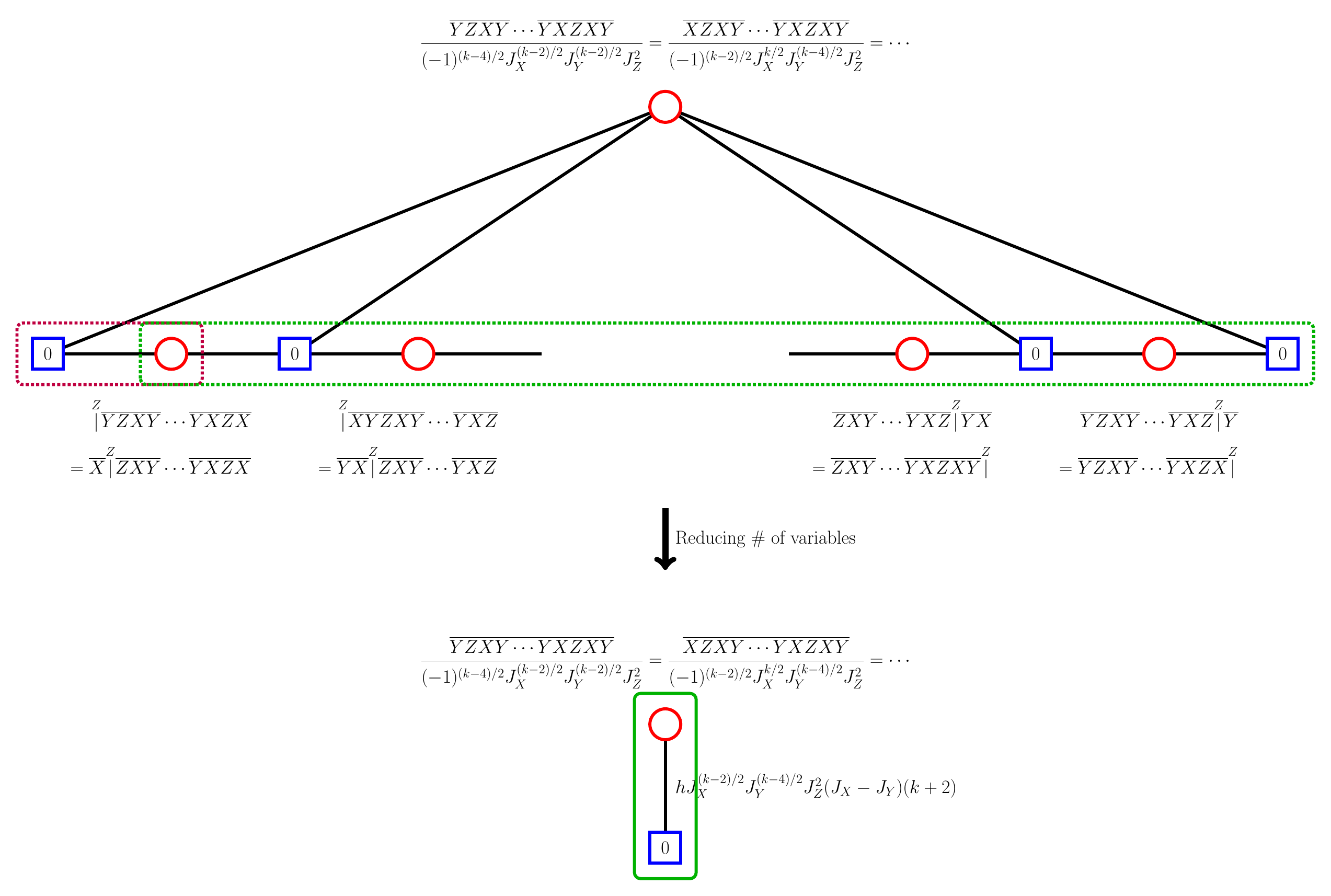}
  \caption{(Top Panel) Two quasi-promising paths in the commutator graph for length $k$ quantities, highlighted by two dotted boxes. (Bottom Panel) By applying the modification indicated by the fourth row in Fig.\ref{fig:graph_mod}, we obtain a subgraph where the $\protect\bsv$ vertex havs only one neighboring $\protect\rcv$ vertex with a non-zero edge coefficient(when $h\neq 0$ and $J_X\neq J_Y$ with $k\geq 6$). This demonstrates the coefficients of all doubling-product operators become zero.}
  \label{fig:appendix_xyz_5}
\end{figure}

To establish zero coeffieicnts for all length $k$ Pauli strings, we need to identify two quasi-promising paths that share a disturbing vertex and intersect at a single vertex. As discussed in the main text, the existence of such two quasi-promising paths generally implies that all vertices have zero coefficients. Specifically, the condition $h J_X^{(k-2)/2}J_Y^{(k-4)/2}J_Z^2 (J_X-J_Y)(k+2)\neq 0$ guarantees that these quasi-promising paths lead to zero coefficients, which holds when $J_X\neq J_Y$ and $k\geq 4$.

In contrast, if $h=0$, which corresponds to the integrable model\cite{nozawa2020explicit}, all edges connecting the disturbing $\rcv$ vertex and $\bsv$ vertices carry zero weight, meaning that the quasi-promising paths cannot be formed.

\begin{figure}[h]
\captionsetup{justification=centering}
\centering
\includegraphics[width=1\textwidth]{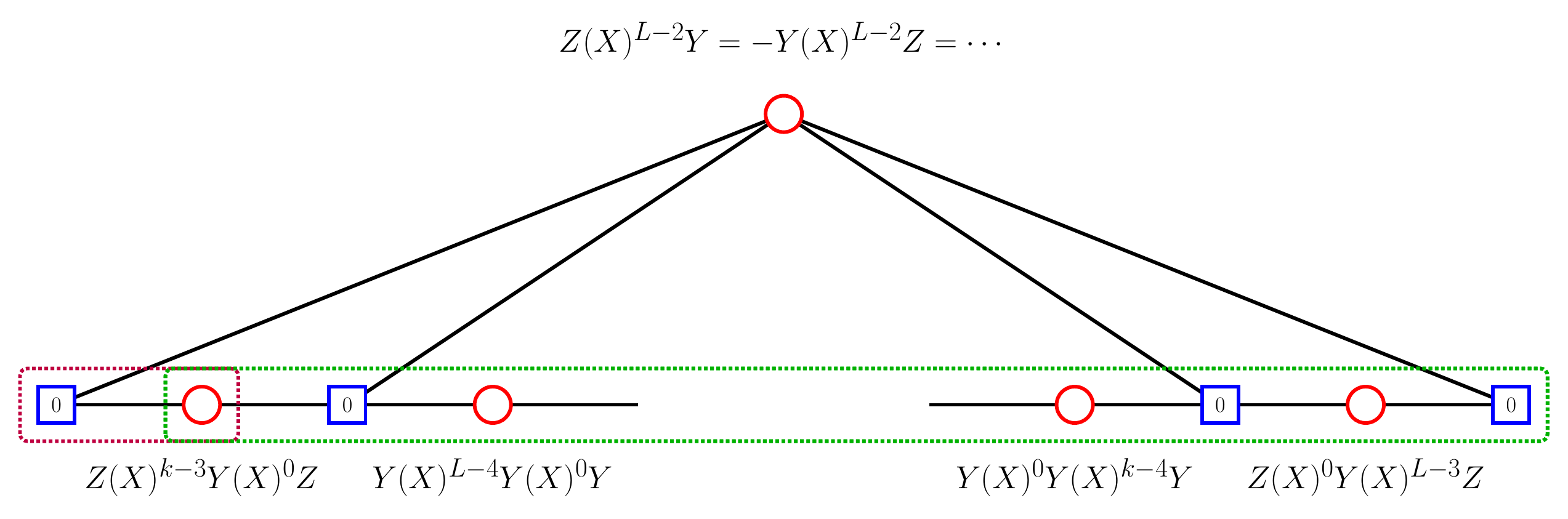}
  \caption{Graph-theoretical representation of part of the proof of nonintegrability of the mixed-field ising chain model in \cite{PhysRevB.109.035123}, illustrating quasi-promising paths.}
  \label{fig:appendix_xyz_5_2}
\end{figure}

It is worth noting that this quasi-promising path technique can also be applied to prove the nonintegrability of the mixed-field ising chain model\cite{PhysRevB.109.035123}, as in Fig.\ref{fig:appendix_xyz_5_2}, emphasizing the universality of the quasi-promising path as a tool for proving nonintegrability.

\bibliographystyle{plain}
\bibliography{biblio.bib}